\documentclass[journal]{IEEEtran}

\usepackage{cite}
\ifCLASSINFOpdf
   \usepackage[pdftex]{graphicx}
\else
   \usepackage[dvips]{graphicx}
\fi
\usepackage{amsmath,bm}
\allowdisplaybreaks[4]
\usepackage{algorithmic}

\usepackage{array}

\ifCLASSOPTIONcompsoc
  \usepackage[caption=false,font=normalsize,labelfont=sf,textfont=sf]{subfig}
\else
  \usepackage[caption=false,font=footnotesize]{subfig}
\fi
\usepackage{url}

\usepackage{amsthm}

\usepackage{amssymb}
\usepackage{mathtools}
\usepackage{cases}
\usepackage{autobreak}

\usepackage{booktabs}
\usepackage{multirow}

\usepackage{color}
\usepackage{hyperref}
\usepackage{cite}

\usepackage{amsmath}
\usepackage{extarrows}

\usepackage{comment}

\usepackage{algorithmic}
\usepackage{algorithm}

\newcommand{\E}{\mathbb{E}}

\newcommand{\cov}{\mathrm{Cov}}
\newcommand{\Var}{\mathrm{Var}}

\DeclareMathOperator{\Tr}{Tr}

\newcommand{\RNum}[1]{\uppercase\expandafter{\romannumeral #1\relax}}

\newtheorem{remark}{Remark}
\newtheorem{theorem}{Theorem}
\newtheorem{lemma}{Lemma}
\newtheorem{proposition}{Proposition}

\begin{document}

\title{Bias for the Trace of the Resolvent and Its Application on Non-Gaussian and Non-centered MIMO Channels}

\author{Xin~Zhang,~\IEEEmembership{Graduate Student Member,~IEEE} and S.H.~Song,~\IEEEmembership{Senior Member,~IEEE} 
\thanks{
This research was supported by HKUST Startup Fund (R9249).

X.~Zhang and S.H.~Song are with the Department
of Electronic and Computer Engineering, the Hong Kong University of Science and Technology. E-mail: xzhangfe@connect.ust.hk, eeshsong@ust.hk.

Copyright (c) 2017 IEEE. Personal use of this material is permitted.  However, permission to use this material for any other purposes must be obtained from the IEEE by sending a request to pubs-permissions@ieee.org.
}
}

\maketitle

\begin{abstract}
The mutual information (MI) of Gaussian multi-input multi-output (MIMO) channels has been evaluated by utilizing random matrix theory (RMT) 
and shown to asymptotically follow Gaussian distribution, where the ergodic mutual information (EMI) converges to a deterministic 
quantity. However, with non-Gaussian channels, there is a bias between the EMI and its deterministic equivalent (DE), whose evaluation is not available in the literature. This bias of the EMI is related to the bias for the trace of the resolvent in large RMT. In this paper, we first derive the bias for the trace of the resolvent, which is further extended to compute the bias for the linear spectral statistics (LSS). Then, we apply the above results on non-Gaussian MIMO channels to determine the bias for the EMI. It is also proved that the bias for the EMI is $-0.5$ times of that for the variance of the MI. Finally, the derived bias is utilized to modify the central limit theory (CLT) and calculate the outage probability. Numerical results show that the modified CLT significantly outperforms previous methods in approximating the distribution of the MI and improves the accuracy for the outage probability evaluation.
\end{abstract}
\begin{IEEEkeywords}
Mutual Information, Non-Gaussian MIMO Channel, Trace of the Resolvent, Random Matrix Theory.
\end{IEEEkeywords}

\section{Introduction}
\label{introduction}
\IEEEPARstart{B}y deploying a large number of antennas at both the transmitter and the receiver, large-scale multiple-input multiple-output (MIMO) systems can achieve high spectral efficiency and wide coverage in an energy-efficient way. However, the presence of the channel randomness and the large system scale bring challenges not only to system design but also performance analysis. The evaluation of mutual information (MI), $C(\sigma^{2})=\log \det(\frac{\bold{H}\bold{H}^{H}}{\sigma^2}+\bold{I})$, where $\sigma^2$ represents the power of the noise, is usually performed by numerical methods due to the difficulty in obtaining a closed-form solution except for some special cases~\cite{kamath2005asymptotic}. In particular, the difficulty comes from the lack of knowledge regarding the distribution of the eigenvalues for the Gram matrix $\bold{H}\bold{H}^{H}$. Fortunately, when the number of transmit and receive antennas grow to be large with the same pace, large random matrix theory (RMT) is a powerful tool and has been utilized to evaluate the MI with an explicit expression. In this paper, we will focus on evaluating the MI of large scale MIMO systems by RMT.

\subsection{Prior Art in Communications}
By utilizing RMT, there have been some works that focused on the evaluation of the MI per antenna, i.e., $I(\sigma^{2})=\frac{C(\sigma^{2})}{N}$ with $N$ denoting the number of receive antennas~\cite{hachem2007deterministic}~\cite{wen2012deterministic}~\cite{zhang2013capacity}~\cite{zhang2021large}, where the asymptotic equivalence of $I(\sigma^{2})$ with its deterministic equivalent (DE) was proved. On the other hand, it has also been shown that RMT could achieve good performance in approximating the ergodic mutual information (EMI), i.e., $\E C(\sigma^2)$, even when the system is not very large. A natural question is whether we can get rid of the factor $\frac{1}{N}$ safely and guarantee the convergence.

The problem was first investigated in~\cite{hachem2008new} under the correlated Gaussian MIMO channel, in which Hachem~\textit{et al.} proved the asymptotic Gaussianity, i.e. central limit theory (CLT) of $C(\sigma^{2})$, using Gaussian tools (integration by parts formula and the Poincar{\`e}-Nash inequality)~\cite{pastur2005simple}. It was shown that, different from $I(\sigma^{2})$, the almost sure convergence of $C(\sigma^{2})$ no longer holds, but $\E C(\sigma^{2})$ can still be approximated by a deterministic quantity $V(\sigma^{2})$ within the error of $O({M}^{-1})$. Rician (non-centered Gaussian) channel was considered in~\cite{dumont2010capacity}, where the authors proved that $\E C(\sigma^{2})$ converges to its DE with the same rate as that in~\cite{hachem2008new}. Then in~\cite{hachem2012clt}, Hachem~\textit{et al.} derived a CLT for non-centered channels using the Stieltjes transform method (or the Bai-Silverstein method)~\cite{bai2008clt}, which is applicable for non-Gaussian fading channels. It was indicated that the non-zero pseudo-variance and fourth-order cumulant will not only cause the bias between the EMI and its DE but also a gap between the variances of the MI for Gaussian and non-Gaussian fading channels, which will be referred to as the bias for the variance in this paper. The non-zero pseudo-variance comes from the non-circular property of the channel. Here, circularity means that $e^{\jmath \theta}X$ has the same distribution as $X$, where $X$ is a complex random variable and $\theta \in \mathbb{R} $ is deterministic. A typical non-circular case is the Hoyt distribution, which has been adopted for the modeling of cellular and satellite channels~\cite{fraidenraich2007mimo}. On the other hand, the non-zero fourth-order cumulant is common in severe fading channels. In~\cite{kammoun2010fluctuations}, the effect of non-zero fourth-order cumulant on the asymptotic distribution of the MI was investigated. The deterministic approximation of the EMI and the analysis for the fluctuation of the MI over non-Gaussian channels have also attracted some attention. In~\cite{bao2015asymptotic}, Bao~\textit{et al.} derived the CLT for the MI of the i.i.d. (independent and identically distributed) channel with non-zero pseudo-variance and fourth-order cumulant. In~\cite{hu2019central}, Hu~\textit{et al.} investigated the CLT for the MI of an elliptically correlated (EC) channel and validated the bias arising from non-Gaussianity with  non-linear correlations. In~\cite{kammoun2012fluctuations}, the CLT for the  signal-to-interference-plus-noise-ratio (SINR) at the linear Wiener receiver over non-centered and non-Gaussian channels was derived. However, the bias for the EMI of non-centered and non-Gaussian MIMO channels is not available in the literature. 

\subsection{Prior Art in RMT}
The MI of non-Gaussian MIMO channels discussed above is a special case of the linear spectral statistics (LSS) of random matrices in RMT, when the function takes the form $f(x)=\log(1 + x/\sigma^2)$~\cite{bao2015asymptotic}~\cite{hu2019central}. On the bias for the LSS, researchers have made some progress for the centered case. In~\cite{bao2015asymptotic}, by following a similar approach as that in~\cite{bai2008clt}, Bao~\textit{et al.} established a CLT of the LSS for sample covariance matrices with complex i.i.d. entries under the non-zero pseudo-variance assumption. In~\cite{najim2016gaussian}, Najim~\textit{et al.} investigated the CLT for the LSS of the centered and correlated matrix, and derived the bias for the mean and variance. For the non-centered case, Banna~\textit{et al.}~\cite{banna2020clt} investigated the CLT for the information-plus-noise matrix, determined the bias for the variance, but left the bias for the mean as a computationally-challenging task. By far, the bias for the mean of the non-centered matrix has not been investigated in the literature, which is related to the missing bias for the EMI in the communication community.

Given the bias for the LSS can be obtained by Cauchy's integral formula with respect to the bias for the trace of the resolvent~\cite{bai2008clt}~\cite{bao2015asymptotic}, we will first evaluate the latter in this paper. Specifically, the bias for the trace of the resolvent is derived and then generalized to compute the bias for the LSS of non-centered random matrices. This resolved the problem from the perspective of RMT. Then, the result is utilized to derive the bias for the EMI of MIMO channels, solving the problem in the communication community. The derived bias is utilized to modify the CLT and calculate the outage probability for MIMO systems. Furthermore, the relation between the bias for the mean and that for the variance is investigated and we prove that the former is $-0.5$ times of the latter. Numerical simulations are performed with different non-Gaussian channel models. The results validate that the bias of the EMI does not vanish when $N$ goes to infinity and confirm the relation between the biases for the mean and variance. Furthermore, it is shown that the CLT, modified by the derived biases, outperforms other methods in approximating the cumulative distribution function (CDF) of the MI.

\subsection{Contributions}
The main contributions of this paper are summarized as follows.

\begin{itemize}
  \item [1)] 
\textbf{RMT contribution}: We derive an approximation of the bias for the resolvent of non-centered random matrices. This result is a necessary complement for the CLT of the LSS for sample covariance matrices of non-centered cases~\cite{banna2020clt}. We also show that the bias for the trace of the resolvent can be written in a derivative form, which is referred to as the alternative expression and contributes to the computation for the LSS. Our results are of great meaning in evaluating the expectation of functional spectral of non-centered random matrices.

 \item [2)] 
 \textbf{Communication contribution}: By applying the result in 1) to MIMO systems, we derive an explicit expression for the bias of the EMI, which complements for the CLT in~\cite{hachem2012clt} and takes previous results~\cite{hachem2012clt}\cite{kammoun2010fluctuations}\cite{bao2015asymptotic} as special cases. The result is then utilized to approximate the distribution of the MI and compute the outage probability. Numerical results show that the modified CLT outperforms previous results. Based on the bias for the resolvent, we also prove that the bias for the mean of the MI is $-0.5$ times of that for the variance. With the modified mean and variance, the  distribution of the MI and the outage probability are approximated with higher accuracy. 
\end{itemize}

\subsection{Paper Outline and Notations}

The rest of this paper is organized as follows. In Section~\ref{sys_model}, we introduce the system model and connect the bias for the EMI 
with the bias for the trace of the resolvent. In Section~\ref{pre_res}, we present the preliminary results in RMT that will be utilized 
in the derivation of this paper. The main mathematical result regarding the bias for the resolvent, which involves two expressions, and its extension to compute the LSS are given in Section~\ref{main_res}. Then, in Section~\ref{main_res_2}, we apply the above result on non-Gaussian MIMO channels to derive the bias for the EMI and determine the relation between the bias for the EMI and that for the variance. The derived bias is also utilized 
to modify the CLT and calculate the outage probability. In Section~\ref{simu}, we perform extensive numerical experiments
to validate the accuracy of the derived bias and demonstrate the better fitness of the modified CLT. Section~\ref{conc} concludes the paper.

We use the boldface upper case letters to represent the matrix such as $\bold{X}$, and its entry at the $i$-th row and the $j$-th column is denoted by $x_{ij}$. The boldface lower case letters represent the column vectors. $(\cdot)^{T}$ and $(\cdot)^{H}$ denote the transpose and Hermitian transpose, respectively. $\Tr(\cdot)$ represents the trace operator and $\overline{(\cdot)}$ denotes the conjugate operator. $\mathrm{diag}(\bold{a})$ represents the diagonal matrix whose diagonal entries are elements of vector $\bold{a}$. $\mathrm{vdiag}(\bold{A})=(A_{11}, A_{22},...,A_{NN})^{T}$, where $\bold{A}$ is an $N$ by $N$ matrix. $\|\bold{A}\|$ denotes the spectral norm of $\bold{A}$. $\E (\cdot)$ represents the expectation operator and $\E^{\frac{1}{2}}(\cdot)$ is equivalent to $\sqrt{\E(\cdot)}$. For brevity of formulars, we assume that the expectation operator has the lowest operation priority and ignore the $()$ over the random variables (RVs), i.e., $\E XY=\E(XY)$. $\Var(\cdot)$ represents the variance and $\Var^{\frac{1}{2}}(\cdot)=\sqrt{\Var(\cdot)}$. $\mathcal{N}(\mu, \sigma^2)$ and $\mathcal{CN}(\mu, \sigma^2)$ denote the Gaussian distribution and circularly complex Gaussian distribution, respectively, with mean $\mu$ and variance $\sigma^2$. $\mathbb{C}$ and $\mathbb{R}^{+}$ represent the set of complex numbers and non-negative real numbers, respectively. $\mathbb{I}(\cdot)$ and $\log(\cdot)$ denote the indicator function and the natural logarithm function, respectively. $ \overset{\mathcal{D}}{\longrightarrow } $ represents convergence in distribution and $\jmath=\sqrt{-1}$. $O(\cdot)$ and $o(\cdot)$ denote the big-$O$ and little-$o$ notations, respectively.

\section{System Model and Problem Formulation}
\label{sys_model}
\subsection{System Model}
Consider a point-to-point MIMO system with $M$ antennas at the transmitter and $N$ antennas at the receiver. 
The received signal $\bold{y}\in \mathbb{C}^{N}$ can be given by
\begin{equation}
\bold{y}=\bold{H}\bold{s}+\bold{n},
\end{equation}
where $\bold{s}\in \mathbb{C}^{M}$ represents the transmitted signal, 
$\bold{H}$ denotes the $N$ by $M$ channel matrix, and $\bold{n}$ is the additive white Gaussian noise (AWGN), whose entries are i.i.d. circular Gaussian random variables with variance $\sigma^2$, i.e. $\E\bold{n}\bold{n}^{H}=\sigma^2\bold{I}$. MI is an essential performance metric. However, its evaluation is challenging and thus attracts great interests~\cite{bao2015asymptotic}~\cite{hu2019central}. Under the assumption $\E\bold{s}\bold{s}^{H}=\bold{I}$, the MI of the concerned MIMO system is given by
\begin{equation}
\label{M_I}
C_{\bold{H}\bold{H}^{H}}(\sigma^2) = \log \det \left(\bold{I}+\frac{\bold{H}\bold{H}^{H}}{\sigma^2} \right),
\end{equation}
which is a random variable due to the randomness of the channel matrix $\bold{H}$. This randomness motivates us to investigate the expectation and fluctuation of $C_{\bold{H}\bold{H}^{H}}(\sigma^2)$, which can be used to compute the average throughput and outage probability.

\subsection{Channel Model}
\label{cha_mod}
In this paper, we consider a non-centered and non-Gaussian channel model with independent antennas, where the channel matrix is given by 
\begin{equation}
\label{matrix_model_a}
\bold{H}=\underbrace{\bold{A}}_\text{LoS component}+\underbrace{\frac{1}{\sqrt{M}}\bold{D}^{\frac{1}{2}}\bold{X}\widetilde{\bold{D}}^{\frac{1}{2}}}_\text{non-LoS component}.
\end{equation}
Here $\bold{A}=\left[a_{ij}\right]$ is a $N$ by $M$ deterministic matrix denoting the LoS component of Rician channel. $\bold{D}$ and $\widetilde{\bold{D}}$ are two deterministic diagonal matrices, $\bold{D}=\mathrm{diag}(d_{1},d_{2},...,d_{N})$, $\widetilde{\bold{D}}=\mathrm{diag}(\widetilde{d}_{1},\widetilde{d}_{2},...,\widetilde{d}_{M})$, representing the variance profile, i.e., $\Var(h_{ij})=\sigma_{ij}^2=d_{i}\widetilde{d}_{j}$. This non-centered matrix $\bold{H}$ is also called signal-plus-noise model~\cite{banna2020clt} when $\bold{D}=\psi^2 \bold{I}$ and $\widetilde{\bold{D}}=\psi^2\bold{I}$ with $\psi \in \mathbb{R}$. The non-identical diagonal entries can be utilized to model the antenna power imbalance~\cite{levin2011multi} or the non-identical channel gain in distributed antenna systems~\cite{wen2012deterministic}~\cite{zhang2013capacity}~\cite{hoydis2010fluctuations}. Note that, with the model considered in~(\ref{matrix_model_a}), the variance profile is separable, i.e., $\sigma_{ij}^2=d_{i}\widetilde{d}_{j}$. The method derived in this paper can also be revised to handle the non-separable case~\cite{hachem2007deterministic}~\cite{kammoun2009central}, in which $\sigma_{ij}^2=d_{i}\widetilde{d}_{j}$ does not hold.

\subsection{Mathematical Formulation}
In the following, we will investigate the evaluation of the MI, $C_{\bold{H}\bold{H}^{H}}(\sigma^2)$, given in~(\ref{M_I}). Denoting the Gram matrix $\bold{B}=\bold{H}\bold{H}^{H}$, the MI can be written as~\cite{hachem2007deterministic}~\cite{dumont2010capacity}
\begin{equation}
\begin{aligned}
\label{c_exp}
 \E C_{\bold{H}\bold{H}^{H}}(\sigma^2) &= \int_{\sigma^2}^{\infty}\frac{N}{\omega}- \E \Tr (\bold{B}+\omega\bold{I})^{-1}  \mathrm{d}\omega
\\
&= \int_{-\infty}^{-\sigma^2}-\frac{N}{z}- \E \Tr (\bold{B}-z\bold{I})^{-1}  \mathrm{d}z.
\end{aligned}
\end{equation}
Therefore, we will focus on the evaluation of the quantity $\E \Tr (\bold{B}-z\bold{I})^{-1}$. When $N$ is large, the evaluation is related to the limiting spectral distribution (LSD) of $\bold{B}$, which is the limit of the empirical spectral distribution (ESD). The ESD of $\bold{B}$ is denoted by
 \begin{equation}
 F_{\bold{B}}(x)=\frac{1}{N}\sum_{i=1}^{N} \mathbb{I}(\lambda_{i}\le x) ,
\end{equation}
and there holds
\begin{equation}
\begin{aligned}
&\Tr(\bold{B}-z\bold{I})^{-1}=N\frac{1}{N}\Tr(\bold{B}-z\bold{I})^{-1}
\\
&= N\int_{\mathbb{R}^{+}}\frac{1}{x-z} \mathrm{d} F_{\bold{B}}(x)=N m_{\bold{B}}(z),
\end{aligned}
\end{equation}
where $m_{\bold{B}}(z)=\int_{\mathbb{R}^{+}}\frac{1}{x-z} \mathrm{d} F_{\bold{B}}(x)$ is the \textit{Stieltjes Transform} of $\bold{B}$ and the matrix $\bold{Q}(z)=\left(\bold{B}-z\bold{I}\right)^{-1}$ is the \textit{resolvent} of $\bold{B}$. The \textit{co-resolvent} of $\bold{B}$ is defined as $\widetilde{\bold{Q}}(z)=\left(\bold{H}^{H}\bold{H}-z\bold{I}\right)^{-1}=\left(\widetilde{\bold{B}}-z\bold{I}\right)^{-1}$~\cite{hachem2012clt}~\cite{hachem2013bilinear}, whose {Stieltjes Transform} is $m_{\widetilde{\bold{B}}}(z)$. It is well known that the convergence holds for the \textit{Stieltjes Transform}~\cite{hachem2007deterministic}~\cite{hachem2012clt},
\begin{equation}
\begin{aligned}
\E m_{\bold{B}}(z) \xrightarrow{M\rightarrow \infty}  \frac{1}{N}\Tr\bold{T}(z),
\\
\E m_{\widetilde{\bold{B}}}(z) \xrightarrow{M\rightarrow \infty}  \frac{1}{M}\Tr\widetilde{\bold{T}}(z),
\end{aligned}
\end{equation}
where $~z\in\mathbb{C} \setminus \mathbb{R}^{+}$. Here, $\bold{T}(z)$ and $\widetilde{\bold{T}}(z)$ are the deterministic approximation of $\E\bold{Q}(z)$ and $\E\widetilde{\bold{Q}}(z)$, respectively, which will be introduced later in~(\ref{basic_eq}).

However, the above result only suffices to give an approximation of the MI per antenna, i.e. $\frac{1}{N}C_{\bold{H}\bold{H}^{H}}(\sigma^2)$. To evaluate $C_{\bold{H}\bold{H}^{H}}(\sigma^2)$, it is natural to investigate the convergence for the trace of the resolvent $\E\Tr \bold{Q}(z)$. It is known that $\frac{1}{N}\Tr\bold{T}(z)$ is a good approximation of $\frac{1}{N}\Tr \bold{Q}(z)$~\cite{hachem2007deterministic}, i.e., when the entries $x_{ij}$s are circularly Gaussian, $\E\Tr \bold{Q}(z) - \Tr\bold{T}(z)\xlongrightarrow[]{M\rightarrow \infty} 0$. However, if the pseudo-variance $\E x^2_{ij}$ or the fourth-order cumulants are non-zero, $\E\Tr \bold{Q}(z) -\Tr\bold{T}(z)\xlongrightarrow[]{M\rightarrow \infty} \mathcal{B}(z)$, which indicates that there will be a bias between $\E C(\sigma^2)$ and its approximation based on $\bold{T}(z)$. In practical systems, $\E x^2_{ij}$ or the fourth-order cumulants are not always zero~\cite{bao2015asymptotic}. The bias term $\mathcal{B}(z)$ for the centered channel ($\bold{A}=\bold{0}$) has been investigated in~\cite{hachem2012clt}~\cite{bao2015asymptotic}~\cite{najim2016gaussian},  but the expression of $\mathcal{B}(z)$ for the non-centered case is not available in literature and will be the focus of this paper. This bias is useful in evaluating the bias of the EMI caused by non-Gaussianity. Next, we will investigate the bias for the trace of the resolvent
\begin{equation}
\mathcal{M}(z)=\E\Tr \bold{Q}(z)-\Tr \bold{T}(z),
\end{equation}
which will then be utilized to evaluate the bias of the EMI for MIMO channels.

\section{Preliminary Results}
\label{pre_res}
To better explain the main results, we first introduce some useful results in RMT that will be utilized in this paper.

\subsection{Assumptions}

\textbf{Assumption A.1} The dimensions $N$ and $M$ go to infinity at the same pace, i.e., $M \rightarrow \infty$,
\begin{equation}
0 < \lim\inf \frac{N}{M} \le c=\frac{N}{M} \le \lim\sup \frac{N}{M} < \infty.
\end{equation}

\textbf{Assumption A.2} The entries of $\bold{X}$ are i.i.d. and $x_{ij}~(1\le i \le N, 1\le j \le M  )$ satisifes
\begin{equation}
\E x_{ij}=0, ~~\E |x_{ij}|^2=1, ~~\E |x_{ij}|^{16} < \infty.
\end{equation}

\textbf{Assumption A.3} The deterministic non-centered component of the channel, $\bold{A}$, has finite spectral norm,
\begin{equation}
\sup_{M \ge 1}\|\bold{A}\| < \infty.
\end{equation}

\textbf{Assumption A.4} The family of deterministic diagonal matrices $\bold{D}$ and $\widetilde{\bold{D}}$ have non-negative entries such that
\begin{equation}
\begin{aligned}
d_{max}&\!=\! \sup_{M \ge 1}\|\bold{D}\| < \infty, ~\widetilde{d}_{max}\!=\! \sup_{M \ge 1}\|\widetilde{\bold{D}}\| < \infty,
\\
d_{min}& \! = \! \inf_{M \ge 1}\frac{1}{M}\Tr \bold{D} > 0, ~ \widetilde{d}_{min}\!=\!\inf_{M \ge 1}\frac{1}{M}\Tr \widetilde{\bold{D}} > 0.
\end{aligned}
\end{equation}
\textbf{A.1} is the asymptotic regime considered for the large-scale system. In this regime, $N$ and $M$ grow to infinity with the same pace so $M \rightarrow \infty $ and $N\rightarrow \infty$ are equivalent. \textbf{A.2} is a general fading constraint on moments instead of distribution, which is well-satisfied by most models like Rician and Hoyt. \textbf{A.3} implies that the columns of $\bold{A}$, i.e., $\bold{a}_{i}$'s, are uniformly bounded in $N$~\cite{kammoun2019asymptotic}~\cite{sanguinetti2018theoretical} and indicates that the rank of the LoS component $\bold{A}$ increases with the number of antennas at the same pace. Although the rank-one LoS is assumed in some works~\cite{jin2007ergodic}~\cite{bolcskei2003impact}, there are scenarios where \textbf{A.3} holds, e.g., short range communications~\cite{bohagen2009spherical} and distributed antenna systems~\cite{wen2012deterministic}~\cite{zhang2013capacity}~\cite{dumont2006cth09}. Specifically, in distributed antenna systems, the antennas of the mobile are collocated but those of the base station (BS) are distant from each other such that the LoS components between each distributed antenna and the mobile antenna arrays are different, which results in a full-rank $\bold{A}$ with high probability~\cite{dumont2006cth09}. \textbf{A.4} implies that the antenna imbalance is finite and the spatial dimensions increase with $N$.

\subsection{Moments Notations}
For ease of illustration, we will use the following symbols to represent some important quantities.
Let $\vartheta$ denote the pseudo-variance of $x_{ij}$, $\kappa$ represent the fourth-order cumulant, and $\zeta$ be the crossed third-order moment, respectively, with
\begin{equation}
\label{moments}
\vartheta=\E x_{ij}^2,~\kappa=\E |x_{ij}|^4-|\vartheta|^2-2,~\zeta=\E |x_{ij}|^2 x_{ij}.
\end{equation}
As mentioned in Section~\ref{introduction}, $\vartheta$ originates from the non-circularity of the channel and it holds true that $0 \le |\vartheta| \le 1 $.

\subsection{Deterministic Approximation}
To approximate $\E \Tr \bold{Q}(z) $, we construct the deterministic approximation of the resolvent $\bold{Q}(z) $ and the co-resolvent $\widetilde{\bold{Q}}(z) $, and denote them by ${\bold{T}}(z)$ and $\widetilde{\bold{T}}(z)$, respectively. In fact, the problem was first resolved in~\cite{hachem2007deterministic} and the approximations were constructed by a system of canonical fixed-point equations. This method originated from~\cite{girko2001theory} and has been widely used in the RMT area~\cite{hachem2007deterministic}~\cite{wen2012deterministic}~\cite{pastur2005simple} for various structures of random matrices. For example, the centered case with $\bold{H}=\bold{R}_{1}^{\frac{1}{2}}\bold{X}\bold{R}_{2}^{\frac{1}{2}}$ was investigated in~\cite{hachem2008new}, where $\bold{R}_{1}$ and $\bold{R}_{2}$ are two positive semi-definite matrices and the approximation of the non-centered case with $\bold{H}=\bold{A}+\bold{R}_{1}^{\frac{1}{2}}\bold{X}\bold{R}_{2}^{\frac{1}{2}}$ was given in~\cite{dumont2010capacity}, both under the assumption that the entries of $\bold{X}$ are i.i.d. and Gaussian distributed. In this paper, we will consider the channel model introduced in~(\ref{matrix_model_a}), where the entries may not be Gaussian. We now introduce the fundamental system of equations to give an approximation for the resolvent $\bold{Q}(z)$.
Given $z\in \mathbb{C} \setminus  \mathbb{R}^{+}$, for $(\delta,\widetilde{\delta})$ and matrices $\bold{T}(z), \widetilde{\bold{T}}(z)$ satisfying the following system of equations,
\begin{equation}
\label{basic_eq}
\left\{
\begin{aligned}
&\delta=\frac{1}{M}\Tr \bold{D}\bold{T}(z)\equiv f(\widetilde{\delta},z), 
\\
 &\widetilde{\delta}=\frac{1}{M}\Tr \widetilde{\bold{D}}\widetilde{\bold{T}}(z)\equiv \widetilde{f}(\delta,z),
\\
&\bold{T}(z)=\left(-z\left( \bold{I}+\widetilde{\delta}\bold{D} \right)+\bold{A}\left( \bold{I}+\delta\widetilde{\bold{D}} \right)^{-1}\bold{A}^{H} \right)^{-1}\!,
\\
&\widetilde{\bold{T}}(z)=\left(-z\left( \bold{I}+{\delta}\widetilde{\bold{D}}\right)+\bold{A}^{H}\left( \bold{I}+\widetilde{\delta}{\bold{D}} \right)^{-1}\bold{A} \right)^{-1}\!,
\end{aligned}
\right.
\end{equation}
there holds 
\begin{equation}
\begin{aligned}
\label{basic_app}
\frac{1}{M}\Tr(\E \bold{Q}(z)  -  \bold{T}(z)) \xlongrightarrow[]{M\xrightarrow \infty }  0,
\\
\frac{1}{M}\Tr(\E \widetilde{\bold{Q}}(z) - \widetilde{\bold{T}}(z))  \xlongrightarrow[]{M\xrightarrow \infty }  0.
\end{aligned}
\end{equation}
\begin{remark} (Solution for the fundamental equations) The existence and uniqueness of the solution for $(\delta, \widetilde{\delta})$ in~(\ref{basic_eq}) have been proved in~\cite{hachem2007deterministic} and can be obtained by Algorithm~\ref{sol_fund}. Then, given $(\delta, \widetilde{\delta})$, we can obtain $\bold{T}(z)$ and $\widetilde{\bold{T}}(z)$.
\begin{algorithm} 
\caption{ Fixed-point algorithm for $(\delta, \widetilde{\delta})$  } 
\label{sol_fund} 
\begin{algorithmic}[1] 
\REQUIRE  $z$, $\bold{A}$, $\bold{D}$, $\widetilde{\bold{D}}$ , $\delta^{(0)}>0$, $\widetilde{\delta}^{(0)}>0$ and set $t=1$.
\REPEAT
\STATE Compute $\delta^{(t)}$ from $\widetilde{\delta}^{(t-1)}$ by $\delta^{(t)}=f(\widetilde{\delta}^{(t-1)},z)$.
\STATE Compute $\widetilde{\delta}^{(t)}$ from ${\delta}^{(t-1)}$ by $\widetilde{\delta}^{(t)}=\widetilde{f}(\delta^{(t-1)},z)$.
\STATE $t \leftarrow  t+1$
\UNTIL Convergence.
\ENSURE  $(\delta, \widetilde{\delta})$.
\end{algorithmic}
\end{algorithm}

\end{remark}

\begin{remark}
(Convergence of the resolvent) The approximation in~(\ref{basic_app}) is a special case of Theorem 2.5 in~\cite{hachem2007deterministic} when the variance profile is separable, and $x_{ij}$s are i.i.d. with finite $4+\varepsilon$ moment. This deterministic equivalent was also derived in~\cite{dumont2010capacity} when the entries are i.i.d. circular complex Gaussian random variables. From Theorem 2 in~\cite{dumont2010capacity}, the following convergence holds true for any given $\bold{U}$ and $\widetilde{\bold{U}}$ with bounded norm:
\begin{equation}
\begin{aligned}
\Tr\bold{U}(\E \bold{Q}(z)-\bold{T}(z)) =O({M}^{-1}),
\\
\Tr\widetilde{\bold{U}}(\E \widetilde{\bold{Q}}(z)-\widetilde{\bold{T}}(z)) =O(M^{-1}).
\end{aligned}
\end{equation}
We can notice that Gaussianity guarantees faster convergence of the Stieltjes transform, i.e., $O(M^{-2})$, when compared with the bound~(\ref{l3_1}) of Lemma~\ref{bound_lemma} in Appendix~\ref{rmt_results}, i.e., $O(M^{-1})$. The integration by parts formula (remark 2.2 in~\cite{pastur2005simple}) and Poincar{\`e}-Nash inequality (Proposition 2.4 in~\cite{pastur2005simple}) for Gaussian random matrices, which are referred to as the Gaussian tools, can be used to prove the $O({M}^{-1})$ convergence speed for the trace of the resolvent. This means that the deterministic approximation $\bold{T}(z)$ is very accurate such that there is no asymptotic bias when the entries are Gaussian. However, if the entries are not Gaussian, it was pointed out in~\cite{najim2016gaussian} that there is a bias related to $\vartheta$ and $\kappa$ when $\bold{H}=\bold{R}_{1}^{\frac{1}{2}}\bold{X}$, where $\bold{R}_{1}$ is a nonnegative definite Hermitian matrix. Similar structure is also presented in~\cite{bao2015asymptotic}~\cite{najim2016gaussian}~\cite{banna2020clt}. However, the expression of the bias with the non-centered and non-Gaussian $\bold{H}$ is still not available in the literature. It was posed as a computationally challenging task in~\cite{banna2020clt}, and will be one of the main contributions of this paper. This contribution is meaningful from the perspectives of both RMT and communication theory. 
\end{remark}

In the following sections, for notational convenience, we will drop the dependencies on $z$ in matrices and use $\omega=-z$ for simplicity. As there will be many complex matrix expressions, some frequently used ones are defined and listed in Table~\ref{var_list} for ease of illustration. Note that with $(\delta, \widetilde{\delta})$ resolved by Algorithm~\ref{sol_fund}, we can compute all the quantities in Table~\ref{var_list}. We now present some important identities of $\bold{T}$ that will be useful for our derivation. From the Sherman-Morrison-Woodbury formula~\cite{horn2012matrix} with respect to the inverse of the perturbed matrix, i.e.,
\begin{equation}
\begin{aligned}
&\left(\bold{A}+\bold{X}\bold{R}\bold{Y}  \right)^{-1}
\\
&=\bold{A}^{-1}-\bold{A}^{-1}\bold{X}\left( \bold{R}^{-1}+\bold{Y}\bold{A}^{-1}\bold{X}\right)^{-1}\bold{Y}\bold{A}^{-1},
\end{aligned}
\end{equation}
we can derive 
\begin{equation}
\begin{aligned}
\label{replace_R}
\widetilde{\bold{T}}
=-z^{-1}\widetilde{\bold{R}}+z^{-1} \widetilde{\bold{R}}\bold{A}^{H}\bold{T}\bold{A} \widetilde{\bold{R}},
\end{aligned}
\end{equation}
and a subsequent result follows
\begin{equation}
\begin{aligned}
\label{replace_tilde_t}
\bold{T}\bold{A}\left(\bold{I}+\delta\widetilde{\bold{D}} \right)^{-1}=\left(\bold{I}+\widetilde{\delta}\bold{D} \right)^{-1}\bold{A}\widetilde{\bold{T}},
\\
\left(\bold{I}+\delta\widetilde{\bold{D}} \right)^{-1}\bold{A}^{H}\bold{T}=\widetilde{\bold{T}}\bold{A}^{H}\left(\bold{I}+\widetilde{\delta}\bold{D} \right)^{-1}.
\end{aligned}
\end{equation}
Furthermore, according to~(\ref{replace_tilde_t}), we have the following identity
\begin{equation}
\begin{aligned}
\label{f_eq}
{F}\!=\! \frac{\Tr \widetilde{\bold{D}}\widetilde{\bold{T}}\bold{A}^{H}{\bold{D}}\bold{R}^{2}\bold{A}\widetilde{\bold{T}}}{M}\!=\!\frac{\Tr \bold{D}\bold{T}\bold{A}\widetilde{\bold{D}}\widetilde{\bold{R}}^2\bold{A}^{H}\bold{T}}{M}.
\end{aligned}
\end{equation}
Some prior RMT results will be utilized in the proof and we put them in Appendix~\ref{rmt_results}.

 \begin{table*}[!htbp]
\centering
\caption{List Of Frequently Used Expressions}
\label{var_list}
\begin{tabular}{|c|c|c|c|c|c|}
\toprule
Symbol& Expression &  Symbol& Expression &Symbol& Expression \\
\midrule
$\bold{R}$ & $(\bold{I}+\widetilde{\delta}\bold{D})^{-1}$
& $\widetilde{\bold{R}}$ & $(\bold{I}+\delta \widetilde{\bold{D}})^{-1}$
& $\bold{S}$& $\mathrm{diag}(t_{11}, t_{22},..., t_{NN})$
\\
$\widetilde{\bold{S}}$ & $\mathrm{diag}(\widetilde{t}_{11}, \widetilde{t}_{22},..., \widetilde{t}_{MM})$
& $\gamma$ & $\frac{1}{M} \Tr \bold{D}  \bold{T} \bold{D}  \bold{T}$ & $\gamma_{T}$ 
& $\frac{1}{M} \Tr \bold{D}  \bold{T} \bold{D}  \bold{T}^{T}$   
\\
$\widetilde{\gamma}$ &  $\frac{1}{M}  \Tr \widetilde{\bold{D}}  \widetilde{\bold{T}} \widetilde{\bold{D}}  \widetilde{\bold{T}}$ 
& $\widetilde{\gamma}_{T}$ &  $\frac{1}{M}  \Tr \widetilde{\bold{D}}  \widetilde{\bold{T}} \widetilde{\bold{D}}  \widetilde{\bold{T}}^{T}$
& $\gamma( \bold{U})$ & $\frac{1}{M}  \Tr \bold{U} \bold{T} \bold{D}  \bold{T}$
\\
$\gamma_{T}( \bold{U})$ & $\frac{1}{M}  \Tr \bold{D} \bold{T} \bold{U}  \bold{T}^{T}$
& $\eta $ & $\frac{1}{M}  \Tr\bold{S}^2\bold{D}^2 $
& $\widetilde{\eta} $ & $\frac{1}{M}\Tr\widetilde{\bold{S}}^2\widetilde{\bold{D}}^2 $
\\
$F$ & $\frac{1}{M}  \Tr \bold{D}  \bold{T} \bold{A}   \bold{\widetilde{R}}^{2}  \bold{\widetilde{D}} \bold{A}^{H}\bold{T}$
& $F_{T}$ & $\frac{1}{M}  \Tr \bold{D}  \bold{T}^{T} \overline{\bold{A}}   \bold{\widetilde{R}}^{2}  \bold{\widetilde{D}} \bold{A}^{H}\bold{T}$
& $\underline{F}_{T}$ & $\frac{1}{M}  \Tr \bold{D}  \bold{T} {\bold{A}}   \bold{\widetilde{R}}^{2}  \bold{\widetilde{D}} \bold{A}^{T}\bold{T}^{T}$ 
\\
$\widetilde{F}_{T}$ &  $\frac{1}{M}  \Tr \widetilde{\bold{D}}  \widetilde{\bold{T}}^{T} {\bold{A}}^{T}   \bold{{R}}^{2}  \bold{{D}} {\bold{A}}\widetilde{\bold{T}}$
&
$\underline{\widetilde{F}}_{T}$ & $\frac{1}{M}  \Tr \widetilde{\bold{D}}  \widetilde{\bold{T}} {\bold{A}}^{H}   \bold{{R}}^{2}  \bold{{D}} \overline{\bold{A}}\widetilde{\bold{T}}^{T}$
& 
$F(\bold{U})$ & $\frac{1}{M}  \Tr \bold{U}  \bold{T} \bold{A}   \bold{\widetilde{R}}^{2}  \bold{\widetilde{D}} \bold{A}^{H}\bold{T}$
\\
 $F_{T}(\bold{U})$ & $\frac{1}{M}  \Tr \bold{U}  \bold{T}^{T} \overline{\bold{A}}   \bold{\widetilde{R}}^{2}  \bold{\widetilde{D}} \bold{A}^{H}\bold{T}$
 &
 $\underline{F}_{T}(\bold{U})$ &  $\frac{1}{M}  \Tr \bold{U}  \bold{T}{\bold{A}}   \bold{\widetilde{R}}^{2}  \bold{\widetilde{D}} \bold{A}^{T}\bold{T}^{T}$
& 
$\Delta$ & $(1-F)^{2}-z^{2}\gamma\widetilde{\gamma}$
\\
 $ \Delta_{T}$ & $(1-\vartheta F_{T})(1-\overline{\vartheta}\underline{F}_{T})-|\vartheta|^{2}z^2\gamma_{T}\widetilde{\gamma}_{T}$
&
$\widetilde{\mathcal{F}}_{T}(\bold{U})$ & $\frac{1}{M}\Tr \widetilde{\bold{D}}{{\widetilde{\bold{T}}^{T}}}\widetilde{\bold{D}} \widetilde{\bold{T}} \bold{A}^{H} \bold{R}^{2} \bold{U}  \bold{A} \widetilde{\bold{T}}$
&
 ${\mathcal{F}}_{T}(\bold{U})$ & $\frac{1}{M}\Tr {\bold{D}}{{{\bold{T}}^{T}}}{\bold{D}} {\bold{T}} \bold{A} \widetilde{\bold{R}}^{2}\bold{U}  \bold{A}^{H} {\bold{T}}$
\\
\bottomrule
\end{tabular}
\end{table*}

\section{Main Result \RNum{1}: Bias for the Trace of the Resolvent}
\label{main_res}
In this section, we focus on the bias for the trace of the resolvent of non-centered random matrices with general random entries (not necessarily Gaussian). Two expressions for the bias are given in Section~\ref{exp_sec_1} and~\ref{exp_sec_2} by Theorem~\ref{bias_trace} and Theorem~\ref{alt_exp}, respectively, which are also generalized to compute the bias for the LSS in Poposition~\ref{linear_p}.

\subsection{The expression for the bias}
\label{exp_sec_1}
In the following, we first give the main result regarding the bias for the trace of the resolvent and then provide the detailed proof. 
\begin{theorem}
\label{bias_trace}
(Bias for the trace of the resolvent) If assumptions \textbf{A.1}-\textbf{A.4} are satisfied, the bias $\mathcal{M}(z)=\Tr( \E\bold{Q}- \bold{T})$ 
can be expressed as
\begin{equation}
\begin{aligned}
&\mathcal{M}(z)
 \xlongrightarrow[]{M\rightarrow \infty} \mathcal{B}(z)=
 \mathcal{B}_{\vartheta}(z)+\mathcal{B}_{\kappa}(z)
\\
&=\mathcal{Y}_{\vartheta}(\bold{I})+(\widetilde{\delta}+z \widetilde{\delta}')\mathcal{Y}_{\vartheta}(\bold{D})+z\delta'\widetilde{\mathcal{Y}}_{\vartheta}(\widetilde{\bold{D}})
\\
&+\mathcal{Y}_{\kappa}(\bold{I})+(\widetilde{\delta}+z \widetilde{\delta}')\mathcal{Y}_{\kappa}(\bold{D})+z\delta'\widetilde{\mathcal{Y}}_{\kappa}(\widetilde{\bold{D}}),
\end{aligned}
\end{equation}
where $\delta'=\frac{\mathrm{d} \delta}{\mathrm{d} z}$ and $\widetilde{\delta}'=\frac{\mathrm{d} \widetilde{\delta}}{\mathrm{d} z}$. $\mathcal{Y}_{\vartheta}(\bold{I})$ and $\mathcal{Y}_{\vartheta}(\bold{D})$ are obtained by taking $\bold{U}=\bold{I}$ and $\bold{U}=\bold{D}$ in~(\ref{y_u_eq_1}), respectively. Similarly, $\mathcal{Y}_{\kappa}(\bold{I})$, $\mathcal{Y}_{\kappa}(\bold{D})$ are determined by taking $\bold{U}=\bold{I}$, $\bold{U}=\bold{D}$ in~(\ref{y_u_eq_2}) while $\widetilde{\mathcal{Y}}_{\vartheta}(\widetilde{\bold{D}})$, $\widetilde{\mathcal{Y}}_{\kappa}(\widetilde{\bold{D}})$ are obtained by taking $\widetilde{\bold{U}}=\widetilde{\bold{D}}$ in~(\ref{y_u_eq_3}) and~(\ref{y_u_eq_4})
 \begin{subequations}\label{y_u_eq}
    \begin{alignat}{4}
    \begin{split}
    \label{y_u_eq_1}
    &\mathcal{Y}_{\vartheta}(\bold{U})\!=\!\frac{1}{{\Delta}_{T}}[\overline{\vartheta} \underline{F}_{T}(\bold{D}\bold{T}\bold{U})(1\!-\!\vartheta F_{T}) 
\!+\!\vartheta {F}_{T}(\bold{U}\bold{T}\bold{D})\times\!
\\
&
(1\!-\!\overline{\vartheta}\underline{ F}_{T}) 
+|\vartheta|^2z \widetilde{\gamma}_{T}\widetilde{\mathcal{F}}_{T}(\bold{U}) 
+|\vartheta|^2 z^2 \widetilde{{\gamma}}_{T} {\gamma}_{T}(\bold{U}\bold{T} \bold{D})
  ],
\end{split}
  \\
\begin{split}
    \label{y_u_eq_2}
&  \mathcal{Y}_{\kappa}(\bold{U})= \frac{z \kappa\eta}{M}\Tr \widetilde{\bold{D}}^2\widetilde{\bold{S}} 
\widetilde{\bold{R}}^{2} \bold{A}^{H}  \bold{T}\bold{U}\bold{T}\bold{A}
\\
&+\frac{\kappa z^2 \widetilde{\eta}}{M}\Tr {\bold{S}}\bold{D}^2 \bold{T}\bold{U}\bold{T}
,
\end{split}
\\
\begin{split}
    \label{y_u_eq_3}
&\widetilde{\mathcal{Y}}_{\vartheta}(\widetilde{\bold{U}})\!=\!\frac{1}{{\Delta}_{T}}[{\vartheta}\widetilde{\underline{F}}_{T}(\widetilde{\bold{D}}\widetilde{\bold{T}}\widetilde{\bold{U}})(1\!-\!\overline{\vartheta} \widetilde{F}_{T}) 
\!+\!\overline{\vartheta} \widetilde{F}_{T}(\widetilde{\bold{U}}\widetilde{\bold{T}}\widetilde{\bold{D}})\times\!
\\
&
(1-{\vartheta}\widetilde{\underline{ F}}_{T})
+|\vartheta|^2z {\gamma}_{T}{\mathcal{F}}_{T}(\widetilde{\bold{U}}) 
+|\vartheta|^2 z^2 {{\gamma}}_{T} \widetilde{\gamma}_{T}(\widetilde{\bold{U}}\widetilde{\bold{T}} \widetilde{\bold{D}})
  ],
  \end{split}
  \\
  \begin{split}
      \label{y_u_eq_4}
&  \widetilde{\mathcal{Y}}_{\kappa}(\widetilde{\bold{U}})=  \frac{z \kappa\widetilde{\eta}}{M}\Tr {\bold{D}}^2{\bold{S}} 
{\bold{R}}^{2} \bold{A}  \widetilde{\bold{T}}\widetilde{\bold{U}}\widetilde{\bold{T}}\bold{A}^{H}
\\
&+
\frac{\kappa z^2 {\eta}}{M}\Tr \widetilde{\bold{S}}\widetilde{\bold{D}}^2\widetilde{\bold{T}}\widetilde{\bold{U}}\widetilde{\bold{T}}.
     \end{split}
    \end{alignat}
  \end{subequations}
The symbols including $F_{T}(\cdot)$, $\widetilde{F}_{T}(\cdot)$, $\gamma_{T}(\cdot)$, $\widetilde{\gamma}_{T}(\cdot)$, $\widetilde{\gamma}_{T}(\cdot)$, $\eta$, $\widetilde{\eta}$ and $\Delta_{T}$ are given in Table~\ref{var_list}.
\end{theorem}

\begin{IEEEproof} 
We will first give the sketch of the proof with four steps:

\textit{Step 1) Auxiliary quantities and decomposition}. We introduce some auxiliary quantities as follows
\begin{equation}
\label{int_eq}
\begin{aligned}
\alpha&=\frac{1}{M}\E\Tr\bold{D}\bold{Q},~~\widetilde{\alpha}=\frac{1}{M}\E\Tr\widetilde{\bold{D}}\widetilde{\bold{Q}},
\\
\bold{C}&=\left(\omega\left( \bold{I}+\widetilde{\alpha}\bold{D} \right)+\bold{A}\left( \bold{I}+\alpha\widetilde{\bold{D}} \right)^{-1}\bold{A}^{H} \right)^{-1},
\\
\widetilde{\bold{C}}&=\left(\omega\left( \bold{I}+{\alpha}\widetilde{\bold{D}} \right)+\bold{A}^{H}\left( \bold{I}+\widetilde{\alpha}{\bold{D}} \right)^{-1}\bold{A} \right)^{-1},
\end{aligned}
\end{equation}
where $\alpha$, $\widetilde{\alpha}$, $\bold{C}$, $\widetilde{\bold{C}}$ correspond to $\delta$, $\widetilde{\delta}$, $\bold{T}$, $\widetilde{\bold{T}}$ in~(\ref{basic_eq}), respectively. $\bold{C}$ can be regarded as the intermediate evaluation between $\E \bold{Q}$ and $\bold{T}$. With $\bold{C}$, we can decompose the bias into two parts
\begin{equation}
\label{decomp}
\mathcal{M}(z)=\Tr(\E\bold{Q}-\bold{C})+\Tr(\bold{C}-\bold{T})=\mathcal{K}(z)+\mathcal{W}(z).
\end{equation}
Then we will show that $\mathcal{W}(z)=\Tr(\bold{C}-\bold{T})$ can be approximated by a linear combination of $M({\alpha}-{\delta})$ and $M(\widetilde{\alpha}-\widetilde{\delta})$.

\textit{Step 2) Further representations} The right hand side (RHS) of~(\ref{decomp}) could further be represented as a linear combination of $\Tr(\E\bold{Q}-\bold{C})$, $\Tr\bold{D}(\E\bold{Q}-\bold{C})$ and $\Tr\widetilde{\bold{D}}(\E\widetilde{\bold{Q}}-\widetilde{\bold{C}})$. Thus, if we can evaluate these three terms, we will be able to solve the whole problem.  

\textit{Step 3) Evaluation for} $\Tr\bold{U}(\E\bold{Q}-\bold{C})$. Instead of handling $\Tr(\E\bold{Q}-\bold{C})$, $\Tr\bold{D}(\E\bold{Q}-\bold{C})$ and $\Tr\widetilde{\bold{D}}(\E\widetilde{\bold{Q}}-\widetilde{\bold{C}})$ respectively, we will evaluate a more general form $\Tr\bold{U}(\E\bold{Q}-\bold{C})$, where $\bold{U}$ is a diagonal matrix with bounded norm. The result will be presented in Lemma~\ref{u_lemma}.

\textit{Step 4) Determine the approximation of $\mathcal{M}(z)$ by combining the results from the previous steps.}

In the following, we will provide the detailed proof.

\underline{\textit{Step 1}}: We use the auxiliary intermediate quantities in~(\ref{int_eq}) to decompose the bias. By the matrix identity $\bold{A}-\bold{B}=\bold{B}(\bold{B}^{-1}-\bold{A}^{-1})\bold{A}$, we have
\begin{equation}
\label{diff_qt}
\begin{aligned}
&\Tr (\E\bold{Q}-\bold{T})=\Tr (\E\bold{Q}-\bold{C})+\Tr (\bold{C}-\bold{T})
\\
&=\Tr (\E\bold{Q}-\bold{C})
- M(\widetilde{\alpha}-\widetilde{\delta}) \frac{\omega}{M}\Tr \bold{C}\bold{D}\bold{T}
\\
&+M({\alpha}-{\delta})\frac{1}{M}\Tr \bold{C}\bold{A}\widetilde{\bold{D}}\widetilde{\bold{R}}\left( \bold{I}+\alpha\widetilde{\bold{D}} \right)^{-1}\bold{A}^{H}\bold{T}
\\
& \xlongequal[]{a} \Tr (\E\bold{Q}-\bold{C})
-  M(\widetilde{\alpha}-\widetilde{\delta}) \frac{\omega}{M}\Tr \bold{D}\bold{T}^{2}
\\
&+M({\alpha}-{\delta})\frac{1}{M}\Tr \bold{T}\bold{A}\widetilde{\bold{D}}\widetilde{\bold{R}}^{2}\bold{A}^{H}\bold{T}+O(M^{-1})
\\
&=\Tr (\E\bold{Q}-\bold{C})-\omega\gamma(\bold{I})M(\widetilde{\alpha}-\widetilde{\delta})+F(\bold{I})M({\alpha}-{\delta})
\\
&+O(M^{-1}),
\end{aligned}
\end{equation}
where step $a$ follows from~(\ref{l3_11}) of Lemma~\ref{bound_lemma} in Appendix~\ref{rmt_results}.
Similarly, we have the following relation
\begin{equation}
\begin{aligned}
 &M({\alpha}-{\delta}) 
 \!=\!\Tr\bold{D} \left(\E\bold{Q}-\bold{C}\right) \! - \!  \omega M(\widetilde{\alpha}-\widetilde{\delta}) \frac{\Tr\bold{D} \bold{T}\bold{D}\bold{T}}{M}
 \\
 &\!+\!M({\alpha}-{\delta})\frac{\Tr \bold{D}\bold{T}\bold{A}\widetilde{\bold{D}}\widetilde{\bold{R}}^{2}\bold{A}^{H}\bold{T}}{M}+\varepsilon_{1},
 \end{aligned}
\end{equation}
and 
 \begin{equation}
\begin{aligned}
  &M(\widetilde{\alpha}-\widetilde{\delta}) 
  \!=\! \Tr \widetilde{\bold{D}}\left(\E\widetilde{\bold{Q}} \!-\!   \widetilde{\bold{C}}\right)\!-\!\omega M({\alpha}\!-\!{\delta}) \frac{\Tr\widetilde{\bold{D}} \widetilde{\bold{T}}\widetilde{\bold{D}}\widetilde{\bold{T}}}{M}
  \\
    &\!+\!M(\widetilde{\alpha}-\widetilde{\delta})\frac{\Tr \widetilde{\bold{D}}\widetilde{\bold{T}}\bold{A}^{H}{\bold{D}}\bold{R}^{2}\bold{A}\widetilde{\bold{T}}}{M}+\varepsilon_{2},
\end{aligned}
\end{equation}
where $\varepsilon_{i},i=1,2$ are of order $O(M^{-1})$. 

\underline{\textit{Step 2}}: In this step, we will represent the bias as a linear combination of $\Tr(\E\bold{Q}-\bold{C})$, $\Tr\bold{D}(\E\bold{Q}-\bold{C})$ and $\Tr\widetilde{\bold{D}}(\E\widetilde{\bold{Q}}-\widetilde{\bold{C}})$. By~(\ref{f_eq}), we can construct the system of equations with respect to $M({\alpha}-{\delta})$ and $M(\widetilde{\alpha}-\widetilde{\delta}) $ as
\begin{equation}
\label{sys_e}
\begin{bmatrix} 
1- F & \omega \gamma \\  
 \omega \widetilde{\gamma}   & 1 -{F}
\end{bmatrix}
\begin{bmatrix} 
 M({\alpha}-{\delta})
 \\
M(\widetilde{\alpha}-\widetilde{\delta}) 
\end{bmatrix}
\!=\!
\begin{bmatrix} 
\Tr \bold{D}(\E\bold{Q}-\bold{C})
 \\
\Tr \widetilde{\bold{D}}(\E\widetilde{\bold{Q}}-\widetilde{\bold{C}})
\end{bmatrix}
\!+\!      \bm{\varepsilon},
\end{equation}
where $\| \bm{\varepsilon} \|=O(M^{-1})$ and $F$ is given in Table~\ref{var_list}. It can be observed that $\Delta$ given in table~\ref{var_list} is the determinant of the coefficient matrix in~(\ref{sys_e}), and $\Delta_{T}$ is the ``conjugate'' of $\Delta$.

By~(\ref{replace_R}) and~(\ref{replace_tilde_t}), we can obtain the coefficient $F(\bold{I})$ in~(\ref{diff_qt}) as,
\begin{equation}
\begin{aligned}
F(\bold{I})&=\frac{1}{M}\Tr \bold{T}\bold{A}  \widetilde{\bold{R}}^{2} \widetilde{\bold{D}}\bold{A}^{H}\bold{T}
= \frac{1}{M}\Tr  \bold{R}^{2}\bold{A}\widetilde{\bold{T}}\widetilde{\bold{D}}\widetilde{\bold{T}}\bold{A}^{H}
\\
&= \frac{1}{M} \Tr {\bold{R}} \bold{A}\widetilde{\bold{T}}\widetilde{\bold{D}}\widetilde{\bold{T}}\bold{A}^{H}
- \widetilde{\delta} F
\\
&=(1-F)\widetilde{\delta} - \omega \widetilde{\gamma}(\bold{I})-   \omega \delta \widetilde{\gamma}.
\end{aligned}
\end{equation}
Expression of the bias can be further simplified by the following lemma.
\begin{lemma}
\label{delta_p}
Denoting $\delta'_{\omega}=\frac{\mathrm{d} \delta}{\mathrm{d} \omega}$ and $\widetilde{\delta}'_{\omega}=\frac{\mathrm{d} \widetilde{\delta}}{\mathrm{d} \omega}$, we have the following system of equations
\begin{equation}
\label{delta_p_eq}
\begin{bmatrix} 
1- F & \omega \gamma \\  
 \omega \widetilde{\gamma}   & 1 -{F}
\end{bmatrix}
\begin{bmatrix} 
 \delta'_{\omega} 
 \\
\widetilde{\delta}'_{\omega} 
\end{bmatrix}
=
\begin{bmatrix} 
-\frac{1}{M}\Tr \bold{D}\bold{T}^{2}-\widetilde{\delta}\gamma
 \\
-\frac{1}{M}\Tr \widetilde{\bold{D}}\widetilde{\bold{T}}^{2}-{\delta}\widetilde{\gamma}
\end{bmatrix}.
\end{equation}
\end{lemma}
\begin{IEEEproof}
The proof of~Lemma \ref{delta_p} is given in Appendix~\ref{proof_delta_p}.
\end{IEEEproof}
By Lemma~\ref{delta_p} and $\delta'=-\delta'_{\omega}$, $\widetilde{\delta}'=-\widetilde{\delta}'_{\omega}$, we have
\begin{equation}
\begin{aligned}
&\frac{1}{\Delta}[\omega^2\gamma(\bold{I}) \widetilde{\gamma}+F(\bold{I})(1-{F})]
=\widetilde{\delta}+
\\
&\frac{\omega}{\Delta}[ -(1-F)(  \widetilde{\gamma}(\bold{I})
+    \delta \widetilde{\gamma}) 
+ \omega \widetilde{\gamma} (\widetilde{\delta}\gamma+\gamma(\bold{I}) )]
\\
&=\widetilde{\delta}+\omega \widetilde{\delta}'_{\omega}=\widetilde{\delta}+z \widetilde{\delta}',
\end{aligned}
\end{equation}
and
\begin{equation}
\label{w_delta}
\begin{aligned}
\frac{1}{\Delta} [\omega \gamma(\bold{I})(1-F)+F(\bold{I})\omega\gamma   ]
=-\omega\delta'_{\omega}=-z\delta'.
\end{aligned}
\end{equation}
Therefore, according to~(\ref{sys_e}) to~(\ref{w_delta}), we have
\begin{equation}
\label{tq_ori}
\begin{aligned}
&\Tr (\E\bold{Q}-\bold{T})=  \Tr (\E\bold{Q}-\bold{C})
\\
&\!+\! \frac{F(\bold{I})}{\Delta}  [(1-{F})\Tr \bold{D}(\E\bold{Q}-\bold{C})
\!-\! \omega  {\gamma} \Tr \widetilde{\bold{D}}(\E\widetilde{\bold{Q}}\!-\!\widetilde{\bold{C}})]
\\
&\!-\! \frac{ \omega \gamma(\bold{I})}{\Delta}[(1-F)\Tr \widetilde{\bold{D}}(\E\widetilde{\bold{Q}}\!-\!\widetilde{\bold{C}})
\!-\! \omega  \widetilde{\gamma} \Tr \bold{D}(\E\bold{Q}-\bold{C})
 ]
 \\
 &=\mathcal{Z}(\bold{I})+(\widetilde{\delta}+z \widetilde{\delta}')\mathcal{Z}(\bold{D})+z\delta'\widetilde{\mathcal{Z}}(\widetilde{\bold{D}})+\varepsilon_{z},
\end{aligned}
\end{equation}
where $\varepsilon_{z}$ is of the order $O(M^{-1})$ and $\mathcal{Z}(\bold{U})=\Tr\bold{U}(\E\bold{Q}-\bold{C})$ and $\widetilde{\mathcal{Z}}(\widetilde{\bold{U}})=\Tr\widetilde{\bold{U}}(\E\widetilde{\bold{Q}}-\widetilde{\bold{C}})$ denote the intermediate bias.

\underline{\textit{Step 3}}: Next, we need to evaluate $\mathcal{Z}(\bold{I})$, $\mathcal{Z}(\bold{D})$ and $\widetilde{\mathcal{Z}}(\widetilde{\bold{D}})$. Instead of computing the three terms respectively, we will obtain a more general expression of $\mathcal{Z}(\bold{U})=  \Tr \bold{U}(\E\bold{Q}-\bold{C})$, where $\bold{U}$ is a deterministic diagonal matrix with bounded spectral norm. The result can be obtained by Lemma~\ref{u_lemma}.
Then, we can obtain $\mathcal{Z}(\bold{I})$, $\mathcal{Z}(\bold{D})$ and $\widetilde{\mathcal{Z}}(\widetilde{\bold{D}})$ by letting $\bold{U}=\bold{I}$, $\bold{U}=\bold{D}$ and $\widetilde{\bold{U}}=\widetilde{\bold{D}}$, respectively.
\begin{lemma} 
\label{u_lemma}
Assume that assumptions \textbf{A.1}-\textbf{A.4} hold and let $\bold{U}$ be a deterministic diagonal matrix with bounded norm. Then for the intermediate bias $\mathcal{Z}(\bold{U})=  \Tr \bold{U}(\E\bold{Q}-\bold{C})$, there holds
\begin{equation}
\label{u_lemma_eq}
\mathcal{Z}(\bold{U}) \xlongrightarrow[]{M\rightarrow \infty}
\mathcal{Y}(\bold{U})=\mathcal{Y}_{\vartheta}(\bold{U})+\mathcal{Y}_{\kappa}(\bold{U}),
\end{equation}
 where $\mathcal{Y}_{\vartheta}(\bold{U})$ and $\mathcal{Y}_{\kappa}(\bold{U})$ are given in~(\ref{y_u_eq}).
\end{lemma} 
\begin{IEEEproof}
The proof of~Lemma \ref{u_lemma} is postponed to Appendix~\ref{lemma_1_proof}.
\end{IEEEproof}
\begin{remark}
This result is also applicable to the adjoint $\widetilde{\mathcal{Z}}(\widetilde{\bold{U}})=  \Tr \widetilde{\bold{U}}(\E\widetilde{\bold{Q}}-\widetilde{\bold{C}})$ by taking $\widetilde{\cdot}$ over each symbol.
If we let $\bold{A}=\bold{0}$, the result degenerates to Lemma 7.1 in~\cite{hachem2012clt}. 
\end{remark}

\underline{\textit{Step 4}}: To derive the final approximation of $\mathcal{M}(z)$, we only need to plug the RHS of~(\ref{u_lemma_eq}) into~(\ref{tq_ori}).
\end{IEEEproof}

\begin{remark} 
Theorem~\ref{bias_trace} provides the gap between the expectation of the trace of the resolvent and that of the approximation $\bold{T}$. We can observe that the bias  can be divided into two parts, which are related to $\vartheta$ and $\kappa$, respectively, indicating that the bias will disappear if $\vartheta$ and $\kappa$ are zero. This means that Gaussianity is only a sufficient condition but not a necessary one for the non-zero bias, but in this paper, we will refer to all cases with non-zero $\vartheta$ and $\kappa$ as the non-Gaussian case. Also, it can be noted that the bias is $O(1)$, which implicates
\begin{equation}
\frac{1}{M}\Tr(\E\bold{Q}-\bold{T}) =O(M^{-1}).
\end{equation}
This bound coincides with the bound~(\ref{l3_1}) in Lemma~\ref{bound_lemma} and the optimal convergence rate is $O(M^{-1})$. Furthermore, by comparing the two bounds, we can obtain that the constant $K$ in~(\ref{l3_1}) is evaluated to be $|\mathcal{B}(z)+o(1)|$. The bound is tighter than that established by the \textit{Generalized Lindeberg Principle} in~\cite{wen2012deterministic}, which is $O(M^{-\frac{1}{2}})$.
\end{remark}

\subsection{Alternative expression for the bias}
\label{exp_sec_2}
The bias could also be represented as the derivative of a simpler expression. 
\begin{theorem}
 \label{alt_exp}
 (Alternative expression) If assumptions \textbf{A.1}-\textbf{A.4} are satisfied, the bias $\mathcal{B}(z)$ 
can be expressed as
\begin{equation}
\label{alt_for}
\begin{aligned}
&\mathcal{M}(z)\xlongrightarrow[]{M\rightarrow \infty}  \mathcal{B}(z)=\mathcal{B}_{\vartheta}(z)+\mathcal{B}_{\kappa}(z)
\\
&=\frac{1}{2} \frac{\mathrm{d} ( -\log\left(\Delta_{T}\right)
+ \frac{\kappa z^2}{M^2}\Tr\bold{D}^2 \bold{S}^2\Tr\widetilde{\bold{D}}^2 \widetilde{\bold{S}}^2  )
}{\mathrm{d}z}.
\end{aligned}
\end{equation}
\end{theorem}
\begin{IEEEproof}
The proof of Theorem~\ref{alt_exp} is given in Appendix~\ref{alt_exp_proof}.
\end{IEEEproof}
Theorem~\ref{bias_trace} and Theorem~\ref{alt_exp} provide more general results than those in~\cite{hachem2012clt} and~\cite{najim2016gaussian}. Specifically, when $\bold{A}=\bold{0}$, the result degenerates to Proposition 2.3 in~\cite{hachem2012clt} and when $\bold{R}=\bold{D}$ and $\bold{I}=\widetilde{\bold{D}}$, the result coincides with the last formula in~\cite[page 1867]{najim2016gaussian}. There are two extreme cases in terms of the bias. When $x_{ij}$s are complex i.i.d. Gaussian, we have $\vartheta=\kappa=0$ and $\mathcal{B}(z)= 0$. When $x_{ij}$s are real, we have $\vartheta=1$. The general case we consider here is the intermediate case with $0< |\vartheta| < 1 $.

Theorem~\ref{bias_trace} and Theorem~\ref{alt_exp} can be extended to compute the LSS of $\bold{H}\bold{H}^{H}$ by Cauchy's integral formula~\cite{bai2008clt}. The non-Gaussianity will not only cause the bias for the mean but also the variance. The bias for the variance of the LSS has been discussed in~\cite{banna2020clt} for non-centered random matrices. Therefore, we only focus on the bias for the mean, which is given in the following proposition.
\begin{proposition}
\label{linear_p}
(Bias for the LSS) If the function $f$ is analytic in the region which contains $[0, u_{+}]$, we have the following approximation for the expectation of the LSS of $\bold{H}\bold{H}^{H}$ with $\bold{H}$ defined in~(\ref{matrix_model_a}),
\begin{equation}
\label{lin_for}
\begin{aligned}
& \E\Tr f(\bold{H}\bold{H}^{H}) \xlongrightarrow[]{M \rightarrow \infty}   \frac{-1}{2\pi \jmath} \int_{\mathcal{C}} f(z)\Tr \bold{T}(z) \mathrm{d}z
\\
&+
\frac{-1}{2\pi \jmath} \int_{\mathcal{C}} f(z) \mathcal{B}(z) \mathrm{d}z
=\mathcal{V}_{f}+\mathcal{B}_{f},
\end{aligned}
\end{equation}
where the contour $\mathcal{C}$ in the analytic region is in the positive direction and contains the interval $[0, u_{+}]$ with $u_{+}=2\| \bold{A} \|^2+2 d_{max}\widetilde{d}_{max}(1+\sqrt{c})^2$.
\end{proposition}
\begin{IEEEproof}
The proof of~Proposition \ref{linear_p} is postponed to Appendix~\ref{lss_proof}.
\end{IEEEproof}
Here $u_{+}$ is not a tight bound for the largest eigenvalue of $\bold{H}\bold{H}^{H}$, which was discussed in~\cite{dozier2007analysis} with a more complex form, but it suffices to guarantee the correctness of the result in~(\ref{lin_for}). It is easy to observe that our model covers the information-plus-noise model, i.e. $\bold{H}=\bold{A}+\sigma \bold{X}$ when $\bold{D}=\sigma\bold{I}$, $\widetilde{\bold{D}}=\sigma\bold{I}$. Therefore, our results can answer the question posed as an interesting and computationally challenging problem in Section 2.4 of~\cite{banna2020clt}.~\cite{banna2020clt} provided discussion on the case with Gaussian entries where the bias vanishes and focused on the computation of the variance of CLT. Proposition~\ref{linear_p} provides a method to compute the mean for the CLT.

\begin{remark}
(\textit{LoS component}) The $\mathcal{V}_{f}$ term in~(\ref{lin_for}) for Gaussian matrices  is only related to the spectral of the LoS component $\bold{A}$ while the bias $\mathcal{B}_{f}$ not only depends on the spectral of $\bold{A}$ but also the singular value decomposition (SVD) of $\bold{A}$ or the eigenvectors of $\bold{A}\bold{A}^{H}$ and $\bold{A}^{H}\bold{A}$. Specifically, when $\bold{D}=\bold{I}$ and $\widetilde{\bold{D}}=\bold{I}$. if we perform SVD for $\bold{A}=\bold{U}\bold{\Lambda}_{\bold{A}}\bold{V}^{H}$, where $\bold{U}$ and $\bold{V}$ are unitary matrices, we have $\bold{A}\bold{A}^{H}=\bold{U}\bold{\Lambda}_{\bold{A}\bold{A}^{H}}\bold{U}^{H}$ and $\bold{A}^{H}\bold{A}=\bold{V}\bold{\Lambda}_{\bold{A}^{H}\bold{A}}\bold{V}^{H}$. Then $\delta$ and $\widetilde{\delta}$ become
 \begin{equation}
 \begin{aligned}
\delta&=\frac{1}{M}\Tr\bold{T}_{\bold{A}}(z) ,
\\
\widetilde{\delta}&= \frac{1}{M}\Tr\widetilde{\bold{T}}_{\bold{A}}(z),
\end{aligned}  
\end{equation}
where $\bold{T}_{\bold{A}}(z)=\left( -z(1+\widetilde{\delta})\bold{I}+\frac{1}{1+\delta}\bold{\Lambda}_{\bold{A}\bold{A}^{H}}  \right)^{-1}$ and $\widetilde{\bold{T}}_{\bold{A}}(z)=\left( -z(1+\delta)\bold{I}+\frac{1}{1+\widetilde{\delta}}\bold{\Lambda}_{\bold{A}^{H}\bold{A}}  \right)^{-1}$.
In this case, $\mathcal{V}_{f}$ is only related to the spectral of $\bold{A}$ and independent of $\bold{U}$ and $\bold{V}$. However, due to the co-existence of $\bold{T}^{T}$, $\overline{\bold{T}}$, $\bold{A}^{T}$ and $\overline{\bold{A}}$ in $\mathcal{B}_{f}$, $\bold{U}$ and $\bold{V}$ will not be compensated in the term related to $\vartheta$. Specifically, in this case, $F_{T}$ and $\gamma_{T}$ in $\mathcal{B}_{f}$ can be written as
 \begin{equation}
 \begin{aligned}
F_{T}&\!=\!\frac{1}{M(1+\delta)^{2}}\Tr\bold{T}_{\bold{A}}(z) \bold{\Lambda}_{\bold{A}} {\bold{V}}^{T}\bold{V}\bold{\Lambda}_{\bold{A}}\bold{T}_{\bold{A}}(z)\bold{U}^{H} \overline{\bold{U}},
\\
\gamma_{T}&\!=\! \frac{1}{M}\Tr \bold{T}_{\bold{A}}(z)
\bold{U}^{T}{\bold{U}} \bold{T}_{\bold{A}}(z)\bold{U}^{H}\overline{\bold{U}}.
\end{aligned}  
\end{equation}
Therefore, the bias $\mathcal{B}_{f}$ is related to the SVD of $\bold{A}$ instead of only the spectral of $\bold{A}$.
\end{remark}

By far, we have obtained the asymptotic expression for the bias of the trace of the resolvent and that for the LSS. Next, we will apply the above results to MIMO channels and determine the bias of the EMI.

\section{Main Result \RNum{2}: Bias for the Ergodic Mutual Information}
\label{main_res_2}
Based on the result in Section~\ref{main_res}, we will analyze the bias of the EMI and make the CLT in~\cite{hachem2012clt} complete with a more accurate mean. The derived results will be compared with those of existing works. Furthermore, an approximation of the outage probability is given by using the modified CLT.

\subsection{Bias for the EMI}
As a direct application of the result in Section~\ref{main_res}, we can obtain the bias of the EMI for the non-centered and non-Gaussian MIMO channels defined in~(\ref{matrix_model_a}). Note that in~\cite{hachem2012clt}, the bias for the centered case is given but the result for the non-centered case is not available. The following CLT provides the asymptotic distribution of the MI.

\begin{proposition}
\label{p_rel}
 (Complement for the CLT of the MI in~\cite{hachem2012clt}) Let $z=-\sigma^2$ and we have $\Delta_{T}=|1-\vartheta{F}_{T}|^2-|\vartheta|^2 \sigma^4{\gamma}_{T}\widetilde{\gamma}_{T}$. The CLT for the MI, $C=C(\sigma^2)$, can be given by 
\begin{equation}
C - V \overset{\mathcal{D}}{\longrightarrow } \mathcal{N}\left(\mathcal{B}_{C}, \Theta\right),
\end{equation}
where
\begin{equation}
\label{bia_c}
\begin{aligned}
\mathcal{B}_{C}&=\frac{1}{2}\log\left(\Delta_{T}\right)
-\frac{ \kappa \sigma^4}{2M^2}\Tr\bold{D}^2 \bold{S}^2\Tr\widetilde{\bold{D}}^2 \widetilde{\bold{S}}^2
\\
&=\mathcal{B}^{\vartheta}_{C}+\mathcal{B}^{\kappa}_{C},
\end{aligned}
\end{equation}
\begin{equation}
V=-\log \det\left( \sigma^2\bold{T} \right)+ \log \det\left(\bold{I}+\delta\widetilde{D}  \right)-M\sigma^2\delta\widetilde{\delta},
\end{equation}
\begin{equation}
\begin{aligned}
\Theta&\!=\!-\log(\Delta)\!+\!(-\log({\Delta}_{T})\!+\! 
\frac{\kappa \sigma^4}{M^2}\Tr\bold{D}^2 \bold{S}^2\Tr\widetilde{\bold{D}}^2 \widetilde{\bold{S}}^2)
\\
&\!=\!\Theta_{G}\!+\!\Theta_{\mathcal{B}}.
\end{aligned}
\end{equation}
$(\delta,\widetilde{\delta})$, $(\bold{T}, \widetilde{\bold{T}})$ are determined by Algorithm~\ref{sol_fund} from the system of equations in~(\ref{basic_eq}) when $z=-\sigma^2$. $\bold{S}$, $\widetilde{\bold{S}}$, $F_{T}$, $\gamma_{T}$ and $\widetilde{\gamma}_{T}$ are given in Table~\ref{var_list}.
Here $V$ and $\Theta_{G}$ are the approximations for the mean and variance of the MI for Gaussian channels. $\mathcal{B}_{C}$ is the bias for the mean induced by non-Gaussianity, which consists of $\mathcal{B}^{\vartheta}_{C}$ and $\mathcal{B}^{\kappa}_{C}$, caused by the non-circularity and severe fading of MIMO channels, respectively. Similar phenomenon happens in the expression of the variance. Importantly, it holds true that
$\underline{-0.5\times \Theta_{\mathcal{B}} = \mathcal{B}_{C}}$.
\end{proposition}

\begin{IEEEproof}
$V$, which corresponds to $\mathcal{V}_{f}$ in Proposition~\ref{linear_p}, has been resolved in~\cite{hachem2007deterministic}. The proof of the asymptotic Gaussianity and the expression of variance $\Theta$ were given in~\cite{hachem2012clt}. Here we only focus on $\mathcal{B}_{C}$, which corresponds to $\mathcal{B}_{f}$ in~(\ref{lin_for}). By~(\ref{c_exp}), we have 
\begin{equation}
\label{c_bias}
 \begin{aligned}
& \mathcal{B}_{C}= \int_{-\infty}^{-\sigma^2}  \Tr (\bold{T}-z\bold{I})^{-1}-   \E \Tr (\bold{B}-z\bold{I})^{-1}  \mathrm{d}z
 \\
 &=\int_{-\infty}^{-\sigma^2} -\mathcal{B}(z)\mathrm{d}z 
\\
&\overset{a}{=}-\frac{1}{2} [ -\log\left(\Delta_{T}\right)
+\frac{\kappa z^2}{M^2}\Tr\bold{D}^2 \bold{S}^2\Tr\widetilde{\bold{D}}^2 \widetilde{\bold{S}}^2) ]|_{z=-\sigma^2}
\\
&=-0.5\times \Theta_{\mathcal{B}},
\end{aligned}
\end{equation}
where step $a$ follows from~(\ref{alt_for}) in~Theorem~\ref{alt_exp}. Alternatively, we can also take $f(x)=\log(1 + x/\sigma^2)$ in Proposition~\ref{linear_p} to obtain the bias for the EMI, which requires the computation of contour integral. We ignore the detailed computation here.
\end{IEEEproof}
It can be observed that although Theorem~\ref{alt_exp} is in a derivative form, it is more convenient for the type of computation involved in Proposition~\ref{p_rel}. Also, note that the variance consists of two parts, which are the variance when $x_{ij}$ are i.i.d. circularly Gaussian RVs, i.e., $\Theta_{G}$ (corresponds to the results in~\cite{hachem2008new} when $\bold{A}=\bold{0}$) and the bias caused by the non-Gaussianity, i.e., $\Theta_{\mathcal{B}}$, which has been resolved in~\cite{hachem2012clt}. More generally, the bias for the variance with respect to the LSS of large random information-plus-noise matrices was investigated in~\cite{banna2020clt}. By Proposition~\ref{p_rel}, an explicit expression of the bias for the mean is provided. When $x_{ij}$s are i.i.d. complex Gaussian, $\vartheta=\kappa=0$ and $\E C - V = o(1)$ which corresponds to the result developed in~\cite{dumont2010capacity} by utilizing Gaussian tools where the variance $\Theta=\Theta_{G}$. 
With the biases for the mean and variance, we can obtain a new CLT with modified mean and variance, whose better approximation performance will be discussed in Section~\ref{simu}. According to Proposition~\ref{p_rel}, we can compute the bias $\mathcal{B}_{C}$ by the following three steps:

\underline{\textit{Step 1}} Compute $(\delta,\widetilde{\delta})$ according to Algorithm~\ref{sol_fund} with the channel parameters $z=-\sigma^2$, $\bold{D}$, $\widetilde{\bold{D}}$ and $\bold{A}$.

\underline{\textit{Step 2}} Get $\bold{T}$ and $\widetilde{\bold{T}}$ by inserting $(\delta,\widetilde{\delta})$ into~(\ref{basic_eq}) and then obtain the related quantities $F_{T}$, $\gamma_{T}$, $\widetilde{\gamma}_{T}$ as defined in Table~\ref{var_list}.

\underline{\textit{Step 3}} Obtain $\mathcal{B}_{C}$ by~(\ref{bia_c}).

\begin{remark} 
The correlated non-Gaussian channel was also analyzed by the \textit{Generalized Lindeberg Principle} in~\cite{wen2012deterministic} and~\cite{zhang2013capacity}. Specifically, the convergence of the case with Gaussian entries was first analyzed. Then, the interpolation technique was used to show that the gap between the Stieltjes Transform for the Gaussian and non-Gaussian cases is $O({M}^{-\frac{1}{2}})$. It follows from Proposition~\ref{p_rel} that the upper bound of the gap could be optimized to be $O(M^{-1})$ when the channels are uncorrelated.
\end{remark}

\subsection{Comparison with existing works on non-Gaussian channels}
To illustrate the generality and correctness of our results, we will compare them with the existing results regarding the bias for the EMI.
\subsubsection{Centered case ($\bold{A}=\bold{0}$~\cite{hachem2012clt})} When $\bold{A}=\bold{0}$, we have $\gamma_{T}=\gamma$, $\widetilde{\gamma}_{T}=\widetilde{\gamma}$, $\Delta_{T}=1-|\vartheta|^2 z^2\gamma\widetilde{\gamma}$ and $\frac{1}{M^2}\Tr\bold{D}^2 \bold{S}^2\Tr\widetilde{\bold{D}}^2 \widetilde{\bold{S}}^2=\gamma\widetilde{\gamma}$.
The bias becomes
 \begin{equation}
\E C-V  \xlongrightarrow[]{M \longrightarrow \infty}  \frac{1}{2}[\log\left(1-|\vartheta|^2 z^2\gamma\widetilde{\gamma}\right)-\kappa z^2 \gamma\widetilde{\gamma}],
\end{equation}
which is consistent with the result for the centered case in~\cite{hachem2012clt}.

\subsubsection{Centered i.i.d. case ($\bold{A}=\bold{0}$, $\bold{D}=\bold{I}$, $\widetilde{\bold{D}}=\bold{I}$~\cite{bao2015asymptotic})} In this case, the CLT will degenerate to the case in~\cite{bao2015asymptotic}, where the constraint on the finite $16$-th order moments will not be needed. Furthermore, it can be shown that the $(-0.5)$ relation between the bias for the mean and that for the variance holds true for this simple case.

\subsubsection{Non-centered circular case ($\vartheta=0$, $\bold{A}\neq \bold{0}$, and $\bold{D}=\bold{I}$, $\widetilde{\bold{D}}=\bold{I}$~\cite{kammoun2010fluctuations})} In this case, the $\mathcal{B}_{\vartheta}$ term will vanish. By Proposition~\ref{p_rel}, the closed form bias can be acquired immediately, where the integral form $\E C -V \xlongrightarrow[]{M \longrightarrow \infty} \int_{\sigma^2}^{\infty}\mathcal{B}_{\kappa}(-\omega)d\omega$ was utilized in~\cite{kammoun2010fluctuations}.

\subsection{Outage probability approximation}
By now, we have modified the CLT for the non-centered model with non-Gaussian and non-circular entries. As an important application, our results can be utilized to 
approximate the outage probability of large MIMO systems. From the definition of outage probability, we have
\begin{equation}
P_{out}(R)=\mathbb{P}(C\le R) \approx 1-Q( \frac{R-\overline{C}}{\sqrt{\Theta}}),
\end{equation}
where $R$ is the threshold for the transmission rate and $\overline{C}=V+\mathcal{B}_{C}$ is the modified mean by the bias derived in this paper. The performance of this modified approximation will be presented in Section~\ref{simu}.

\section{Numerical Results}
\label{simu}
In this section, we will compare the theoretical results with the empirical ones to show the accuracy of the derived results. 

\subsection{Simulation setting}
In the simulation, we consider the down-link distributed antenna systems where the antennas of the BS are far from each other but those of the mobile are collocated as a uniform linear array (ULA). As a result, the steering vectors of the ULA towards different transmit antennas are uncorrelated and the LoS component is full-rank. In particular, the steering vector from the mobile ULA towards the $i$-th transmit antenna is given by 
\begin{equation}
\left[1, e^{\jmath kd \sin(\phi_{i})}, ..., e^{\jmath (N-1) kd \sin(\phi_{i})}\right]^{T},
\end{equation}
where $k=\frac{2\pi }{L}$, with $L$ denoting the wavelength and $d$ is the inter-element spacing, which is set to be equal to $L$. $\phi_{i}$ is the angle of arrival (AoA) of the $i$-th transmit antenna. Denoting $\alpha_{i}=2\pi\sin(\phi_i)$, the LoS component can be written as
\begin{equation}
{\bold{A}}=\left[\bold{a}(\alpha_{1}),\bold{a}(\alpha_{2}),...,\bold{a}(\alpha_{M}) \right],
\end{equation}
where $\bold{a}(\alpha)=\left[ 1, e^{\jmath \alpha},...,e^{\jmath (N-1)\alpha} \right]^{T}$. Without loss of generality, $\alpha_{m}$'s are set to be $\alpha_{m}=\frac{2\pi m}{N},~m=0,1,...,M-1$. This model was also used in the analysis of non-centered channels~\cite{dumont2010capacity}~\cite{kammoun2010fluctuations}~\cite{kammoun2012fluctuations}.

To simulate both the non-Gaussianity and non-circularity of the non-LoS component, we consider the following channel model. Let $x_{n,m}=X=Z_{r} +\jmath Z_{i}=r\sigma_{r}cos(\phi) +\jmath r\sigma_{i}  sin(\phi)$, $1\le n \le N, 1\le m \le M$, where $\sigma_{r}^2$ and $\sigma_{i}^2$ are two deterministic parameters to control the non-circularity of the channel, similar as the ones for Hoyt distribution~\cite{kumar2010random}. $r\ge0$ and $\phi$ are two independent real random variables, where $\phi$ is uniformly distributed in $[0,2\pi)$. When $r$ follows the Rayleigh distribution, $|Z|$ will follow the Hoyt distribution~\cite{adali2011complex}. In fact, the generation process of the model can be regarded as a two-step way: a non-Gaussian circular model is generated by $Y=rcos(\phi)+\jmath r sin(\phi)$ first and then weights are added to the real and imaginary parts of $Y$ to generate the non-circularity. In this case, the pseudo-variance $\vartheta$ and the fourth cumulant $\kappa$ are related to $\sigma_r$ and $\sigma_i$ with
\begin{equation}
\begin{aligned}
\vartheta&= \E X^2 =  \E Z_{r}^2 \!-\!  \E Z_{i}^2 \!+\! 2\jmath \E (Z_{r}Z_{i}) = \frac{\sigma_{r}^2 \!-\! \sigma_{i}^2}{2}\E r^2,
\\
\kappa&=(\E |X|^4  \!-\!  |\vartheta|^2 -2)=\E Z_{r}^4 \!+\! \E Z_{i}^4 \!+\! 4 \E Z_{r}^2 \E Z_{i}^2 
\\
&=( \frac{3}{8} \sigma_{r}^{4} \!+\!  \frac{3}{8}  \sigma_{i}^{4} \!+\!  \frac{2}{8} \sigma_{r}^{2}\sigma_{i}^{2})\E r^4 \!-\!  |\vartheta|^2\!-\!2
\\
&
\equiv P(\E r^4, \vartheta ).
\end{aligned}
\end{equation}
Meanwhile, there holds
\begin{equation}
\E X=0,
\end{equation}
\begin{equation}
\label{e_x}
\E |X|=\E r \E \sqrt{\sigma^2_{r} -2\vartheta \mathrm{sin}^{2}(\phi) }=f(\vartheta)\E r ,
\end{equation}
and
\begin{equation}
\E |X|^2=\frac{\sigma^2_r +\sigma_i^2 }{2} \E  r^2=1,
\end{equation}
where $f(\vartheta)=\frac{2\sigma_{r}}{\pi}\mathcal{E}(\sqrt{\frac{2\vartheta}{\sigma^2_{r} }})$ and
$\mathcal{E}(k)= \int_{0}^{\frac{\pi}{2}}\sqrt{1-k^2 \mathrm{sin}^2(x)}  \mathrm{d}x$ is the complete elliptic integral of the second kind.

In order to guarantee $\E |X|^2=1$, we set $\E r^2=1$ and $\sigma_r^2+\sigma_i^2=2$. The coefficient of variation (CV) is defined as
\begin{equation}
CV=\sqrt{\frac{\Var(|X|)}{(\E|X|)^2}}=\sqrt{\frac{1}{(\E|X|)^2}-1},
\end{equation}
which represents the severity of the fading. 
Since $\E (\sqrt{\cdot})\le \sqrt{\E(\cdot) } $, we have
\begin{equation}
f(\vartheta) \le\sqrt{ {\sigma^2_{r}-\frac{\sigma^2_{r}- \sigma^2_{i}}{2}}  }   =\sqrt{ \frac{\sigma^2_{r}+\sigma^2_{i}}{2}  }=1.
\end{equation}
Thus, we can conclude that the non-circularity will increase the CV because $f(\vartheta)\le 1$ in~(\ref{e_x}) although $\E |X|^2$ is a constant taking value $1$. 

We consider three widely used models for $r$, whose parameters are given in table~\ref{para_table}.
 \begin{table*}[!htbp]
\centering
\caption{Properties Of Non-circular Non-gaussian Channels}
\label{para_table}
\begin{tabular}{|c|c|c|c|}
\hline
Distribution of $r$ & Weibull & Log-normal & Nakagami-m \\
\hline
p.d.f $(r\ge 0)$& $ f_{\lambda,k}(r)=\frac{k}{\lambda}(\frac{r}{\lambda})^{k-1}e^{-(\frac{r}{\lambda})^{k}} $ 
& $f_{\mu,\sigma}(r)=\frac{1}{r\sqrt{2\pi}\sigma } e^{-\frac{\mathrm{ln}( r)-\mu}{2\sigma^2}} $ 
& $f_{m,\omega}(r)=\frac{2m^m}{\Gamma(m)\Omega^m}r^{2m-1} e^{-\frac{m}{\Omega}r^2} $   
\\
\hline
Parameter setting 
& $\lambda=\sqrt{\frac{1}{\Gamma(1+\frac{2}{k})}}$ 
& $\mu=-\sigma^2$ 
& $\Omega=1$
\\
\hline
Fourth-order cumulant &  $P( \frac{ \Gamma(1+\frac{4}{k})}{\Gamma(1+\frac{2}{k})^2},\vartheta)   $ 
&$P( e^{4\sigma^2},\vartheta)$  &  
$ P(1+\frac{1}{m},\vartheta) $
\\
\hline
$CV$ & $ \sqrt{ \frac{\Gamma(1+\frac{2}{k}) }{f(\vartheta)^2 \Gamma(1+\frac{1}{k})^2 } -1}  $
&
$\sqrt{ \frac{e^{\sigma^2}}{f(\vartheta)^2 } -1}$
&
$\sqrt{ \frac{m \Gamma(m)^2}{f(\vartheta)^2\Gamma(m+\frac{1}{2})^2 } -1}$
\\
\hline
\end{tabular}
\end{table*}
The non-centered channel can be given by
\begin{equation}
\bold{H}=\frac{1}{\sqrt{M}} (\sqrt{\frac{K}{K+1}}{\bold{A}}+\sqrt{\frac{1}{K+1}}\bold{D}^{\frac{1}{2}}\bold{X}\widetilde{\bold{D}}^{\frac{1}{2}}),
\end{equation}
where $\frac{1}{\sqrt{M}}$ aims to make the spectral norm of the LoS matrix bounded and $K$ is the Rician factor. 

\subsection{The bias used in the CLT}
\label{simu62}
\begin{figure*}[h] 
    \centering
  \subfloat[\label{pdf_2}]{
       \includegraphics[width=0.3\linewidth]{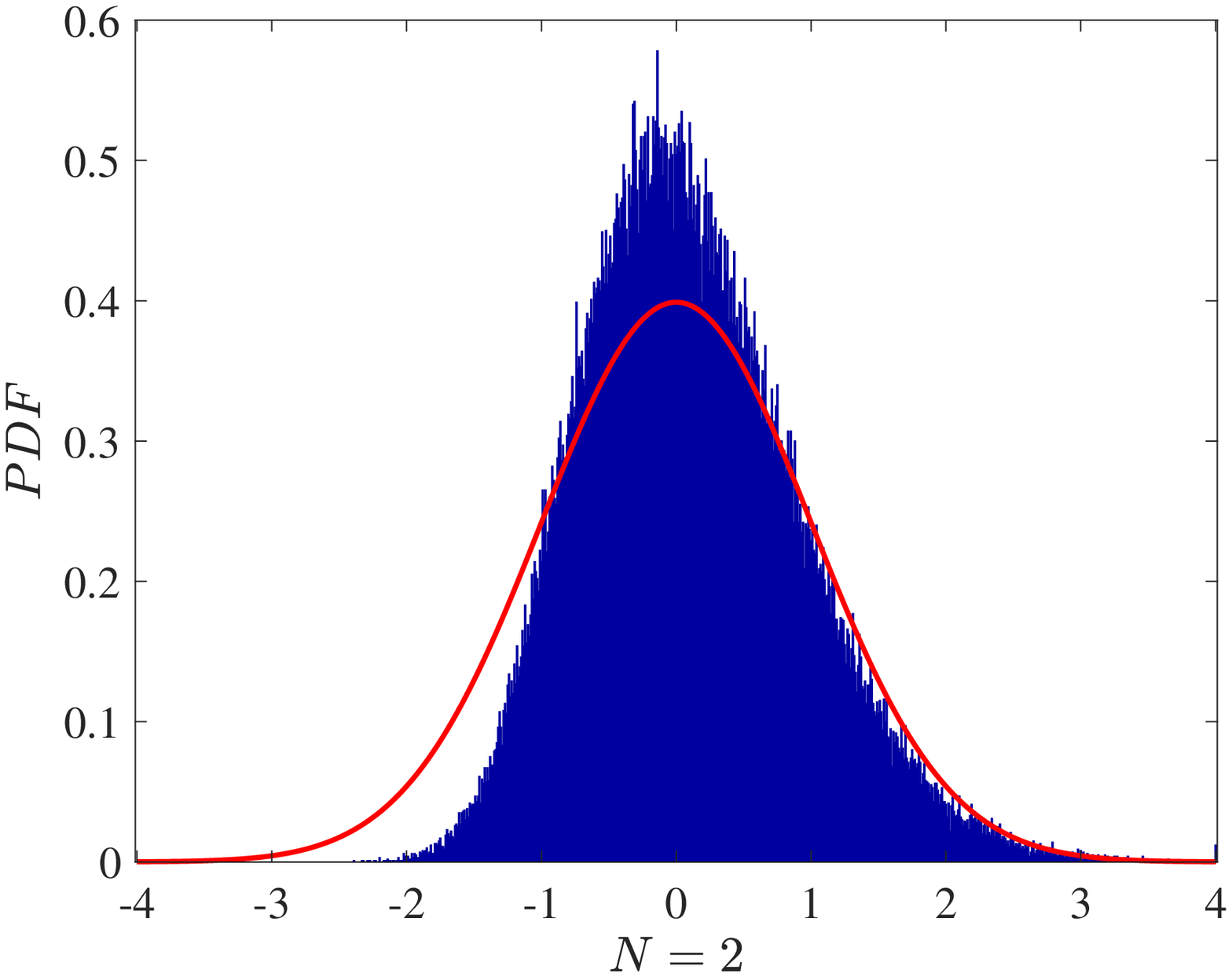}}
  \subfloat[\label{pdf_4}]{
        \includegraphics[width=0.3\linewidth]{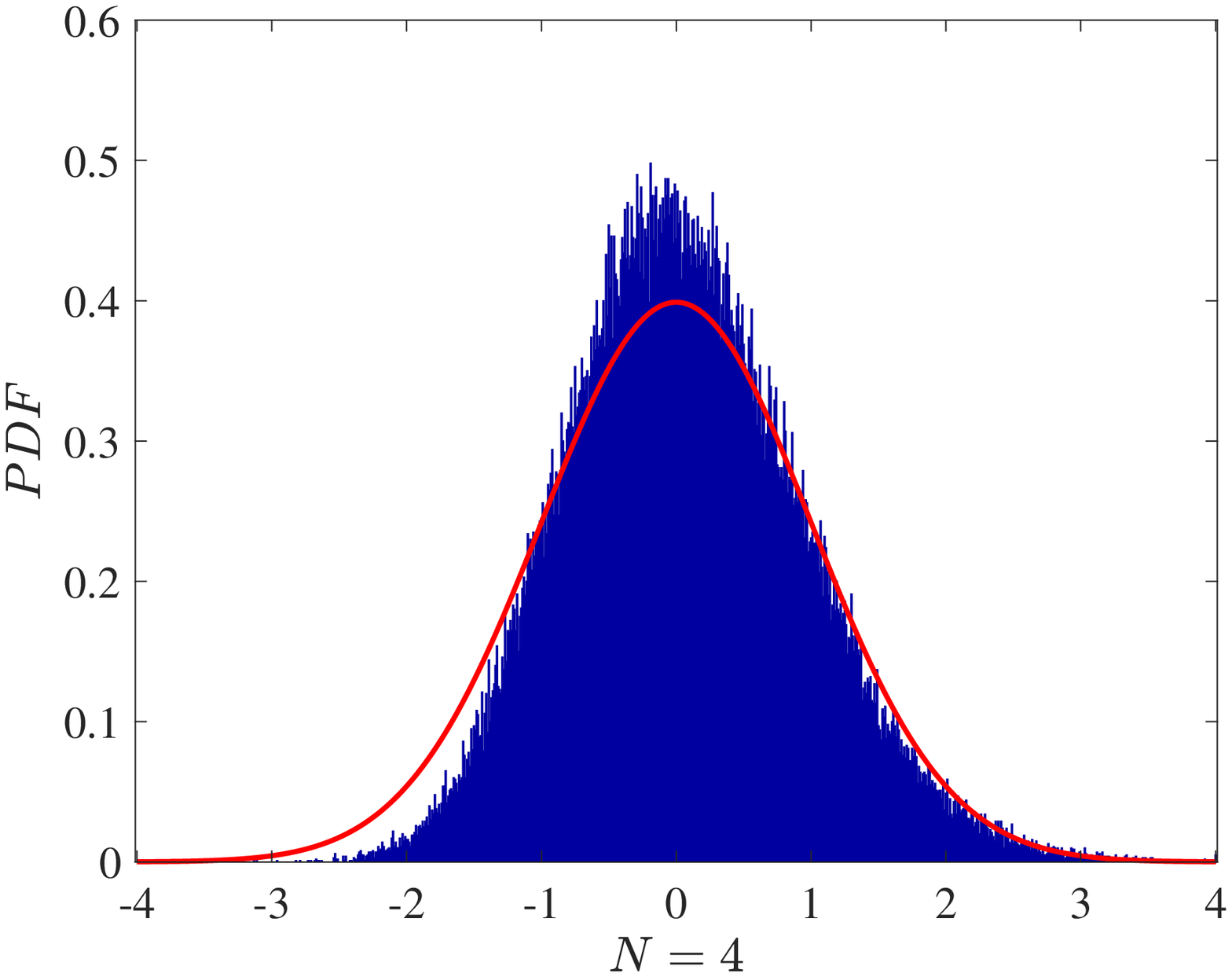}}
  \subfloat[\label{pdf_8}]{
        \includegraphics[width=0.3\linewidth]{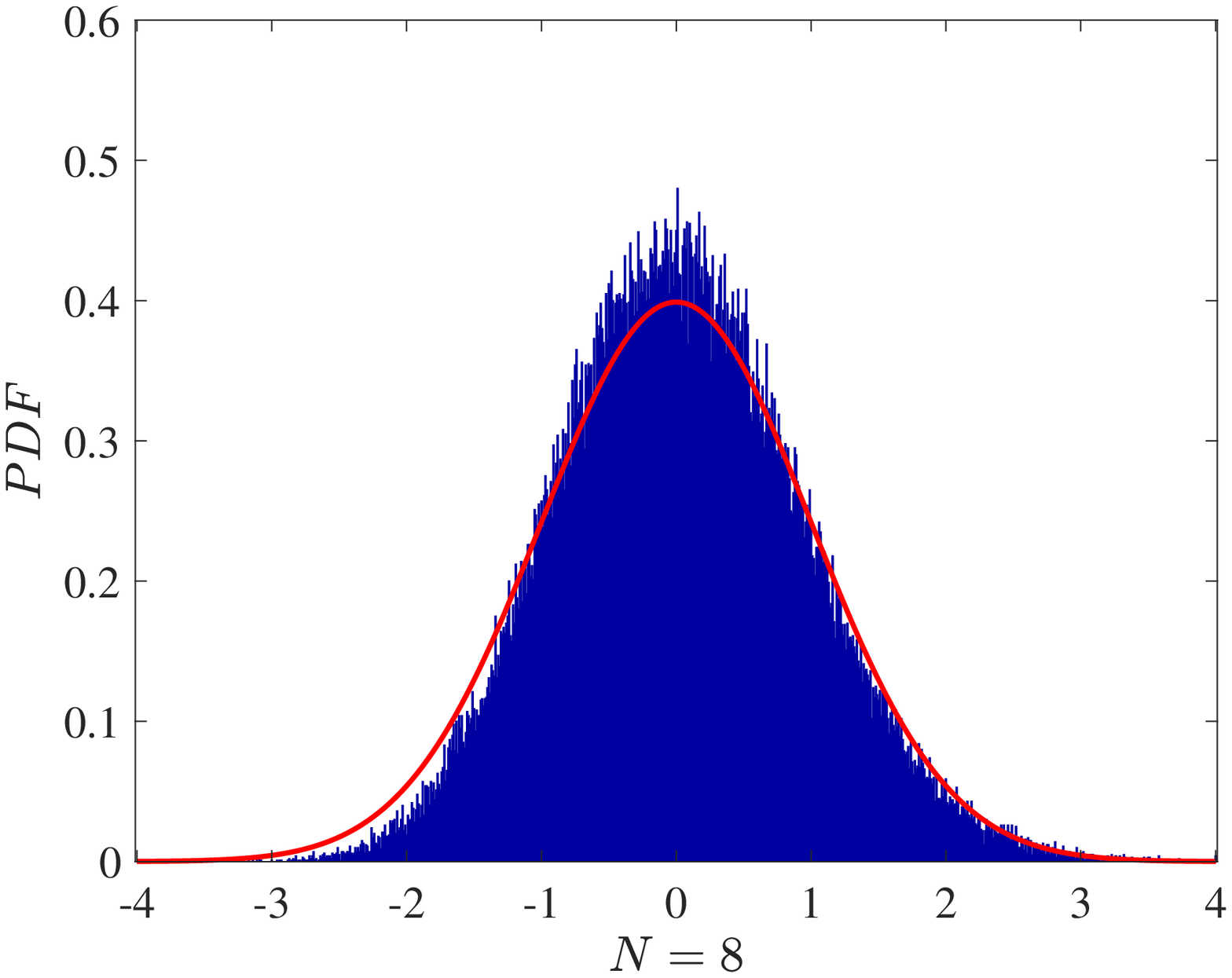}}
    \hfill
  \subfloat[\label{pdf_16}]{
        \includegraphics[width=0.3\linewidth]{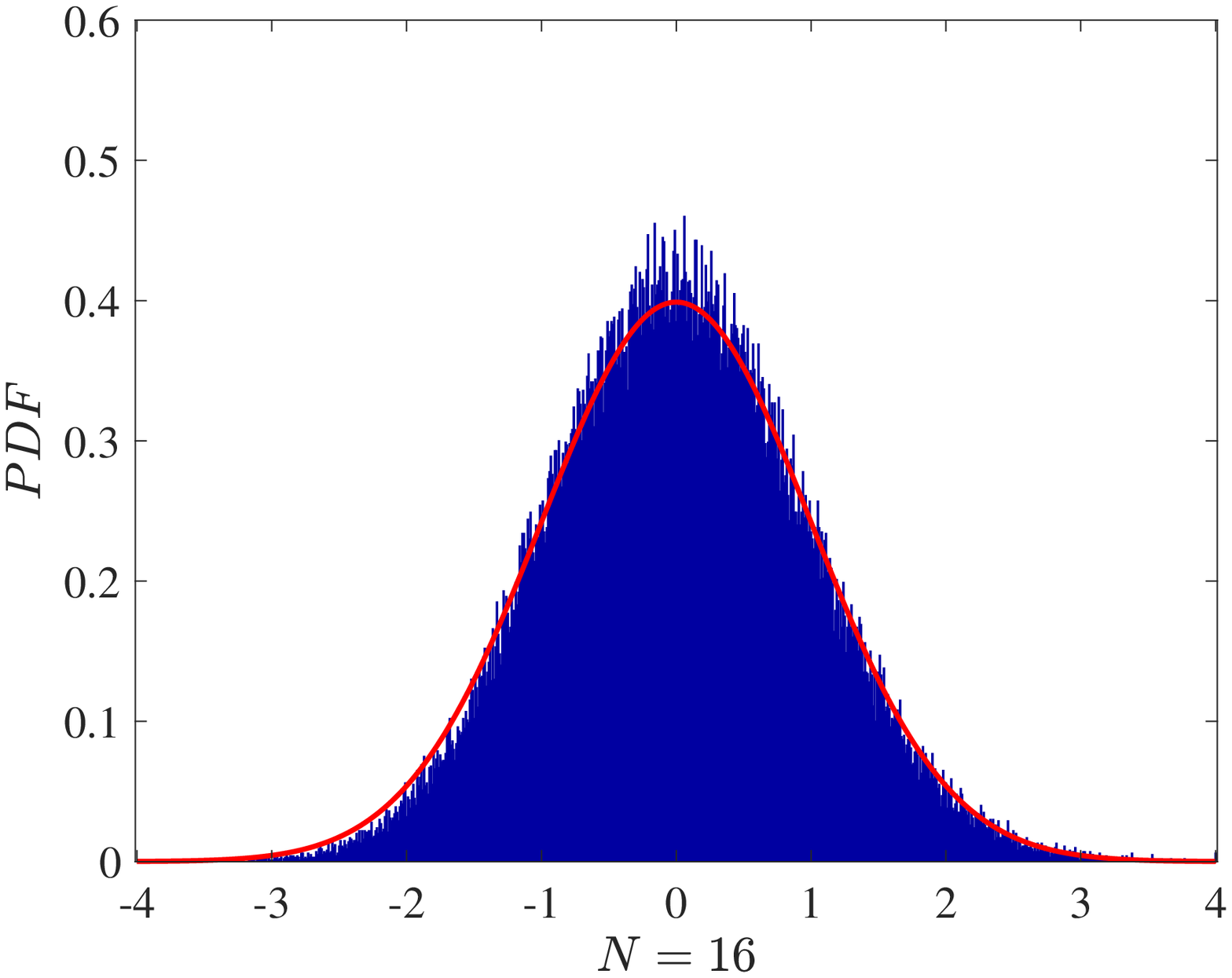}}
    \subfloat[\label{pdf_32}]{
        \includegraphics[width=0.3\linewidth]{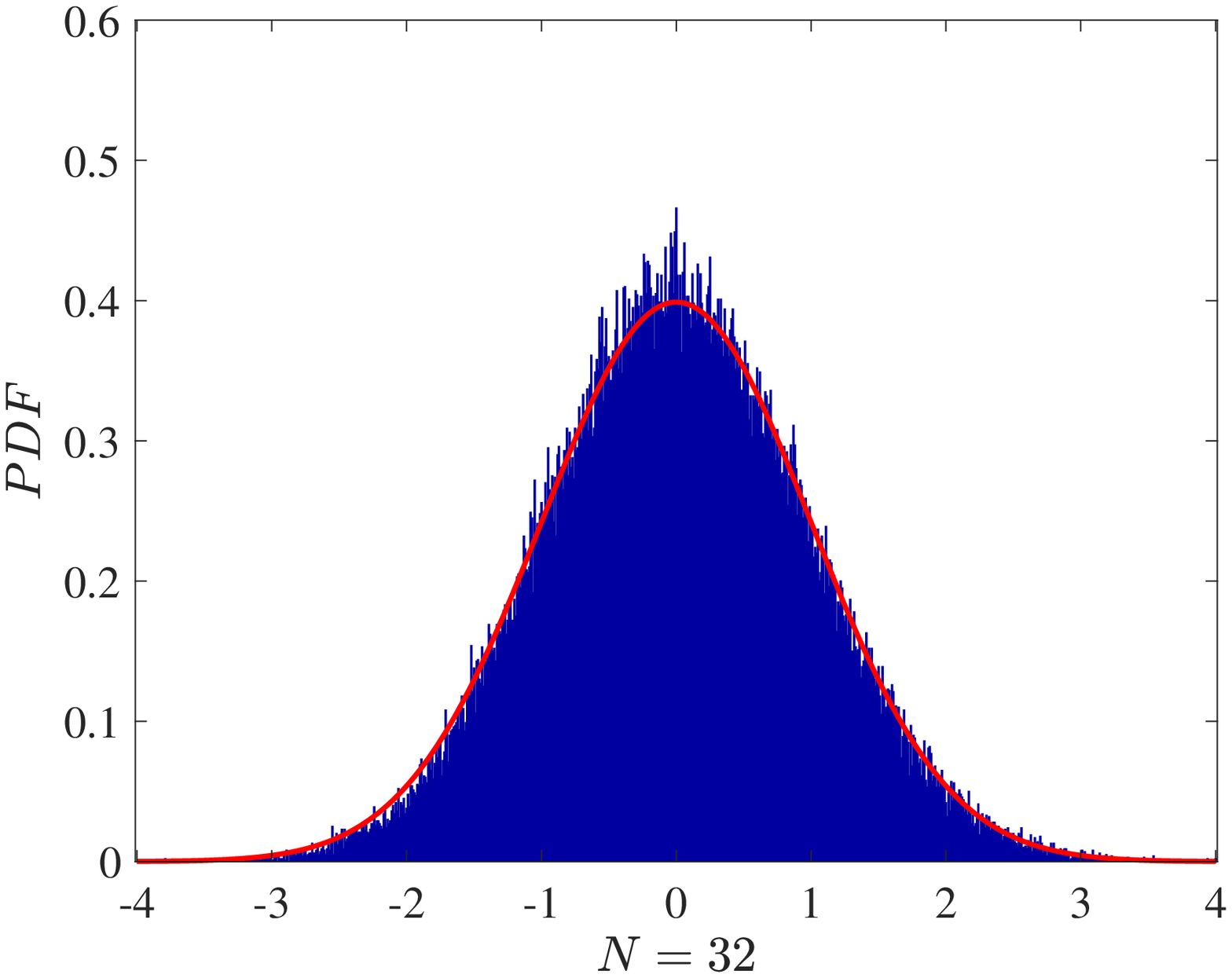}}  
        \subfloat[\label{pdf_64}]{
        \includegraphics[width=0.3\linewidth]{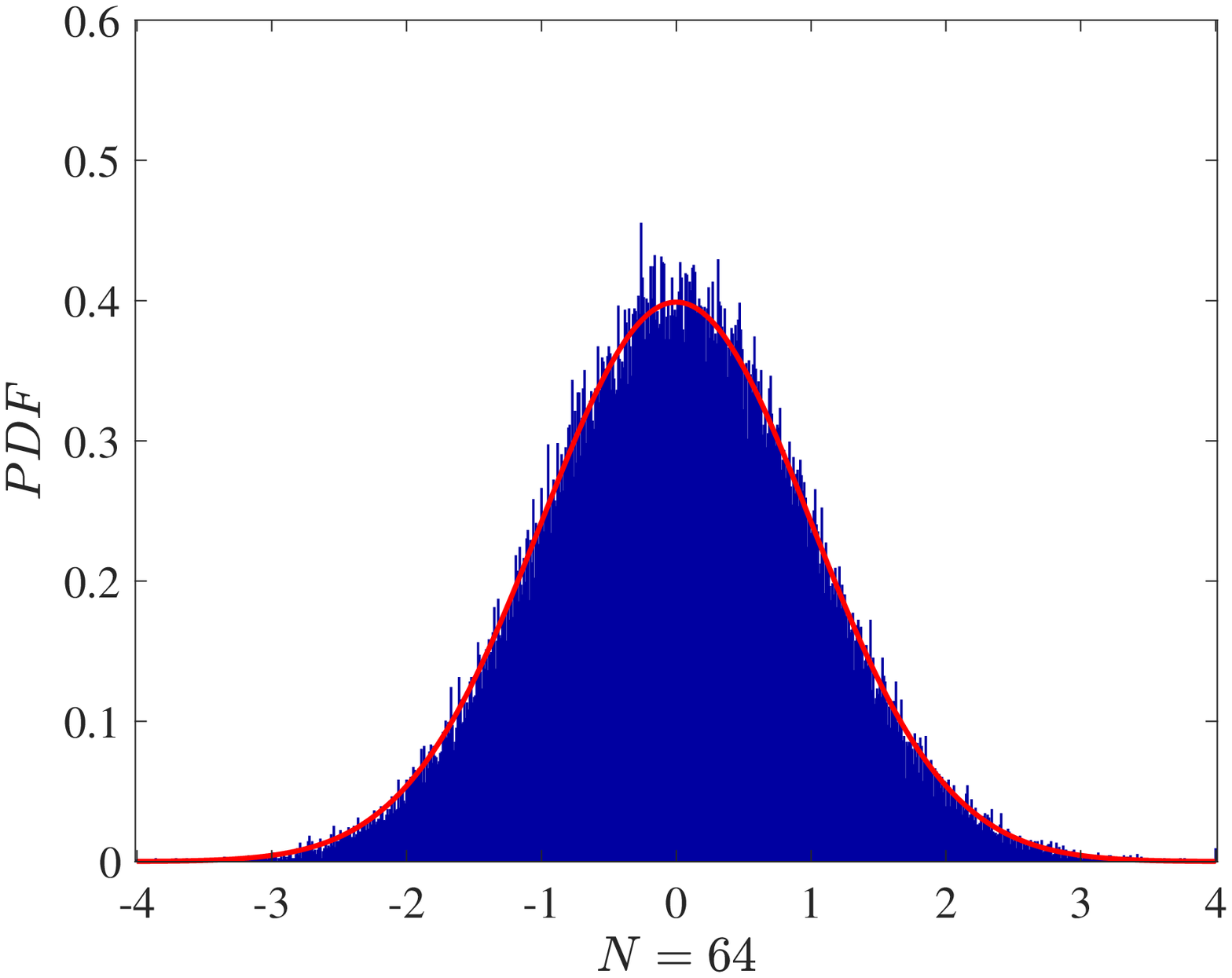}}      
\caption{The fitness of the CLT with modified mean and variance.} \label{pdf_N}
\end{figure*}

We first validate the accuracy of the modified mean and variance by utilizing them to approximate the distribution of the MI with CLT. Here, we set $\vartheta=0.6$ ($\sigma_r^2=1.6$, $\sigma_i^2=0.4$), $\sigma^2=0.2$, and $c=\frac{N}{M}=0.5$. $r$ follows the Weibull distribution with parameter $k=1$ and the number of realizations is $50000$. In Fig.~\ref{pdf_2} to Fig.~\ref{pdf_64}, we compare the empirical PDFs (normalized histogram) of the normalized MI ${\frac{(C-\overline{C})}{\sqrt{\Theta}}}$ (plotted in blue) with the standard Gaussian distribution (plotted in red). It can be observed that with the modified mean and variance, the CLT provides an asymptotically accurate approximation for the fluctuation of the MI, which validates the accuracy of the bias.

\subsection{Biases for the mean and variance}
\begin{figure}[!ht]
    \centering
    \includegraphics[width=1.0\linewidth]{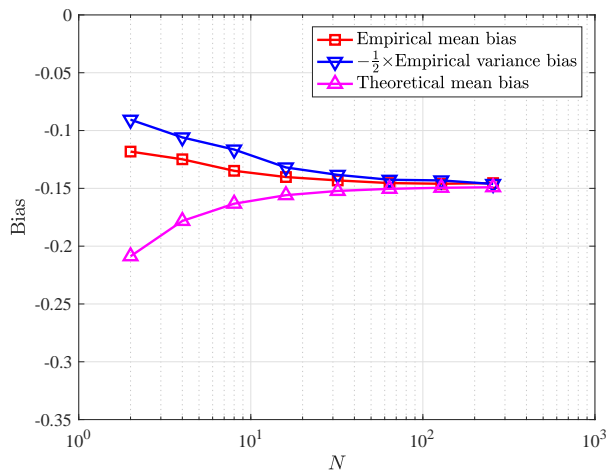}
  \caption{Numerical evaluation of the bias.}
  \label{bias_accuracy}
\end{figure}
Fig.~\ref{bias_accuracy} shows the derived bias for the mean and the empirical biases for both the mean and variance, where the settings are the same as those in Fig.~\ref{pdf_N}. It can be observed that the biases do not converge to zero but approach a constant  when $N=cM$ grows larger. The result also validates the relation that the bias of the mean is $-0.5$ times of that for the variance, which is proved in Proposition~\ref{p_rel}.

\subsection{Impact of the biases}
\begin{figure}[!ht] 
    \centering
    \includegraphics[width=1.0\linewidth]{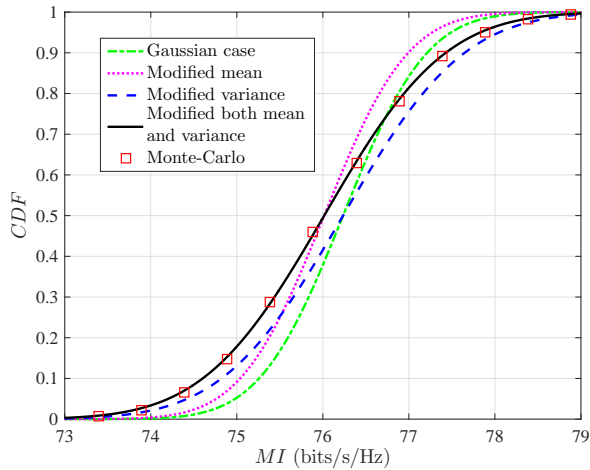}
  \caption{Numerical evaluation of different CLTs for the MI.}
  \label{cdf_opt}     
\end{figure}
\begin{figure}[!ht]
    \centering
    \includegraphics[width=1.0\linewidth]{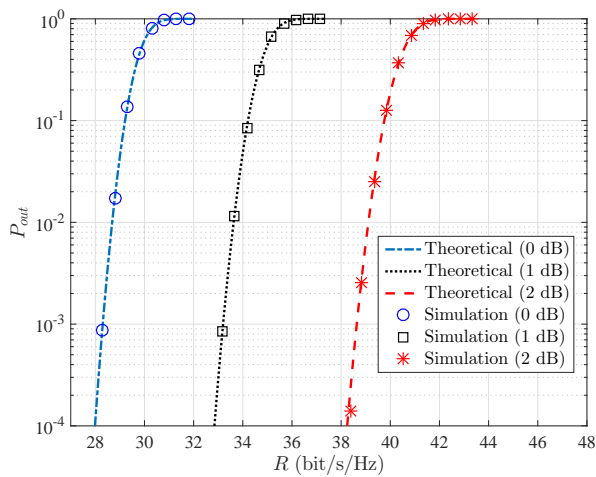}
      \caption{Outage probability determined with the modified CLT.}
      \label{outage_R}
\end{figure}
In Fig.~\ref{cdf_opt}, we compare the empirical CDF and CLT of the MI for four different cases, i.e., CLT for Gaussian channels, CLT for non-Gaussian channels with the mean modified by its bias, CLT for non-Gaussian channels with the variance modified by its bias, and the one in Proposition~\ref{p_rel}, where both the mean and variance are modified. The simulation is performed with the settings $N=32$, $c=0.5$, $\sigma^2=0.2$, $\vartheta=0.6$ and the numder of realizations is $10^{5}$. It can be observed that the bias of the mean corresponds to a shift of the distribution and, when both the biases for the mean and variance are fixed, the theoretical result matches the empirical one very well. To further illustrate the accuracy of the derived result and its application in practical communication systems, we show in Fig.~\ref{outage_R} the outage probabilities with different SNR (signal-to-noise ratio) values. It can be observed that the modified CLT provides accurate estimation of the outage probability.

\subsection{Impact of CV}
\begin{figure}[!ht]
    \centering
    \includegraphics[width=1.0\linewidth]{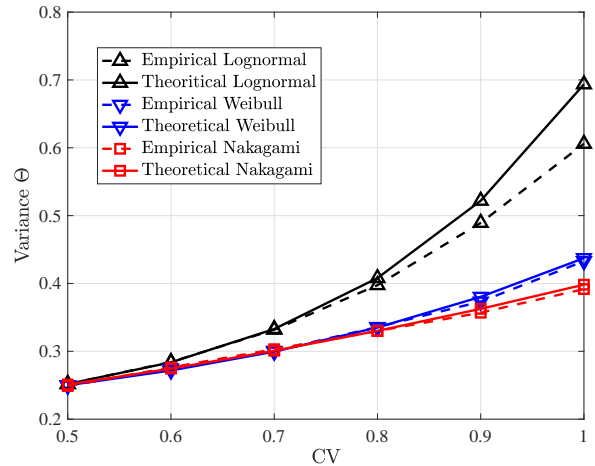}
      \caption{Impact of CV on estimation accuracy.}
      \label{cv_var}
\end{figure}
Finally, we show the approximation accuracy with different fading channels. Here we consider three types of distributions, including Lognormal, Nakagami-m, and Weibull as shown in Table~\ref{para_table}. We set $\sigma^2=0.2$ and the number realizations is $50000$. In Fig.~\ref{cv_var}, we plot both the empirical and theoretical variances with respect to CV, ranging from 0.5 to 1. Note that a larger CV corresponds to a larger variance, i.e., more severe fading. It can be observed that the estimation is less accurate for more severe fading channels.

\section{conclusion}
\label{conc}

In this paper, we investigated the bias of the EMI for MIMO channels, caused by non-Gaussianity and non-circularity. For that 
purpose, we first derived an explicit expression for the bias of the resolvent and generalized it for the LSS of non-centered random matrices, which resolves the computationally-challenging problem mentioned in~\cite{banna2020clt}. With the new result, we also derived a tighter approximation for the mean of the LSS. By applying the above results to MIMO channels, we calculated the bias of the EMI for non-centered and non-Gaussian MIMO channels, which includes the previous results for Gaussian and centered channels as special cases. Furthermore, we showed that the bias for the mean is $-0.5$ times of that for the variance. The derived biases for the mean and variance were utilized to provide a modified CLT and calculate the outage probability. Numerical results validated the accuracy of the derived biases and their effectiveness in evaluating the MI and outage probability of MIMO systems. Our results represent one step forward in evaluating the LSS of non-centered random matrices (RMT perspective) and the MI of MIMO channels (communication perspective). The analysis for the general correlated and low-rank MIMO channels requires further investigation.

\begin{appendices}
\section{Some useful results}
\label{rmt_results}

Here we introduce some important results that will be utilized in the proof of Lemma~\ref{u_lemma} in Appendix~\ref{lemma_1_proof}. The following two lemmas, i.e., Lemma~\ref{bound_lemma} and Lemma~\ref{rank_lemma}, will be used to evaluate the approximation error between $\bold{T}(z)$ and $\bold{Q}(z)$.
\begin{lemma}(Error bounds for trace and bilinear form of $\bold{Q}$~\cite{hachem2012clt},~\cite{hachem2013bilinear}) 
\label{bound_lemma}
Let $\bold{u}\in \mathbb{C}^{N}$ and $\bold{v} \in \mathbb{C}^{N}$ be two deterministic vectors with bounded Euclidean norm
\begin{equation}
\|\bold{u} \|_2 < \infty, ~\|\bold{v} \|_2 < \infty,
\end{equation}
and $\bold{U}\in \mathbb{C}^{N\times N} $ be a deterministic matrix with bounded spectral norm. If assumptions \textbf{A.1}-\textbf{A.4} hold, then 
given $\| \bold{U}\|< \infty$, 
there exist constants $K$, $K'$, $K_{p}$ and $K_{p}'$ such that the following bounds hold
\\
(i) bound for trace:
 \begin{equation} 
  \label{l3_1}     
     \frac{1}{M}|\Tr \bold{U}(\E\bold{Q}(z)-\bold{T}(z)) | \le \frac{K}{M},
\end{equation}
 \begin{equation} 
  \label{l3_11}     
     \frac{1}{M}|\Tr \bold{U}(\E\bold{Q}(z)-\bold{C}(z)) | \le \frac{K'}{M},
\end{equation}
(ii) bound for variance:
 \begin{equation}
 \label{l3_2}    
 \Var (\Tr\bold{U}\bold{Q}(z)) \le K',
\end{equation}
(iii) bound for the bias of the bilinear form: for $p \in \left[1,2 \right]$,
\begin{equation}
\E |\bold{u}^{H}(\bold{Q}(z)-\bold{T}(z))\bold{v} |^{2p} \le \frac{K_p}{M^p},
\end{equation}
\begin{equation}
\label{q_eq_bi}
\E |\bold{u}^{H}(\bold{Q}(z)-\E\bold{Q}(z))\bold{v} |^{2p} \le \frac{K'_p}{M^p}.
\end{equation}

\end{lemma}

The following lemma gives an estimate for a rank-one perturbation of the resolvent. Specially, if we take $\bold{u}=\bold{v}=\bold{e}_{i}$, where $\bold{e}_{i}$ denotes the column vector whose $i$-th entry is $1$ and others are zero, \textit{(iii)} will become the bound for diagonal elements
\begin{equation}
\label{q_t_bound}
\E |q_{ii}(z)-t_{ii}(z) |^{2p} \le \frac{K_p}{M^p},
\end{equation}
\begin{equation}
\label{q_e_bound}
\E |q_{ii}(z)-\E q_{ii}(z) |^{2p} \le \frac{K'_p}{M^p}.
\end{equation}

\begin{lemma} 
\label{rank_lemma}
(Bound for rank one perturbations, Lemma 2.6 in~\cite{silverstein1995empirical}) For any matrix $\bold{A}$ and $z\in \mathbb{C}  \setminus \mathbb{R}^{+}$, the resolvent $\bold{Q}(z)$ and the perturbed resolvent $\bold{Q}_{j}(z)=\left(-z\bold{I}+\bold{H}\bold{H}^{H}-\bold{h}_{j} \bold{h}_{j}^{H}\right)^{-1}$ satisfy
\begin{equation}
\label{rank_one}
|\Tr\bold{A}(\bold{Q}_{j}(z)-\bold{Q}(z)) |\le \frac{\|\bold{A} \|}{dist(z,\mathbb{R}^{+})}.
\end{equation}
\end{lemma}

We need the following lemma to handle the variance and covariance of errors.
\begin{lemma}
\label{lemma_ext}
 (Expansion of covariance of two quadratic forms, Eq. (3.20) in~\cite{hachem2012clt}) Let $\bold{z}=\bold{a}+ \frac{1}{\sqrt{N}}\bold{D}^{\frac{1}{2}}\bold{x} $, where $\bold{x}=(X_{1},X_{2},...,X_{N})^{T}$, is a random vector with i.i.d. entries, $\bold{D}$ is a diagonal non-negative matrix and $\bold{a} \in \mathbb{C}^{N}$ is a deterministic vector. Assuming that $\bold{\Gamma}  $ and $\bold{\Lambda}$ are two $N\times N$ deterministic matrices, the covariance of the quadratic forms $\bold{z}^{H}\bold{\Gamma}\bold{z}$ and $\bold{z}^{H}\bold{\Lambda}\bold{z}$ is given by
\begin{equation}
\label{qua_ext}
\begin{aligned}
& \E (\bold{z}^{H}\bold{\Gamma}\bold{z} - \E \bold{z}^{H}\bold{\Gamma}\bold{z}  ) 
(\bold{z}^{H}\bold{\Lambda}\bold{z} - \E \bold{z}^{H}\bold{\Lambda}\bold{z}  )  
\\
&=\frac{1}{N^2}\Tr \bold{\Gamma}\bold{D}\bold{\Lambda}\bold{D}
\!+\!  \frac{1}{N} \bold{a}^{H}\bold{\Gamma}\bold{D}\bold{\Lambda}\bold{a}
\!+\!  \frac{1}{N} \bold{a}^{H}\bold{\Lambda}\bold{D}\bold{\Gamma}\bold{a}
\\
&\!+\!   \frac{|\vartheta|^2}{N^2}\Tr \bold{\Gamma}\bold{D}\bold{\Lambda}^{T}\bold{D}
\!+\!  \frac{\vartheta}{N} \bold{a}^{H}\bold{\Lambda}\bold{D}\bold{\Gamma}^{T}\overline{\bold{a}}
\!+\! \frac{\overline{\vartheta}}{N} \bold{a}^{T}\bold{\Lambda}^{T}\bold{D}\bold{\Gamma}\bold{a}
\\
&\!+\! \frac{\zeta}{N^{\frac{3}{2}}} \bold{a}^{H} \bold{\Lambda} \bold{D}^{\frac{3}{2}}\mathrm{vdiag}(\bold{\Gamma}) 
\!+\! \frac{\overline{\zeta}}{N^{\frac{3}{2}}} \bold{a}^{H} \bold{\Gamma} \bold{D}^{\frac{3}{2}}\mathrm{vdiag}(\bold{\Lambda}) 
\\
&\!+\! \frac{\kappa}{N^2}\Tr\bold{D}^2 \mathrm{diag}(\bold{\Lambda} )\mathrm{diag}(\bold{\Gamma} ),
\end{aligned}
\end{equation}
where $\vartheta$, $\kappa$ and $\zeta$ are given in~(\ref{moments}).
\end{lemma}
It can be observed from (\ref{qua_ext}) that the $\vartheta$ and $\kappa$ related terms will not vanish if $\vartheta$ and $\kappa$ are non-zero. (\ref{qua_ext}) will be used to evaluate small quantities.

 \begin{table*}[!htbp]
\centering
\caption{Expressions for Lemma~\ref{t_lemma} and~\ref{ct_lemma}}
\label{apen_var}
\begin{tabular}{|c|c|c|c|c|c|c|}
\toprule
  & Symbol& Expression &  Symbol& Expression& Symbol& Expression \\
  \midrule
  \multirow{3}{*}{Expectations} &
 $\underline{\alpha}_{T}$ & $ \frac{1}{M}\E\Tr\bold{Q}^{T}\bold{D}\bold{Q}\bold{D} $
&
$\underline{\beta}_{T}$ & $\frac{1}{M}\sum\limits_{i=1}^{M}\omega^2 \widetilde{d}_{i}\widetilde{t}^2_{ii}\E\bold{a}^{T}_{i}\bold{Q}^{T}_{i}\bold{D}\bold{Q}_{i}\bold{a}_{i}  $
&$\alpha_{T}(\bold{U})$
&
$
\frac{1}{M}\E\Tr\bold{Q} \bold{U}\bold{Q}^{T}\bold{D} $
\\
&
${\beta}_{T}(\bold{U})$ & $\frac{1}{M}\sum\limits_{i=1}^{M}\omega^2 \widetilde{d}_{i}\widetilde{t}^2_{ii}\E\bold{a}^{H}_{i}\bold{Q}_{i}\bold{U}\bold{Q}^{T}_{i}\overline{\bold{a}}_{i}  $
&
&
&
&
\\
\midrule
 \multirow{1}{*}{Deterministic quantity}
& $G$
&
$ \sum\limits_{i,j=1, i \ne j}^{M}\frac{\widetilde{d}_{i}\widetilde{d}_{j}|\bold{a}_{j}^{H}{\bold{T}}{\bold{a}}_{i}|^2}{M^2(1+\delta\widetilde{d}_{i})^2(1+\delta\widetilde{d}_{j})^2}$
&
${G}_{T}$
&
$ \sum\limits_{i,j=1, i \ne j}^{M}\frac{\widetilde{d}_{i}\widetilde{d}_{j}(\bold{a}_{j}^{H}{\bold{T}}{\bold{a}}_{i})^2}{M^2(1+\delta\widetilde{d}_{i})^2(1+\delta\widetilde{d}_{j})^2}$
&$L$
&
$\frac{\omega^2}{M}\sum\limits_{i=1}^{M} \widetilde{t}_{ii}^{2}\widetilde{d}^2_{i}$
\\
\bottomrule
\end{tabular}
 \end{table*}

The following two lemmas give the evaluation of  $\alpha_{T}(\bold{U})$, $\beta_{T}(\bold{U})$, $\underline{ \alpha}_{T}$ and $\underline{\beta}_{T}$, which are defined in Table~\ref{apen_var} and will be used in the proof of Lemma~\ref{u_lemma}.
\begin{lemma}
\label{t_lemma}
The approximation of $\underline{\alpha}_{T}$ and $\underline{\beta}_{T}$ is determined by solving the following system of equations:
\begin{equation}
\begin{bmatrix} 
1- \overline{\vartheta} \underline{F}_{T} &  -{\vartheta}{{G}}_{T}-|\vartheta|^2 \underline{F}_{T}L
\\
 -\overline{\vartheta}\gamma_{T}     & 1 -\vartheta {{F}}_{T}- |\vartheta|^2\gamma_{T} L
\end{bmatrix}
\begin{bmatrix} 
\underline{ \beta}_{T}
 \\
\underline{\alpha}_{T}
\end{bmatrix}
=
\begin{bmatrix}
\underline{F}_{T}
\\ 
\gamma_{T}
\end{bmatrix}
+\bm{\varepsilon},
\end{equation}
where $\|\bm{\varepsilon}\| = O(M^{-\frac{1}{2}})$ and other symbols are given in Table~\ref{apen_var}.
\end{lemma}

\begin{lemma} 
\label{ct_lemma}
Let $\bold{U} \in \mathbb{C}^{N\times N}$ be a deterministic matrix with bounded spectral norm. Then the approximation of ${\alpha}_{T}(\bold{U})$ and ${\beta}_{T}(\bold{U})$ is determined by solving the following system of equations:
\begin{equation}
\begin{bmatrix} 
1 \!-\! \vartheta F_{T} &  -\overline{\vartheta}{G}_{T}\!-\! |\vartheta|^2 F_{T}L    \\  
-\vartheta\gamma_{T}     & 1 \!-\! \overline{\vartheta}\underline{F}_{T} \!-\!|\vartheta|^2 \gamma_{T}  L
\end{bmatrix}
\begin{bmatrix} 
 \beta_{T}(\bold{U})
 \\
\alpha_{T}(\bold{U})
\end{bmatrix}
\!=\!
\begin{bmatrix} 
F_{T}(\bold{U})
 \\
\gamma_{T}(\bold{U})
\end{bmatrix}
\!+\! \bm{\varepsilon}',
\end{equation} 
where $\|\bm{\varepsilon}'\| = O(M^{-\frac{1}{2}})$ and other symbols are given in Table~\ref{apen_var}.
\end{lemma}

The proofs of Lemma~\ref{t_lemma} and Lemma~\ref{ct_lemma} are omitted here as they are similar to those in~\cite{hachem2012clt},~\cite{kammoun2019asymptotic}. Given $G_{T}+L=\omega^2\widetilde{\gamma}_{T}$, we can show that the determinant of the coefficient matrix is $\Delta_{T}$. Given the range of $z=-\omega$ can be generalized to $z \in \mathbb{C}\setminus \mathbb{R}^{+}$ and $\Delta_{T}$ will not vanish~\cite{hachem2012clt} under the assumptions \textbf{A.3} and \textbf{A.4}, the evaluations for $\underline{ \alpha}_{T}$, $\underline{ \beta}_{T}$ exist.

\section{Proof of Lemma~\ref{delta_p}}
\label{proof_delta_p}
\begin{IEEEproof}
Taking the derivative of $\omega$ over the identity $\delta=\frac{1}{M}\Tr\bold{D}\bold{T}$, we have
\begin{equation}
\begin{aligned}
\delta'_\omega&=-\frac{1}{M}( \Tr \bold{D}\bold{T}^{2}
 + \widetilde{\delta}  \Tr \bold{D}\bold{T} \bold{D}\bold{T}
 +\omega\widetilde{\delta}'_{\omega}  \Tr \bold{D}\bold{T} \bold{D}\bold{T}
 \\
 & -\delta'_\omega \Tr \bold{D}\bold{T} \bold{A}\widetilde{\bold{R}}^{2}\widetilde{\bold{D}}\bold{A}^{H} \bold{T} 
 )
 \\
 &=-( \frac{1}{M}\Tr \bold{D}\bold{T}^{2}
 + \widetilde{\delta}  \gamma
 +\omega\widetilde{\delta}'_{\omega}  \gamma
- \delta'_{\omega} F
 ).
\end{aligned}
\end{equation}
By applying the same operations to $\widetilde{\delta}=\frac{1}{M}\Tr\widetilde{\bold{D}}\widetilde{\bold{T}}$, we can obtain the other equation in~(\ref{delta_p_eq}).
\end{IEEEproof}

\section{Proof of Lemma~\ref{u_lemma}}
\label{lemma_1_proof}

We will first introduce some quantities, that will be widely used in the proof.

\textit{1. Perturbed resolvent:} The resolvent $\bold{Q}(z)$ perturbed by the rank-one matrix $\bold{h}_{j} \bold{h}_{j}^{H}$ is given by
\begin{equation}
\bold{Q}_{j}(z)=\left(-z\bold{I}+\bold{H}\bold{H}^{H}-\bold{h}_{j} \bold{h}_{j}^{H}\right)^{-1},
\end{equation}
where $\bold{h}_{j}$ is the $j$-th column of $\bold{H}$. In fact, $\bold{Q}_{j}(z)$ is equal to $\bold{Q}(z)$ perturbed by the rank-one matrix $\bold{h}_{j}\bold{h}^{H}_{j}$. The approximation for $\E \bold{Q}_{j}(z)$, i.e. $\bold{T}_{j}(z)$, is given by
\begin{equation}
\label{t_jz}
\bold{T}_{j}(z)=\left(-z\left( \bold{I}+\widetilde{\delta}\bold{D} \right)+\bold{A}_{j}\left( \bold{I}+\delta\widetilde{\bold{D}}_{j} \right)^{-1}\bold{A}^{H}_{j} \right)^{-1}\!,
\end{equation}
where we can get $\bold{A}_{j}$ from $\bold{A}$ by removing the $j$-th column. The same holds for $\widetilde{\bold{D}}_{j}$.

\textit{2. Diagonal element of the co-resolvent:}
The  $j$-th element on the diagonal of $\widetilde{\bold{Q}}$, $\widetilde{q}_{jj}(z)$, can be given by
\begin{equation}
\label{q_jj}
\widetilde{q}_{jj}(z)=\frac{-1}{z(1+  \bold{h}_{j}^{H}\bold{Q}_{j} \bold{h}_{j} )}.
\end{equation}
By~(\ref{t_jz}), the approximation for $\widetilde{q}_{jj}(z)$ can be given by
\begin{equation}
\widetilde{t}_{jj}(z)=\frac{-1}{z(1+  \bold{a}_{j}^{H}\bold{T}_{j} \bold{a}_{j}+\widetilde{d}_{j}\delta )},
\end{equation}
and given a vector $\bold{b}$, there holds (Eq. 3.11 in~\cite{hachem2012clt})
\begin{equation}
\label{t_tj}
-z\widetilde{t}_{jj}(z)\bold{a}_{j}^{H}\bold{T}_{j}(z)\bold{b}=\frac{\bold{a}_{j}^{H}\bold{T}(z)\bold{b} }{1+\widetilde{d}_{j}\delta}.
\end{equation}

\textit{3. Intermediate approximation of the diagonal element and its error:} The intermediate approximation $\widetilde{b}_{j}(z)$ for $\widetilde{q}_{jj}(z)$ can be given by 
\begin{equation}
\widetilde{b}_{j}(z)=\frac{-1}{z(1+  \bold{a}_{j}^{H}\bold{Q}_{j}(z) \bold{a}_{j}+\frac{\widetilde{d}_{j}}{M}\Tr\bold{D}\bold{Q}_{j}(z) )}.
\end{equation}
The error $e_{j}(z)$ is
\begin{equation}
e_{j}(z)= \bold{h}_{j}^{H}  \bold{Q}_{j}(z)\bold{h}_{j}-\bold{a}_{j}^{H}\bold{Q}_{j} \bold{a}_{j}-\frac{\widetilde{d}_{j}}{M}\Tr\bold{D}\bold{Q}_{j}(z),
\end{equation}
which is obtained by taking expectation over $\bold{h}_{j}$ in the denominator of~(\ref{q_jj}) and is bounded by
\begin{equation}
\label{bound_e}
\E |e_{j}(z)|^p = O(M^{-\frac{p}{2}}),
\end{equation}
for $p \ge 2$. Furthermore, we have the following relation
\begin{equation}
\begin{aligned}
\label{replace_q}
\widetilde{q}_{jj}(z)&=\widetilde{b}_{j}(z)+z\widetilde{q}_{jj}(z)\widetilde{b}_{j}(z)e_{j}(z)
\\
&=\widetilde{b}_{j}(z)+z\widetilde{b}^{2}_{j}(z)e_{j}(z)+z^2\widetilde{q}_{jj}(z)\widetilde{b}^{2}_{j}(z)e^2_{j}(z).
\end{aligned}
\end{equation}
This identity is essential in the derivation of the bias, in which we will use the intermediate quantities $\widetilde{b}_{j}(z)$ and the error ${e}_{j}(z)$ to replace 
$\widetilde{q}_{jj}(z)$, and the terms with high order of ${e}_{j}(z)$ will vanish when $M$ becomes large.

\vspace{-0.5cm}

\subsection{Sketch of proof}
From Lemma~\ref{lemma_ext}, we know that the bias should be related to $\vartheta$ and $\kappa$. The basic idea of the proof is to expand $\widetilde{q}_{jj}$ by~(\ref{replace_q}) and use~(\ref{qua_ext}) to handle the expectation over small quantities. The proof can be summarized as:

\textit{Step 1. The expectation-form expression related to $\vartheta$ and $\kappa$:} We show that the terms unrelated to $\vartheta$ and $\kappa$ can be ignored asymptotically, i.e. $\mathcal{Z}(\bold{U})\xlongrightarrow[]{M\rightarrow \infty}\mathcal{Z}_{\theta}(\bold{U})+\mathcal{Z}_{\kappa}(\bold{U})$.

\textit{Step 2. The deterministic expression:} Then we compute the asymptotical expression of $\mathcal{Z}_{\theta}(\bold{U})$ and $\mathcal{Z}_{\kappa}(\bold{U})$, which can be represented by the deterministic quantities in Table~\ref{var_list}.

\vspace{-0.3cm}

\subsection{Details of proof}
\begin{IEEEproof} 
\subsubsection{The expectation-form related to $\vartheta$ and $\kappa$}

First, by the matrix identity $\bold{A}-\bold{B}=\bold{B}(\bold{B}^{-1}-\bold{A}^{-1})\bold{A}$, we have
\begin{equation}
\label{root_eq}
\begin{aligned}
&-\mathcal{Z}(\bold{U})=\E \Tr\bold{U}\left(\bold{C}-\E \bold{Q} \right)=\E \Tr\bold{U}\bold{C} \bold{H}\bold{H}^{H} \bold{Q}
\\
&- \omega\widetilde{\alpha} \Tr\bold{U}\bold{C}\bold{D} \bold{Q}  - \Tr\bold{U}\bold{C}\bold{A}\left( \bold{I}+\alpha\widetilde{\bold{D}} \right)^{-1}\bold{A}^{H}    \bold{Q} 
\\
&=X_1 + X_2 + X_3.
\end{aligned}
\end{equation}
We will first handle the term $X_1$, which can be decomposed as
\begin{equation}
\begin{aligned}
& X_1 = \E\sum_{i=1}^{M}  \bold{h}^{H}_{i}\bold{Q}\bold{U}\bold{C}\bold{h}_{i}
\xlongequal[]{a} \E\sum_{i=1}^{M} \omega\widetilde{q}_{ii}  \bold{h}^{H}_{i}\bold{Q}_{i}\bold{U}\bold{C}\bold{h}_{i}
\\
 &\xlongequal[]{b}  \E\sum_{i=1}^{M} \omega\widetilde{b}_{i}  \bold{h}^{H}_{i}\bold{Q}_{i}\bold{U}\bold{C}\bold{h}_{i}
\! -\! \E\sum_{i=1}^{M} \omega^2\widetilde{b}_{i}^2 e_{i}  \bold{h}^{H}_{i}\bold{Q}_{i}\bold{U}\bold{C}\bold{h}_{i}
 \\
& \!+\! \E\sum_{i=1}^{M} \omega^3 \widetilde{q}_{ii}\widetilde{b}_{i}^2 e^{2}_{i}  \bold{h}^{H}_{i}\bold{Q}_{i}\bold{U}\bold{C}\bold{h}_{i}
=X_{1,1} + X_{1,2} + X_{1,3},
\end{aligned}
\end{equation}
where step $a$ follows from the following equality~\cite{bai2008clt}
\begin{equation}
\label{hq_product}
\bold{h}_{j}^{H}\bold{Q}(z)=\frac{\bold{h}_{j}^{H}\bold{Q}_{j}(z)}{1+\bold{h}_{j}^{H}\bold{Q}_{j}(z)\bold{h}_{j}}=-z\widetilde{q}_{jj}\bold{h}_{j}^{H}\bold{Q}_{j}(z),
\end{equation}
and step $b$ follows from~(\ref{replace_q}). In the following, we will first handle $X_{1,2}$ and $X_{1,3}$, and leave the evaluation of $X_{1,1}$ together with $X_{2}$ and $X_{3}$.

Since $\E (X-\E X)Y=\E(X-\E X)(Y-\E Y)$, we can use Lemma~\ref{lemma_ext} to expand the term $X_{1,2}$ as shown in~(\ref{eq_x12}) at the top of the next page, 
\begin{figure*}[t!]
\begin{equation}
\label{eq_x12}
\begin{aligned}
& X_{1,2} = -\E\sum_{i=1}^{M} \omega^2\widetilde{b}_{i}^2 e_{i}  (\bold{h}^{H}_{i}\bold{Q}_{i}\bold{U}\bold{C}\bold{h}_{i}
- \frac{\widetilde{d}_{i}}{M} \Tr \bold{D}\bold{Q}_{i}\bold{U}\bold{C}
- \bold{a}^{H}_{i}\bold{Q}_{i}\bold{U}\bold{C}\bold{a}_{i} )
=\E \sum_{i=1}^{M} -\frac{\omega^2\widetilde{d}^{2}_{i} \widetilde{b}_{i}^2 }{M^2}\Tr\bold{Q}_{i}\bold{D}\bold{Q}_{i}\bold{U}\bold{C} \bold{D}
\\
&- \frac{|\vartheta|^2\omega^2\widetilde{d}^{2}_{i}  \widetilde{b}_{i}^2}{M^2}\Tr\bold{Q}_{i}\bold{U}\bold{C} \bold{D} {\bold{Q}}^{T}_{i}\bold{D}
-\frac{\omega^2\widetilde{d}_{i}\widetilde{b}_{i}^2 }{M}\bold{a}^{H}_{i}\bold{Q}_{i}\bold{D}\bold{Q}_{i}\bold{U}\bold{C} \bold{a}_{i} 
-\frac{\omega^2\widetilde{d}_{i}\widetilde{b}_{i}^2 }{M} \bold{a}^{H}_{i}\bold{Q}_{i}\bold{U}\bold{C}\bold{D} \bold{Q}_{i}\bold{a}_{i}  
\\
&- \frac{\vartheta\omega^2  \widetilde{b}_{i}^2 \widetilde{d}_{i} }{M} \bold{a}^{H}_{i}\bold{Q}_{i}\bold{U}\bold{C}\bold{D}{ \bold{Q}}^{T}_{i}\overline{\bold{a}_{i}} 
-\frac{\overline{\vartheta}\omega^2  \widetilde{b}_{i}^2 \widetilde{d}_{i} }{M} \bold{a}^{T}_{i}  { \bold{Q}}^{T}_{i}\bold{D}  \bold{Q}_{i}\bold{U}\bold{C}\bold{a}_{i} 
-  \frac{\kappa\omega^2\widetilde{d}_{i}^2 \widetilde{b}_{i}^2}{M^2}  \sum_{j=1}^{N}d^{2}_{j} \left[ \bold{Q}_{i}\right]_{jj} \left[ \bold{Q}_{i}\bold{U}\bold{C}  \right]_{jj}
+\varepsilon_{1,2}
\\
&=X_{1,2,1}+X_{1,2,2}+X_{1,2,3}+X_{1,2,4}+X_{1,2,5}
+X_{1,2,6}
+X_{1,2,7}+\varepsilon_{1,2},
\end{aligned}
\end{equation}
\hrulefill
\vspace{-0.3cm}
 \end{figure*}
where
\begin{equation}
\begin{aligned}
\label{ep_1_2_dis}
&|\varepsilon_{1,2}|=|\E M^{-\frac{3}{2}} \sum_{i=1}^{M} \zeta  \bold{a}^{H}_{i}\bold{Q}_{i}\bold{U}\bold{C}  \bold{D}^{\frac{3}{2}}\mathrm{vdiag}(\bold{Q}_{i})\bold{a}_{i}  
\\
&
+\zeta \E \bold{a}^{H}_{i}\bold{Q}_{i} \bold{D}^{\frac{3}{2}}\mathrm{vdiag}(\bold{Q}_{i}\bold{U}\bold{C} )\bold{a}_{i}
\\
&+\overline{\zeta}  \E \bold{a}^{H}_{i}\bold{C}^{T}\bold{U} \bold{Q}^{T}_{i} \bold{D}^{\frac{3}{2}}\mathrm{vdiag}(\bold{Q}_{i})\bold{a}_{i}  
\\
&+\overline{\zeta} \E \bold{a}^{H}_{i}\mathrm{vdiag}(\bold{Q}^{T}_{i}) \bold{D}^{\frac{3}{2}}\bold{Q}_{i}\bold{U}\bold{C}\bold{a}_{i}
 | =O(M^{-\frac{1}{2}}).
 \end{aligned}
 \end{equation}
Note (\ref{ep_1_2_dis}) can be verified by a similar approach as that in~\cite{hachem2012clt}. Then we turn to the term $X_{1,3}$ and obtain
\begin{equation}
\label{eq_x13}
\begin{aligned}
&X_{1,3} 
= \E\sum_{i=1}^{M} \omega^3 \widetilde{b}_{i}^3 e^{2}_{i} ( \bold{a}^{H}_{i}\bold{Q}_{i}\bold{U}\bold{C}\bold{a}_{i}
+\frac{\widetilde{d}_{i}}{M}\Tr \bold{D}  \bold{Q}_{i}\bold{U}\bold{C})
\\
&+\varepsilon_{1,3,1}+\varepsilon_{1,3,2}
=X_{1,3,1}+X_{1,3,2}+\varepsilon_{1,3,1}+\varepsilon_{1,3,2},
\end{aligned}
\end{equation}
where by~(\ref{bound_e}), we have
\begin{equation}
\begin{aligned}
&\E | \bold{h}^{H}_{i}\bold{Q}_{i}\bold{U}\bold{C}\bold{h}_{i}|^2 \le 2 \E |e_{i}|^2 +2 \E |\frac{\widetilde{d}_{i}}{M}\Tr \bold{D}\bold{Q}_{i}\bold{U}\bold{C} |^2
\\
&+2\E |\bold{a}^{H}_{i} \bold{Q}_{i}\bold{U}\bold{C}\bold{a}_{i} |^2 =O(1),
\end{aligned}
\vspace{-0.1cm}
\end{equation}
\begin{equation}
\begin{aligned}
&|\varepsilon_{1,3,1}|=|-\E \sum_{i=1}^{M} \omega^4 \widetilde{b}_{i}^4\widetilde{q}_{ii} e^{3}_{i}   \bold{h}^{H}_{i}\bold{Q}_{i}\bold{U}\bold{C}\bold{h}_{i}| 
\\
&\le K\sum_{i=1}^{M} \E^{\frac{1}{2}}(|e^6_{i}|)\E^{\frac{1}{2}}(| \bold{h}^{H}_{i}\bold{Q}_{i}\bold{U}\bold{C}\bold{h}_{i}|^2)  =O(M^{-\frac{1}{2}}),
\end{aligned}
\vspace{-0.1cm}
\end{equation}
\begin{equation}
\begin{aligned}
& |\varepsilon_{1,3,2}| =  |\E \sum_{i=1}^{M} \omega^3 \widetilde{b}_{i}^3 e^{2}_{i} ( \bold{h}^{H}_{i}\bold{Q}_{i}\bold{U}\bold{C}\bold{h}_{i}-
\\
&
\E_{\bold{h}_{i}}  \bold{h}^{H}_{i}\bold{Q}_{i}\bold{U}\bold{C}\bold{h}_{i}) | 
 \le \frac{K}{\sqrt{M}} \sum_{i=1}^{M} \E^{\frac{1}{2}}(|e^4_{i}|)
=O(M^{-\frac{1}{2}}),
\end{aligned}
\vspace{-0.2cm}
\end{equation}
given $\E|\bold{h}^{H}_{i}\bold{Q}_{i}\bold{U}\bold{C}\bold{h}_{i}-\!\E_{\bold{h}_{i}}  \bold{h}^{H}_{i}\bold{Q}_{i}\bold{U}\bold{C}\bold{h}_{i}|^2=O(M^{-1})$ by Lemma~\ref{lemma_ext}. By far, we have completed the evaluation of $X_{1,2}$ and $X_{1,3}$. Next, we will handle the remaining terms
\begin{equation}
\label{eq_w12}
\begin{aligned}
&\quad X_{1,1} + X_2 + X_3
\\
&=\sum_{i=1}^{M}    \frac{\omega \widetilde{d}_{i}}{M}  (\E\widetilde{b}_{i} \Tr \bold{D}\bold{Q}_{i}\bold{U}\bold{C}
- \E\widetilde{q}_{ii} \E \Tr \bold{D}\bold{Q}\bold{U}\bold{C} )
\\
&
 +[\E ( \omega \widetilde{b}_{i}\bold{a}_{i}^{H}\bold{Q}_{i}\bold{U}\bold{C}\bold{a}_{i} 
-\frac{ \bold{a}_{i}^{H}\bold{Q}\bold{U}\bold{C}\bold{a}_{i}}{1+\alpha\widetilde{d}_{i}} )]
= W_1 + W_2.
\end{aligned}
\end{equation}
By the rank-one perturbation identity~\cite{bai2008clt}, i.e.
\begin{equation}
\label{q_qi}
\begin{aligned}
\bold{Q}(z)&=\bold{Q}_{j}(z)-\frac{\bold{Q}_{j}(z) \bold{h}_{j} \bold{h}_{j}^{H}  \bold{Q}_{j}(z)}{1+ \bold{h}_{j}^{H} \bold{Q}_{j}(z) \bold{h}_{j} }
\\
&=\bold{Q}_{j}(z)+z\widetilde{q}_{jj}(z)\bold{Q}_{j}(z) \bold{h}_{j} \bold{h}_{j}^{H}  \bold{Q}_{j}(z),
\end{aligned}
\end{equation}
and the identify about $\widetilde{q}_{jj}$ in~(\ref{replace_q}), we have~(\ref{eq_w1}) at top of the next page about term $W_1$, 
\begin{figure*}[ht]
\begin{equation}
\label{eq_w1}
\begin{aligned}
W_1 &= \sum_{i=1}^{M}\E   ( \widetilde{b}_{i} -\E\widetilde{q}_{ii} )\frac{ \omega \widetilde{d}_{i}}{M} \Tr \bold{D}\bold{Q}_{i}\bold{U}\bold{C} 
+\E\widetilde{q}_{ii} \E \widetilde{q}_{ii} \frac{\omega^{2} \widetilde{d}_{i}}{M} \Tr \bold{D}\bold{Q}_{i}\bold{h}_{i}\bold{h}^{H}_{i} \bold{Q}_{i}\bold{U}\bold{C} 
=\E\sum_{i=1}^{M}   ( \widetilde{q}_{ii} -\E\widetilde{q}_{ii} )\frac{ \omega \widetilde{d}_{i}}{M} \Tr \bold{D}\bold{Q}_{i}\bold{U}\bold{C}  
\\
&\underbrace{-\omega^{3}\widetilde{d}_{i}\E\widetilde{b}^{3}_{i} e_i^{2}\frac{1}{M} \Tr \bold{D}\bold{Q}_{i}\bold{U}\bold{C}}_{=-X_{1,3,2}} + \varepsilon_{x_{1,3,2}}
+\underbrace{\omega^{2} \widetilde{d}^{2}_{i}   \E \widetilde{q}_{ii} \E \frac{\widetilde{b}_{i}}{M^{2}}  \Tr \bold{D}\bold{Q}_{i}\bold{D} \bold{Q}_{i}\bold{U}\bold{C}}_{\approx -X_{1,2,1}}
+\underbrace{\omega^{2} \widetilde{d}_{i} \E \widetilde{q}_{ii}   \frac{\E\widetilde{b}_{i}\bold{a}_{i}^{H}\bold{Q}_{i}\bold{U}\bold{C}\bold{D}\bold{Q}_{i}\bold{a}_{i}}{M}}_{\approx -X_{1,2,4}}+\varepsilon_{w_{1}}
\\
&=\varepsilon_{W_{1}}+W_{1,1}+\varepsilon_{x_{1,3,2}}+W_{1,2}+W_{1,3}+\varepsilon_{w_{1}},
\end{aligned}
\end{equation}
\hrulefill
\vspace{-0.5cm}
\end{figure*}
where by~(\ref{l3_2}) of Lemma~\ref{bound_lemma},
\begin{equation}
\begin{aligned}
&|\varepsilon_{x_{1,3,2}}|=|\E\sum_{i=1}^{M} \omega^{3}\widetilde{d}_{i}\E\widetilde{b}^{2}_{i}(\widetilde{q}_{ii}-\widetilde{b}_{i}) e_i^{2}\frac{1}{M} \Tr \bold{D}\bold{Q}_{i}\bold{U}\bold{C} | 
\\
&\le K  M^{-\frac{1}{2}}=O(M^{-\frac{1}{2}}),
\end{aligned}
\vspace{-0.2cm}
\end{equation}
\begin{equation}
\begin{aligned}
\!|\varepsilon_{W_{1}}|\! \le \! \frac{K}{M} \! \sum_{i=1}^{M}\! \Var^{\frac{1}{2}}(\widetilde{q}_{ii}) \Var^{\frac{1}{2}}(\Tr \bold{D}\bold{Q}_{i}\bold{U}\bold{C}) 
\!=\!O(M^{-\frac{1}{2}}),\!\!\! 
\end{aligned}
\end{equation}
\begin{equation}
\begin{aligned}
& |\varepsilon_{w_{1}}| 
=|\sum_{i=1}^{M}\frac{\omega^{2} \widetilde{d}^{2}_{i}\E \widetilde{q}_{ii}}{M}    \E( \widetilde{q}_{ii}-\widetilde{b}_{i}) \Tr \bold{D}\bold{Q}_{i}\bold{h}_{i}\bold{h}^{H}_{i} \bold{Q}_{i}\bold{U}\bold{C} 
  |
\\
&
\le \frac{K_{1}}{M} \sum_{i=1}^{M}  \E^{\frac{1}{2}}| e_{i}|^2\E^{\frac{1}{2}}| \bold{h}^{H}_{i} \bold{Q}_{i}\bold{U}\bold{C}\bold{D}\bold{Q}_{i}\bold{h}_{i} |^2
=O({M^{-\frac{1}{2}}}).
\end{aligned}
\end{equation}
$W_{1,2} \approx -X_{1,2,1}$ follows from
\begin{equation}
\begin{aligned}
-X_{1,2,1}&=  \sum_{i=1}^{M} \frac{1}{M^2}\E(\widetilde{q}_{ii}-\E  \widetilde{q}_{ii})  \widetilde{b}_{i}\Tr \bold{D}\bold{Q}_{i}\bold{D} \bold{Q}_{i}\bold{U}\bold{C}
\\
&+W_{1,2} = \varepsilon_{w}+W_{1,2} ,
\end{aligned}
\end{equation}
where $|\varepsilon_{w}| \le   \sum_{i=1}^{M}  \frac{K}{M} {\E^{\frac{1}{2}} |\widetilde{q}_{ii}-\E  \widetilde{q}_{ii}|^2  }=O(M^{-\frac{1}{2}}) $, 
given~(\ref{l3_2}) of Lemma~\ref{bound_lemma} and $\|\bold{Q}_{i}\|, \|\bold{U} \|, \|\bold{C} \|, \|\bold{D} \| < \infty $. $W_{1,3} \approx -X_{1,2,4}$ can be derived similarly. 

By far, we have completed the evaluation of $W_{1}$ in~(\ref{eq_w1}) and will turn to the term $W_{2}$. By $ \widetilde{b}_{i}=\omega\widetilde{q}_{ii}+\omega\widetilde{b}_{i}^{2}e_{i}-\omega^2\widetilde{b}_{i}^{2}\widetilde{q}_{ii}e^{2}_{i}$, there holds
\begin{equation}
\label{eq_w2}
\begin{aligned}
& W_2 \!=\! \sum_{i=1}^{M}  ( \omega \E\widetilde{q}_{ii}\bold{a}_{i}^{H}\bold{Q}_{i}\bold{U}\bold{C}\bold{a}_{i} 
\!-\!\frac{1}{1\!+\!\alpha\widetilde{d}_{i}} \E\bold{a}_{i}^{H}\bold{Q}_{i}\bold{U}\bold{C}\bold{a}_{i})
\\
&
\underbrace{- \omega^{3} \E\widetilde{b}^{3}_{i}e^{2}_{i}\bold{a}_{i}^{H}\bold{Q}_{i}\bold{U}\bold{C}\bold{a}_{i}}_{=-X_{1,3,1}}
\!+\! \frac{\E \omega\widetilde{q}_{ii}\bold{h}_{i}^{H}\bold{Q}_{i}\bold{U}\bold{C}\bold{a}_{i} \bold{a}_{i}^{H}\bold{Q}_{i}\bold{h}_{i}}{1\!+\!\alpha\widetilde{d}_{i}}
\\
&
=W_{2,1}+W_{2,2}+W_{2,3}.
\end{aligned}
\end{equation}
Furthermore, by~(\ref{q_qi}), the computation of $W_{2,1}$ is given in~(\ref{eq_w21}) at the top of the next page, 
\begin{figure*}[ht]
\begin{equation}
\label{eq_w21}
\begin{aligned}
& W_{2,1}
= \sum_{i=1}^{M}  \E (\omega \widetilde{q}_{ii}-\frac{1}{1+\alpha\widetilde{d}_{i}} )\bold{a}_{i}^{H}\bold{Q}_{i}\bold{U}\bold{C}\bold{a}_{i} 
= \sum_{i=1}^{M}    \frac{\omega}{1+\alpha\widetilde{d}_{i}} \E \widetilde{q}_{ii}(\frac{\widetilde{d}_{i}}{M}\Tr \bold{D}\E \bold{Q}_{i} -\frac{\widetilde{d}_{i}}{M}\Tr \bold{D} \bold{Q}_{i} 
+ \frac{\widetilde{d}_{i}}{M}\Tr \bold{D} \bold{Q}_{i} 
+\bold{a}^{H}_{i}  \bold{Q}_{i}\bold{a}_{i}
\\
&
-\bold{h}_{i}^{H}\bold{Q}_{i} \bold{h}_{i}  
-\bold{a}^{H}_{i}  \bold{Q}_{i}\bold{a}_{i}
- \frac{\omega\widetilde{d}_{i}}{M} \E (\widetilde{q}_{ii} \bold{h}^{H}_{i} \bold{Q}_{i}\bold{D}\bold{Q}_{i}\bold{h}_{i} ))\bold{a}_{i}^{H}\bold{Q}_{i}\bold{U}\bold{C}\bold{a}_{i} 
=-\sum_{i=1}^{M}    \frac{\omega\E \widetilde{q}_{ii}\bold{a}^{H}_{i}  \bold{Q}_{i}\bold{a}_{i}  \bold{a}_{i}^{H}\bold{Q}_{i}\bold{U}\bold{C}\bold{a}_{i}}{1+\alpha\widetilde{d}_{i}} 
\\
&
-\sum_{i=1}^{M}    \frac{\omega \E \widetilde{q}_{ii}e_{i} \bold{a}_{i}^{H}\bold{Q}_{i}\bold{U}\bold{C}\bold{a}_{i}}{1+\alpha\widetilde{d}_{i}}
-\frac{\omega^{2}\widetilde{d}_{i} \E (\widetilde{q}_{ii} \bold{h}^{H}_{i} \bold{Q}_{i}\bold{D}\bold{Q}_{i}\bold{h}_{i} )\E \widetilde{q}_{i}\bold{a}_{i}^{H}\bold{Q}_{i}\bold{U}\bold{C}\bold{a}_{i} }{M(1+\alpha\widetilde{d}_{i})}
+\varepsilon_{W_{2,1}}
=W_{2,1,1}+W_{2,1,2}+W_{2,1,3}+\varepsilon_{W_{2,1}},
\end{aligned}
\end{equation}
\hrulefill
\vspace{-0.4cm}
\end{figure*}
where 
\begin{equation}
\begin{aligned}
&|\varepsilon_{W_{2,1}}| =  \sum_{i=1}^{M}\frac{\omega \widetilde{d}_{i}\cov(\Tr\bold{D}\bold{Q}_{i}, \widetilde{q}_{ii}\bold{a}_{i}^{H}\bold{Q}_{i}\bold{U}\bold{C}\bold{a}_{i} )}{M(1+\alpha\widetilde{d}_{i})}\le
\\
& \sum_{i=1}^{M} \frac{\omega \widetilde{d}_{i}\Var^{\frac{1}{2}}(\widetilde{q}_{ii}\bold{a}_{i}^{H}\bold{Q}_{i}\bold{U}\bold{C}\bold{a}_{i} )
 \Var^{\frac{1}{2}}(\Tr \bold{D} \bold{Q}_{i})}{M(1+\alpha\widetilde{d}_{i})} =O(M^{-\frac{1}{2}}).
\end{aligned}
\end{equation}
because given~(\ref{l3_1}),~(\ref{l3_2}) and~(\ref{q_e_bound}) of Lemma~\ref{bound_lemma} and the Cauchy-Schwarz inequality, we have
\begin{equation}
\begin{aligned}
&\Var(\widetilde{q}_{ii}\bold{a}_{i}^{H}\bold{Q}_{i}\bold{U}\bold{C}\bold{a}_{i})
\\
&=\E | \widetilde{q}_{ii}\bold{a}_{i}^{H}\bold{Q}_{i}\bold{U}\bold{C}\bold{a}_{i} 
- \bold{a}_{i}^{H}\bold{Q}_{i}\bold{U}\bold{C}\bold{a}_{i} (\E \widetilde{q}_{ii})
\\
&
+ \bold{a}_{i}^{H}\bold{Q}_{i}\bold{U}\bold{C}\bold{a}_{i}( \E \widetilde{q}_{ii})
- (\E\widetilde{q}_{ii}) ( \E\bold{a}_{i}^{H}\bold{Q}_{i}\bold{U}\bold{C}\bold{a}_{i}) 
\\
 &- \mathrm{Cov}(\widetilde{q}_{ii}, \bold{a}_{i}^{H}\bold{Q}_{i}\bold{U}\bold{C}\bold{a}_{i} )|^2
\\
& \le 3 [\E |\widetilde{q}_{ii}-\E \widetilde{q}_{ii}|^2| \bold{a}_{i}^{H}\bold{Q}_{i}\bold{U}\bold{C}\bold{a}_{i}|^2
\\
&
+|\E(\widetilde{q}_{ii})|^2\E |\bold{a}_{i}^{H}\bold{Q}_{i}\bold{U}\bold{C}\bold{a}_{i}-\E \bold{a}_{i}^{H}\bold{Q}_{i}\bold{U}\bold{C}\bold{a}_{i}|^2 
\\
&+ \E^{\frac{1}{2}}(|\bold{a}_{i}^{H}\bold{Q}_{i}\bold{U}\bold{C}\bold{a}_{i}-\E\bold{a}_{i}^{H}\bold{Q}_{i}\bold{U}\bold{C}\bold{a}_{i}|^4)
\times
\\
&\quad \E^{\frac{1}{2}}(|\widetilde{q}_{ii}-\E\widetilde{q}_{ii}|^4)  
]
=O(M^{-1}).
\end{aligned}
\end{equation}
Given~(\ref{replace_q}) and $\E {e}_{i}\bold{a}^{H}_{i}  \bold{Q}_{i}\bold{a}_{i}  \bold{a}_{i}^{H}\bold{Q}_{i}\bold{U}\bold{C}\bold{a}_{i}=0 $, there holds
\begin{equation}
\label{eq_w211}
\begin{aligned}
&W_{2,1,1}
\!=\!-\E\sum_{i=1}^{M}    \frac{\omega \widetilde{b}_{i}}{1+\alpha\widetilde{d}_{i}}  \bold{a}^{H}_{i}  \bold{Q}_{i}\bold{a}_{i}  \bold{a}_{i}^{H}\bold{Q}_{i}\bold{U}\bold{C}\bold{a}_{i}
\\
&
\!-\! \frac{\omega^{3}\widetilde{b}^{3}_{i}e^{2}_{i}}{1+\alpha\widetilde{d}_{i}}  \bold{a}^{H}_{i}  \bold{Q}_{i}\bold{a}_{i}  \bold{a}_{i}^{H}\bold{Q}_{i}\bold{U}\bold{C}\bold{a}_{i}
=W_{2,1,1,1}+W_{2,1,1,2}.
\end{aligned}
\end{equation}
By expanding $e_{i}^2$ using~(\ref{qua_ext}), the term $W_{2,1,2}$ can be decomposed as
\begin{equation}
\label{eq_w212}
\begin{aligned}
&W_{2,1,2}
=\sum_{i=1}^{M}\E\frac{\omega^{2}\widetilde{b}^{2}_{i}\bold{a}_{i}^{H}\bold{Q}_{i}\bold{U}\bold{C}\bold{a}_{i} }{1+\alpha\widetilde{d}_{i}} ( \frac{\widetilde{d}^{2}_{i}}{M^{2}}\Tr\bold{D}\bold{Q}_{i} \bold{D}\bold{Q}_{i}
\\
&
+ \frac{2\widetilde{d}_{i}}{M}\bold{a}_{i}^{H} \bold{Q}_{i} \bold{D}\bold{Q}_{i}\bold{a}_{i} 
+
 \frac{|\vartheta|^{2}\widetilde{d}^{2}_{i}}{M^{2}}\Tr\bold{D}\bold{Q}_{i} \bold{D}{\bold{Q}}_{i}^{T}
\\
&+\frac{\vartheta\widetilde{d}_{i}}{M}  \bold{a}_{i}^{H} \bold{Q}_{i} \bold{D}{\bold{Q}}^{T}_{i}\overline{\bold{a}}_{i} 
+\frac{\overline{\vartheta}\widetilde{d}_{i}}{M}\bold{a}_{i}^{T} {\bold{Q}}_{i}^{T} \bold{D}\bold{Q}_{i} {\bold{a}}_{i} 
\\
&
+\frac{\kappa}{M^{2}}\sum_{j=1}^{N} \widetilde{d}_{i}^{2}{d}_{j}^{2}[\bold{Q}_{i}]^{2}_{jj}
)+\varepsilon_{W_{2,1,2}}
\\
&=W_{2,1,2,1}+W_{2,1,2,2}+W_{2,1,2,others}+\varepsilon_{W_{2,1,2}},
\end{aligned}
\end{equation}
where $\varepsilon_{W_{2,1,2}}=O(M^{-\frac{1}{2}})$. This can be proved by a similar technique utilized to derive $\varepsilon_{1,2}$ in~(\ref{ep_1_2_dis}). Hence, the terms related to $\zeta$  vanish. 
For $W_{2,1,3}$, we have
\begin{equation}
\label{eq_w213}
\begin{aligned}
&W_{2,1,3}
\!=\!\sum_{i=1}^{M}  \underbrace{-\frac{\omega^{2}\widetilde{d}^{2}_{i} \E \widetilde{b}_{i}  \Tr \bold{D} \bold{Q}_{i}\bold{D}\bold{Q}_{i}\E\widetilde{b}_{i} \bold{a}_{i}^{H}\bold{Q}_{i}\bold{U}\bold{C}\bold{a}_{i} }{M^2(1+\alpha\widetilde{d}_{i})}}_{\approx -W_{2,1,2,1}}
\\
&
\underbrace{-  \frac{ \omega^{2}\widetilde{d}_{i}\E\widetilde{b}_{i} \bold{a}_{i}^{H}\bold{Q}_{i}\bold{U}\bold{C}\bold{a}_{i}\E \widetilde{b}_{i}  \bold{a}_{i}^{H} \bold{Q}_{i}\bold{D}\bold{Q}_{i} \bold{a}_{i}}{M(1+\alpha\widetilde{d}_{i})}  }_{\approx -0.5\times W_{2,1,2,2}}
+O(M^{-\frac{1}{2}}) 
\\
&
=W_{2,1,3,1}+W_{2,1,3,2}+O(M^{-\frac{1}{2}}).
\end{aligned}
\end{equation}
Also, similar to~(\ref{t_tj}), notice that
\begin{equation}
\label{ata_b}
\begin{aligned}
&\E (1-\omega\widetilde{b}_{i} \bold{a}_{i}^{H}\bold{Q}_{i}\bold{a}_{i})
=\E \frac{1+\frac{\widetilde{d}_{i}}{M}\Tr\bold{D}\bold{Q}_{i}}{1+\frac{\widetilde{d}_{i}}{M}\Tr\bold{D}\bold{Q}_{i}+\bold{a}_{i}^{H}\bold{Q}_{i}\bold{a}_{i}}
\\
&=\E\omega\widetilde{b}_{i}(1+\alpha \widetilde{d}_{i})+\varepsilon_{b},
\end{aligned}
\end{equation}
where $|\varepsilon_{b}|\le \frac{1}{M}\E |\Tr\bold{D}( \bold{Q}_{i} -\bold{Q})| =O(M^{-1}) $ results from~(\ref{rank_one}) of Lemma~\ref{rank_lemma}. 

By far, we have completed the evaluation of $W_{2,1}$ in~(\ref{eq_w2}). Given $W_{2,2}$ can be cancelled by $X_{1,3,1}$, we now turn to the evaluation of $W_{2,3}$. By replacing $\widetilde{q}_{ii}$ using~(\ref{replace_q}) and further combining~(\ref{qua_ext}),~(\ref{ata_b}), we obtain~(\ref{eq_w23}) about $W_{2,3}$ at the top of the next page,
\begin{figure*}[!t]
\begin{equation}
\label{eq_w23}
\begin{aligned}
&W_{2,3}
= \sum_{i=1}^{M}\E \frac{\omega\widetilde{b}_{i}}{1+\alpha\widetilde{d}_{i}}(\frac{\widetilde{d}_{i}}{M}\bold{a}_{i}^{H} \bold{Q}_{i}\bold{D} \bold{Q}_{i}\bold{U}\bold{C}\bold{a}_{i}+\bold{a}_{i}^{H} \bold{Q}_{i}\bold{a}_{i} \bold{a}_{i}^{H}\bold{Q}_{i}\bold{U}\bold{C}\bold{a}_{i} )
-  \frac{\omega^{2}\widetilde{b}^{2}_{i}\widetilde{d}_{i}}{M(1+\alpha\widetilde{d}_{i})}
(\bold{a}_{i}^{H} \bold{Q}_{i}\bold{D} \bold{Q}_{i}\bold{U}\bold{C}\bold{a}_{i} \bold{a}_{i}^{H}\bold{Q}_{i}\bold{a}_{i}
\\
&+\bold{a}_{i}^{H} \bold{Q}_{i}\bold{U}\bold{C}\bold{a}_{i} \bold{a}_{i}^{H}\bold{Q}_{i} \bold{D} \bold{Q}_{i} \bold{a}_{i} 
+\overline{\vartheta}\bold{a}_{i}^{T} {\bold{Q}}^{T}_{i}\bold{D} \bold{Q}_{i}\bold{U}\bold{C}\bold{a}_{i} \bold{a}_{i}^{H}\bold{Q}_{i}{\bold{a}}_{i}+\vartheta\bold{a}_{i}^{H} \bold{Q}_{i}\bold{U}\bold{C}\bold{a}_{i} \bold{a}_{i}^{H}\bold{Q}_{i} \bold{D}{\bold{Q}}^{T}_{i} \overline{\bold{a}}_{i}  
)
\\
&+ {\frac{\omega^3 \widetilde{b}^{3}_{i}e_{i}^{2}}{1+\alpha\widetilde{d}_{i}}\bold{a}_{i}^{H}\bold{Q}_{i}\bold{U}\bold{C}\bold{a}_{i} \bold{a}_{i}^{H}\bold{Q}_{i}\bold{a}_{i}}
= \sum_{i=1}^{M}\E \underbrace{\frac{\omega^{2}\widetilde{b}^{2}_{i}\widetilde{d}_{i}}{M}\bold{a}_{i}^{H} \bold{Q}_{i}\bold{D} \bold{Q}_{i}\bold{U}\bold{C}\bold{a}_{i}}_{=-X_{1,2,3}}
+ \underbrace{\frac{\omega\widetilde{b}_{i}}{1+\alpha\widetilde{d}_{i}}\bold{a}_{i}^{H} \bold{Q}_{i}\bold{a}_{i} \bold{a}_{i}^{H}\bold{Q}_{i}\bold{U}\bold{C}\bold{a}_{i}}_{=-W_{2,1,1,1}}
\\
&
-\underbrace{  \frac{\omega^{2}\widetilde{b}^{2}_{i}\widetilde{d}_{i}}{1+\alpha\widetilde{d}_{i}}
(\frac{1}{M}\bold{a}_{i}^{H} \bold{Q}_{i}\bold{U}\bold{C}\bold{a}_{i} \bold{a}_{i}^{H}\bold{Q}_{i} \bold{D} \bold{Q}_{i} \bold{a}_{i} }_{ \approx 0.5\times W_{2,1,2,2}}
+\frac{\overline{\vartheta}}{M}\bold{a}_{i}^{T} {\bold{Q}}^{T}_{i}\bold{D} \bold{Q}_{i}\bold{U}\bold{C}\bold{a}_{i} \bold{a}_{i}^{H}\bold{Q}_{i}{\bold{a}}_{i}+\frac{\vartheta}{M}\bold{a}_{i}^{H} \bold{Q}_{i}\bold{U}\bold{C}\bold{a}_{i} \bold{a}_{i}^{H}\bold{Q}_{i} \bold{D}{\bold{Q}}^{T}_{i} \overline{\bold{a}}_{i}  
)
\\
&+ \underbrace{\frac{\omega^3 \widetilde{b}^{3}_{i}e_{i}^{2}}{1+\alpha\widetilde{d}_{i}}\bold{a}_{i}^{H}\bold{Q}_{i}\bold{U}\bold{C}\bold{a}_{i} \bold{a}_{i}^{H}\bold{Q}_{i}\bold{a}_{i}}_{=-W_{2,1,1,2}}+\varepsilon_{W_{2,3}},
\end{aligned}
\end{equation}
\hrulefill
\vspace{-0.5cm}
\end{figure*}
where $\varepsilon_{W_{2,3}}=O(M^{-\frac{1}{2}})$ is the summation of the $\zeta$ and $\kappa$ related terms.

By substituting~(\ref{eq_w211}),~(\ref{eq_w212}) and~(\ref{eq_w213}) into~(\ref{eq_w21}), we can obtain $W_{2,1}$. 
By combining this result with $W_{2,2}$, and $W_{2,3}$, we can obtain $W_{2}$ from~(\ref{eq_w2}). Similarly, by combining $W_{1}$ and $W_{2}$, we can obtain $X_{1,1}+X_{2}+X_{3}$ in~(\ref{eq_w12}). Finally, by substituting this result, $X_{1,2}$ in~(\ref{eq_x12}), and $X_{1,3}$ in~(\ref{eq_x13}) into~(\ref{root_eq}), the terms irrelevant with $\vartheta$ and $\kappa$ will be cancelled. Since it is safe to replace $\alpha$ with $\delta$ by~(\ref{l3_1}), the expression for the bias $\mathcal{Z}(\bold{U})$ is given in~(\ref{bias_init}) at the top of the next page,
\begin{figure*}[!t]
\begin{equation}
\label{bias_init}
\begin{aligned}
&-\mathcal{Z}(\bold{U})=\varepsilon_{u}-\mathcal{Z}_{\vartheta}(\bold{U})-\mathcal{Z}_{\kappa}(\bold{U})
=\varepsilon_{u}+\sum_{i=1}^{M}
\E
\underbrace{- \frac{|\vartheta|^2\omega^2\widetilde{d}^{2}_{i} \widetilde{b}_{i}^2}{M^2}\Tr\bold{Q}_{i}\bold{U}\bold{C} \bold{D} {\bold{Q}}^{T}_{i}\bold{D}
}_{-Z_{1}}
\underbrace{ - \frac{\vartheta\omega^2  \widetilde{b}_{i}^2 \widetilde{d}_{i} }{M} \bold{a}^{H}_{i}\bold{Q}_{i}\bold{U}\bold{C}\bold{D}{ \bold{Q}}^{T}_{i}\overline{\bold{a}_{i}}}_{-Z_{2}}
\\
&\underbrace{ + \frac{\overline{\vartheta}\omega^{2}\widetilde{d}_{i}\widetilde{b}^{2}_{i}  }{M(1+\delta\widetilde{d}_{i})} \bold{a}_{i}^{H}\bold{Q}_{i}\bold{U}\bold{C}\bold{a}_{i}
\bold{a}_{i}^{T} {\bold{Q}}^{T}_{i} \bold{D}\bold{Q}_{i} {\bold{a}}_{i}   
-\frac{\overline{\vartheta}\omega^{2}\widetilde{d}_{i}\widetilde{b}^{2}_{i}}{M(1+\delta\widetilde{d}_{i})}\bold{a}_{i}^{T} {\bold{Q}}^{T}_{i}\bold{D} \bold{Q}_{i}\bold{U}\bold{C}\bold{a}_{i} \bold{a}_{i}^{H}\bold{Q}_{i}{\bold{a}}_{i}
-\frac{\overline{\vartheta}\omega^2  \widetilde{b}_{i}^2 \widetilde{d}_{i} }{M} \bold{a}^{T}_{i}  { \bold{Q}}^{T}_{i}\bold{D}  \bold{Q}_{i}\bold{U}\bold{C}\bold{a}_{i} }_{-Z_{3,4,5,6}}
\\
&
\underbrace{ \!+
 \frac{|\vartheta|^{2}\omega^{2} \widetilde{d}^{2}_{i}\widetilde{b}^{2}_{i}}{M^2(1+\delta\widetilde{d}_{i})}\bold{a}_{i}^{H}\bold{Q}_{i}\bold{U}\bold{C}\bold{a}_{i}\Tr\bold{D}\bold{Q}_{i} \bold{D}{\bold{Q}}_{i}^{T}}_{-Z_{3,4,5,6}}
\underbrace{\!+\! \sum_{j=1}^{N} \!\frac{\kappa  \omega^{2}\widetilde{d}_{i}^{2}{d}_{j}^{2} \widetilde{b}^{2}_{i}}{M^{2}(1+\delta\widetilde{d}_{i})}  \bold{a}_{i}^{H}\bold{Q}_{i}\bold{U}\bold{C}\bold{a}_{i}[\bold{Q}_{i}]^{2}_{jj}
\!-\!  \frac{\kappa\omega^2 \widetilde{d}_{i}^2\widetilde{b}_{i}^2}{M^2} \sum_{j=1}^{N}d^{2}_{j} \left[ \bold{Q}_{i}\right]_{jj} \left[ \bold{Q}_{i}\bold{U}\bold{C}  \right]_{jj}
}_{-\mathcal{Z}_{\kappa}(\bold{U})},
\end{aligned}
\end{equation}
\hrulefill
\vspace{-0.5cm}
\end{figure*}
and $\varepsilon_{u}=O(M^{-\frac{1}{2}})$. Next, we will represent~(\ref{bias_init}) using the deterministic quantities in Table~\ref{var_list}.

\subsubsection{The deterministic expression}
 
It can be observed from~(\ref{bias_init}) that there are six terms related to $\vartheta$. In the following, we will first evaluate the last four terms by setting up an equation. Specifically, we will utilize two methods to evaluate $\E \bold{a}^{T}_{i}  { \bold{Q}}^{T}\bold{D}  \bold{Q}\bold{U}\bold{C}\bold{a}_{i}$.

\textit{Method 1)} We perform decomposition to obtain
\begin{equation}
\begin{aligned}
&\E \bold{a}^{T}_{i}  { \bold{Q}}^{T}\bold{D}  \bold{Q}\bold{U}\bold{C}\bold{a}_{i}
= \bold{a}^{T}_{i}  { \bold{T}}^{T}\bold{D}  \bold{T}\bold{U}\bold{C}\bold{a}_{i}
\\
&+\omega\widetilde{\delta}\E \bold{a}^{T}_{i}  { \bold{T}}^{T}\bold{D}  {\bold{Q}}^{T}\bold{D}  \bold{Q}\bold{U}\bold{C}\bold{a}_{i}
\\
&+ \sum_{j=1}^{M} \frac{\E \bold{a}^{T}_{i}  { \bold{T}}^{T} \overline{\bold{a}}_{j}
\bold{a}_{j}^{T}{\bold{Q}}^{T} \bold{D}  \bold{Q}\bold{U}\bold{C}\bold{a}_{i} }{1+\delta\widetilde{d}_{j}}  
\\
&-\sum_{j=1}^{M} \E   \bold{a}^{T}_{i}  { \bold{T}}^{T} \overline{\bold{h}}_{j}
\bold{h}_{j}^{T}{\bold{Q}}^{T}  \bold{D}  \bold{Q}\bold{U}\bold{C}\bold{a}_{i}
\\
&= \bold{a}^{T}_{i}  { \bold{T}}^{T}\bold{D}  \bold{T}\bold{U}\bold{C}\bold{a}_{i}
+\omega\widetilde{\delta}\E \bold{a}^{T}_{i}  { \bold{T}}^{T}\bold{D}  {\bold{Q}}^{T}\bold{D}  \bold{Q}\bold{U}\bold{C}\bold{a}_{i}
\\
&+K+E.
\end{aligned}
\end{equation}
By~(\ref{q_qi}), we have
\begin{equation}
\begin{aligned}
K&= \sum_{j=1}^{M} \E\frac{1}{1+\delta\widetilde{d}_{j}}  \bold{a}^{T}_{i}  { \bold{T}}^{T} \overline{\bold{a}}_{j}
\bold{a}_{j}^{T}{\bold{Q}}_{j}^{T} \bold{D}  \bold{Q}\bold{U}\bold{C}\bold{a}_{i}
\\
&-\frac{\omega  \widetilde{t}_{jj}}{1+\delta\widetilde{d}_{j}} \bold{a}^{T}_{i}  { \bold{T}^{T}} \overline{\bold{a}}_{j}
\bold{a}_{j}^{T}{\bold{T}}_{j}^{T}\overline{\bold{a}}_{j} \bold{h}_{j}^{T}{\bold{Q}}_{j}^{T}   \bold{D}  \bold{Q}\bold{U}\bold{C}\bold{a}_{i}
\\
&+\varepsilon_{K_1}+\varepsilon_{K_2}+\varepsilon_{K_3},
\end{aligned}
\end{equation}
where
\vspace{-0.2cm}
\begin{equation}
\begin{aligned}
\varepsilon_{K_1}&\!=\!-\!\sum_{i=1}^{M}\!\E \bold{a}^{T}_{i} \bold{T}^{T}\overline{\bold{A}}  \bold{\Lambda}\bold{H}^{T}\bold{D}\bold{Q}\bold{U}\bold{C} \bold{a}_{i},
\\
\bold{\Lambda}&\!=\!\mathrm{diag}(\omega(\widetilde{q}_{jj}-\widetilde{t}_{jj})(1+\bold{h}_{j}^{T}\bold{Q}^{T}_{j}\overline{\bold{h}}_{j})\bold{a}^{T}_{j}{\bold{Q}}^{T}_{j}\overline{\bold{h}}_{j} ),
\\
\varepsilon_{K_2}&\!=\!-\!\sum_{i=1}^{M}\!\frac{\E\omega  \widetilde{t}_{jj}\bold{a}^{T}_{i}  { \bold{T}^{T}} \overline{\bold{a}}_{j}
\bold{a}_{j}^{T}{\bold{T}}_{j}^{T}\overline{\bold{y}}_{j} \bold{h}_{j}^{T}{\bold{Q}}_{j}^{T}   \bold{D}  \bold{Q}\bold{U}\bold{C}\bold{a}_{i} }{1+\delta\widetilde{d}_{j}}\!,
\\
\varepsilon_{K_3}&\!=\!-\sum_{i=1}^{M}\!  \frac{\E\omega  \widetilde{t}_{jj}\bold{a}^{T}_{i}  { \bold{T}^{T}} \overline{\bold{a}}_{j}
\bold{a}_{j}^{T} ({\bold{Q}}_{j}^{T} \!-\! {\bold{T}}_{j}^{T})\overline{\bold{a}}_{j} \bold{h}_{j}^{T}{\bold{Q}}_{j}^{T}   \bold{D}  \bold{Q}\bold{U}\bold{C}\bold{a}_{i}}{1+\delta\widetilde{d}_{j}}.
\end{aligned}
\end{equation}
Given assumptions \textbf{A.1}, \textbf{A.2} and the finite bound of the matrices, we have
\begin{equation}
\begin{aligned}
&|\varepsilon_{K_1}|\le M \E  \| \bold{a}^{T}_{i} \bold{T}^{T} \bold{A}\bold{\Lambda}  \| 
\\
&
\le M \sum_{j=1}^{M}\sqrt{|\bold{a}^{T}_{i} \bold{T}^{T}\overline{\bold{a}}_{i}|^2\E| \lambda_{j}|^2  } =O(M^{-\frac{1}{2}}),
\end{aligned}
\end{equation}
and $|\varepsilon_{K_2}|=O(M^{-\frac{1}{2}})$. $|\varepsilon_{K_3}|=O(M^{-\frac{1}{2}})$ can be handled similarly. In the following derivations, we will omit the discussions over $\varepsilon$. By a similar approach, we expand $\bold{Q}$ using~(\ref{q_qi}) and bound the error using Lemma~(\ref{rank_lemma}) to obtain
\begin{equation}
\begin{aligned}
K&=\E \sum_{j=1}^{M}  \omega  \widetilde{t}_{jj} \bold{a}^{T}_{i}  { \bold{T}}^{T} \overline{\bold{a}}_{j}
\bold{a}_{j}^{T}{\bold{Q}}^{T}_{j}  \bold{D}  \bold{Q}\bold{U}\bold{C}\bold{a}_{i}
\\
&+\frac{\vartheta}{M}\Tr\bold{D}{\bold{Q}}^{T}\bold{D}\bold{Q}\frac{1}{(1+\delta\widetilde{d}_{j})^2}\omega \widetilde{t}_{jj}\widetilde{d}_{j}
 \bold{a}_{i}^{T}{\bold{T}}^{T}\overline{\bold{a}}_{j} \times
 \\
 &\bold{a}_{j}^{T}
 {\bold{T}}^{T}_{j}\overline{\bold{a}}_{j}\bold{a}_{j}^{H}{\bold{T}}\bold{U}\bold{C}{\bold{a}}_{i}
 +O(M^{-\frac{1}{2}}),
\end{aligned}
\end{equation}
and
\begin{equation}
\begin{aligned}
&E
=\E \sum_{j=1}^{M}  -\omega  \widetilde{t}_{jj} \bold{a}^{T}_{i}  { \bold{T}}^{T} \overline{\bold{a}}_{j}
\bold{a}_{j}^{T}{\bold{Q}}^{T}_{j}  \bold{D}  \bold{Q}\bold{U}\bold{C}\bold{a}_{i}
\\
&- \omega \widetilde{\delta}  \bold{a}^{T}_{i}{ \bold{T}}^{T}\bold{D}{\bold{Q}}^{T}\bold{D}\bold{Q}\bold{U}\bold{C}\bold{a}_{i}
\\
&+\frac{|\vartheta|^2}{M}\Tr\bold{D}{\bold{Q}}^{T}\bold{D}\bold{Q}\bold{a}_{i}^{T}{\bold{T}}^{T}\bold{D}{\bold{T}}\bold{U}\bold{C}{\bold{a}}_{i}\sum_{j=1}^{M}\omega^{2}\widetilde{d}^2_{j}\widetilde{t}^2_{jj}
\\
&+\frac{\vartheta}{M}\Tr\bold{D}{\bold{Q}}^{T}\bold{D}\bold{Q}
\frac{\omega  \widetilde{t}_{jj} \widetilde{d}_{j}  \bold{a}_{i}^{T}{\bold{T}}^{T}\overline{\bold{a}}_{j}{\bold{a}}^{H}_{j}\bold{T} \bold{U}\bold{C}\bold{a}_{i}}{1\!+\!\delta\widetilde{d}_{j}}
\\
&+ \frac{\overline{\vartheta}}{M}\bold{a}_{i}^{T}{\bold{T}}^{T}\bold{D}\bold{T}\bold{U}\bold{C}\bold{a}_{i} \sum_{j=1}^{M} \omega^{2}\widetilde{d}_{j} \widetilde{t}_{jj}^{2} \bold{a}^{T}_{j}  { \bold{Q}}_{j}^{T}\bold{D}  \bold{Q}_{j}\bold{a}_{j}
\\
&+O(M^{-\frac{1}{2}}).
\end{aligned}
\end{equation}

Therefore, by~(\ref{t_tj}), we have
\begin{equation}
\label{left_aqa}
\begin{aligned}
&\E \bold{a}^{T}_{i}  { \bold{Q}}^{T}\bold{D}  \bold{Q}\bold{U}\bold{C}\bold{a}_{i}
=\E \bold{a}^{T}_{i}  { \bold{T}}^{T}\bold{D}  \bold{T}\bold{U}\bold{C}\bold{a}_{i}
\\
&+\frac{1}{M}\Tr\bold{D}{\bold{Q}}^{T}\bold{D}\bold{Q}(|\vartheta|^2\bold{a}_{i}^{T}{\bold{T}}^{T}\bold{D}{\bold{T}}\bold{U}\bold{C}{\bold{a}}_{i}\sum_{j=1}^{M}\omega^{2}\widetilde{d}^2_{j}\widetilde{t}^2_{jj}
\\
&+\frac{\vartheta\widetilde{d}_{j} }{(1+\delta\widetilde{d}_{j})^2}\bold{a}_{i}^{T}{\bold{T}}^{T}\overline{\bold{a}}_{j} \bold{a}_{j}^{H}{\bold{T}}\bold{U}\bold{C}{\bold{a}}_{i} )
\\
&+\frac{\overline{\vartheta}}{M} \bold{a}_{i}^{T}{\bold{T}}^{T}\bold{D}\bold{T}\bold{U}\bold{C}\bold{a}_{i} \sum_{j=1}^{M} 
\omega^{2}\widetilde{d}_{j} \widetilde{t}_{jj}^{2} \bold{a}^{T}_{j}  { \bold{Q}}_{j}^{T}\bold{D}  \bold{Q}_{j}\bold{a}_{j}
\\
&+O(M^{-\frac{1}{2}}).
\end{aligned}
\end{equation}

\textit{Method 2)} By~(\ref{q_qi}), we replace both $\bold{Q}$ in $\bold{a}^{T}_{i}  { \bold{Q}}^{T} \bold{D}  \bold{Q}\bold{U}\bold{C}\bold{a}_{i}$ with $\bold{Q}_{i}$ and the discussions of $\varepsilon$ are omitted. Then, we have~(\ref{right_aqa}) about the evaluation of $\E \bold{a}^{T}_{i}  { \bold{Q}}^{T} \bold{D}  \bold{Q}\bold{U}\bold{C}\bold{a}_{i}$ at the top of the next page.
\begin{figure*}[t!]
\begin{equation}
\label{right_aqa}
\begin{aligned}
&\E \bold{a}^{T}_{i}  { \bold{Q}}^{T} \bold{D}  \bold{Q}\bold{U}\bold{C}\bold{a}_{i}
=\E [\bold{a}^{T}_{i}  { \bold{Q}}_{i}^{T}\bold{D}  \bold{Q}_{i}\bold{U}\bold{C}\bold{a}_{i}
-\omega  \widetilde{t}_{ii} \bold{a}^{T}_{i}  { \bold{T}}_{i}^{T} \overline{\bold{a}}_{i}\bold{a}^{T}_{i}  { \bold{Q}}_{i}^{T}\bold{D}  \bold{Q}_{i}\bold{U}\bold{C}\bold{a}_{i}
-\omega \widetilde{t}_{ii} \bold{a}^{H}_{i}  { \bold{T}_{i}} \bold{U}\bold{C}{\bold{a}}_{i}\bold{a}^{T}_{i}  { \bold{Q}}_{i}^{T} \bold{D}  \bold{Q}_{i}\bold{a}_{i}
\\
&
+ \omega^2 \widetilde{t}^2_{ii}\bold{a}^{H}_{i}  { \bold{T}_{i}} \bold{U}\bold{C}{\bold{a}}_{i} \bold{a}^{T}_{i}  { \bold{T}}_{i}^{T} \overline{\bold{a}}_{i} \bold{a}^{T}_{i}  {\bold{Q}}_{i}^{T}\bold{D}  \bold{Q}_{i}\bold{a}_{i}
+ \vartheta\omega^2 \widetilde{t}^2_{ii}\bold{a}^{H}_{i}  { \bold{T}_{i}} \bold{U}\bold{C}{\bold{a}}_{i} \bold{a}^{T}_{i}  { \bold{T}}_{i}^{T} \overline{\bold{a}}_{i} \frac{\widetilde{d}_{i}}{M} \Tr\bold{D}{\bold{Q}}^{T}\bold{D}{\bold{Q}}]+O(M^{-\frac{1}{2}})
\\
&=\E [\omega  \widetilde{t}_{ii}(1+ \delta\widetilde{d}_{i} )  \bold{a}^{T}_{i}  { \bold{Q}}_{i}^{T}\bold{D}  \bold{Q}_{i}\bold{U}\bold{C}\bold{a}_{i}
-\omega^2 \widetilde{t}^2_{ii}(1+\delta \widetilde{d}_{i} ) \bold{a}^{H}_{i}  { \bold{T}_{i}} \bold{U}\bold{C}{\bold{a}}_{i}\bold{a}^{T}_{i}  { \bold{Q}}_{i}^{T} \bold{D}  \bold{Q}_{i}\bold{a}_{i}
\\
&+\vartheta\omega  \widetilde{t}_{ii} (1-\omega \widetilde{t}_{ii} (1+\delta \widetilde{d}_{i}) ) \bold{a}^{H}_{i}  { \bold{T}_{i}} \bold{U}\bold{C}{\bold{a}}_{i}  \frac{\widetilde{d}_{i}}{M} \Tr\bold{D}{\bold{Q}}^{T}\bold{D}{\bold{Q}}]+O(M^{-\frac{1}{2}}).
\end{aligned}
\end{equation}
\hrulefill
\vspace{-0.5cm}
\end{figure*}

By multiplying~(\ref{left_aqa}) and~(\ref{right_aqa}) with $\frac{\overline{\vartheta}\widetilde{d}_{i}}{(1+\delta\widetilde{d}_{i})^2}$ and making the RHS of them equal, we can determine the summation of the last four terms for $\mathcal{Z}_{\vartheta}(\bold{U})$ and define it as $Z_{3,4,5,6}$ in~(\ref{all_z1}) at the top of the next page.
\begin{figure*}[t!]
\begin{equation}
\label{all_z1}
\begin{aligned}
&Z_{3,4,5,6}\!=\!\E\sum_{i=1}^{M}
\frac{\overline{\vartheta} \omega^2  \widetilde{b}_{i}^2 \widetilde{d}_{i} }{M}\bold{a}^{T}_{i}  { \bold{Q}}_{i}^{T}\bold{D}  \bold{Q}_{i}\bold{U}\bold{C}\bold{a}_{i} 
\!+\!\frac{ \overline{\vartheta}\omega^{2}\widetilde{b}^{2}_{i}\widetilde{d}_{i}}{M(1\!+\!\delta\widetilde{d}_{i})}\bold{a}_{i}^{T} {\bold{Q}}_{i}^{T}\bold{D} \bold{Q}_{i}\bold{U}\bold{C}\bold{a}_{i} \bold{a}_{i}^{H}\bold{Q}_{i}{\bold{a}}_{i}
\!-\!\frac{\overline{\vartheta} \omega^{2} \widetilde{d}_{i}\widetilde{b}^{2}_{i}}{M(1\!+\!\alpha\widetilde{d}_{i})} \bold{a}_{i}^{H}\bold{Q}_{i}\bold{U}\bold{C}\bold{a}_{i}
\bold{a}_{i}^{T} {\bold{Q}}^{T}_{i} \bold{D}\bold{Q}_{i} {\bold{a}}_{i} 
\\
&
\!-\! \frac{|\vartheta|^2\omega^{2}\widetilde{d}^{2}_{i}\widetilde{b}^{2}_{i}}{M^{2}(1\!+\!\delta\widetilde{d}_{i})}\bold{a}_{i}^{H}\bold{Q}_{i}\bold{U}\bold{C}\bold{a}_{i}\Tr\bold{D}\bold{Q}_{i} \bold{D}{\bold{Q}}_{i}^{T}
\!=\!\E\sum_{i=1}^{M}
 \frac{\overline{\vartheta}\omega\widetilde{d}_{i}\widetilde{t}_{ii}}{M(1\!+\!\delta\widetilde{d}_{i})}\bold{a}_{i}^{T} {\bold{Q}}_{i}^{T}\bold{D} \bold{Q}_{i}\bold{U}\bold{C}\bold{a}_{i} 
\!-\! \frac{\overline{\vartheta} \omega^{2}\widetilde{d}_{i}\widetilde{t}^{2}_{ii}}{M(1\!+\!\delta\widetilde{d}_{i})} \bold{a}_{i}^{H}\bold{Q}_{i}\bold{U}\bold{C}\bold{a}_{i}
\bold{a}_{i}^{T} {\bold{Q}}^{T}_{i} \bold{D}\bold{Q}_{i} {\bold{a}}_{i} 
\\
&\!-\! \frac{|\vartheta|^2\omega^{2}\widetilde{d}^{2}_{i}\widetilde{t}^{2}_{ii}}{M^2(1\!+\!\delta\widetilde{d}_{i})} \bold{a}_{i}^{H}\bold{Q}_{i}\bold{U}\bold{C}\bold{a}_{i}\Tr\bold{D}\bold{Q}_{i} \bold{D}{\bold{Q}}_{i}^{T}
\!+\!\varepsilon_{Z_{3,4,5,6}}
\!=\!
\E \sum_{i=1}^{M} \frac{\overline{\vartheta}\widetilde{d}_{i}}{(1\!+\!\delta\widetilde{d}_{i})^2} \bold{a}^{T}_{i}  { \bold{T}}^{T}\bold{D}  \bold{T}\bold{U}\bold{T}\bold{a}_{i}
+\frac{\overline{\vartheta}}{M}\Tr\bold{D}{\bold{Q}}^{T}\bold{D}\bold{Q}\times
\\
&
[\frac{|\vartheta|^2\widetilde{d}_{j}}{(1\!+\!\delta\widetilde{d}_{j})^2}\bold{a}_{i}^{T}{\bold{T}}^{T}\bold{D}{\bold{T}}\bold{U}\bold{T}{\bold{a}}_{i}\sum_{j=1}^{M}\omega^{2}\widetilde{d}^2_{j}\widetilde{t}^2_{jj}
\!+\!
\frac{\vartheta\widetilde{d}_{i}}{(1\!+\!\delta\widetilde{d}_{i})^2}\sum_{j=1}^{M}\frac{\widetilde{d}_{j}}{(1\!+\!\delta\widetilde{d}_{j})^2} \bold{a}_{i}^{T}{\bold{T}}^{T}\overline{\bold{a}}_{j} \bold{a}_{j}^{H}{\bold{T}}\bold{U}\bold{T}{\bold{a}}_{i} 
\!-\!\frac{\vartheta\widetilde{d}^2_{i}}{(1\!+\!\delta\widetilde{d}_{i})^3}\bold{a}_{i}^{H}{\bold{T}}\bold{U}\bold{T}{\bold{a}}_{i} 
]
\\
&+ \frac{\widetilde{d}_{i}}{(1+\delta\widetilde{d}_{i})^2}\bold{a}_{i}^{T}{\bold{T}}^{T}\bold{D}\bold{T}\bold{U}\bold{T}\bold{a}_{i} \frac{\overline{\vartheta}^2}{M}\sum_{j=1}^{M} \omega^{2}\widetilde{d}_{j} \widetilde{t}_{jj}^{2} \bold{a}^{T}_{j}  { \bold{Q}}_{j}^{T}\bold{D}  \bold{Q}_{j}\bold{a}_{j}
+\varepsilon_{Z_{3,4,5,6}}
= \overline{\vartheta} [\underline{F}_{T}(\bold{D}\bold{T}\bold{U})(1+\overline{\vartheta}\underline{\beta}_{T})
\\
&
+\alpha_{T}(|\vartheta|^2 \underline{F}_{T}(\bold{D}\bold{T}\bold{U})L -\omega \vartheta \widetilde{\mathcal{F}}_{T}(\bold{U})  )
]+\varepsilon'_{Z_{3,4,5,6}}
\overset{\text{a}}{=}\frac{\overline{\vartheta} }{{\Delta}_{T}}[\underline{F}_{T}(\bold{D}\bold{T}\bold{U})(1-\vartheta F_{T}) - \omega\vartheta  \gamma_{T}\widetilde{\mathcal{F}}_{T}(\bold{U})  
]+\varepsilon''_{Z_{3,4,5,6}},
\end{aligned}
\end{equation}
\hrule
\vspace{-0.5cm}
\end{figure*} 
Here $F_{T}(\cdot)$, $\gamma_{T}(\cdot)$, $\Delta_{T}$ and $\widetilde{F}(\cdot)$ are given in Table~\ref{var_list}. $\alpha_{T}$, $\underline{\beta}_{T}$, $L$ are given in  Table~\ref{apen_var}. Given (\ref{l3_11}), we can replace $\bold{C}$ by $\bold{T}$ safely. Step $a$ in~(\ref{all_z1}) follows from Lemma~\ref{t_lemma} and 
\begin{equation}
\begin{aligned}
&\sum_{j=1}^{M}\frac{\widetilde{d}_{i}\widetilde{d}_{j}\bold{a}_{i}^{T}{\bold{T}}^{T}\overline{\bold{a}}_{j} \bold{a}_{j}^{H}{\bold{T}}\bold{U}\bold{T}{\bold{a}}_{i}}{M(1+\delta\widetilde{d}_{i})^2(1+\delta\widetilde{d}_{j})^2} 
-\frac{1}{M}\Tr   \bold{A}^{H}\bold{T}{\bold{A}}\widetilde{\bold{R}}^3\widetilde{\bold{D}}^2
\\
&=\frac{1}{M}\Tr\bold{A}^{T}\bold{T}^{T}\overline{\bold{A}}\widetilde{\bold{R}}^2\widetilde{\bold{D}} \bold{A}^{H}\bold{T}\bold{U}\bold{T} {\bold{A}}\widetilde{\bold{R}}^2\widetilde{\bold{D}}
\\
&-\frac{1}{M}\Tr   \bold{A}^{H}\bold{T}{\bold{A}}\widetilde{\bold{R}}^3\widetilde{\bold{D}}^2
\\
&=-\frac{\omega}{M}\Tr\widetilde{\bold{R}}\widetilde{\bold{D}}  \bold{A}^{H}\bold{T}\bold{U}\bold{T}{\bold{A}}\widetilde{\bold{R}}\widetilde{\bold{D}} \widetilde{\bold{T}}^{T}=-\omega \widetilde{\mathcal{F}}_{T}(\bold{U}),
\end{aligned}
\end{equation}
which can be derived by~(\ref{replace_R}) and~(\ref{replace_tilde_t}).

By Lemma~\ref{ct_lemma}, we can determine the second term of $\mathcal{Z}_{\vartheta}(\bold{U})$ in~(\ref{bias_init}) as
\begin{equation}
\label{eq_z2}
\begin{aligned}
Z_2&=\sum_{i=1}^{M}\E \frac{\vartheta \omega^{2}\widetilde{d}_{i}\widetilde{b}_{i}^{2}}{M}\bold{a}^{H}_{i}\bold{Q}_{i}\bold{U}\bold{C}\bold{D}\bold{Q}^{T}_{i}\overline{\bold{a}}_{i}
\\
&=\frac{\vartheta}{\Delta_{T}}[{F}_{T}(\bold{U}\bold{T}\bold{D})(1-\overline{\vartheta}{\underline{F}}_{T}-{\gamma_{T}}|\vartheta|^2L)
\\
&+{\gamma}_{T}(\bold{U}\bold{T}\bold{D}) \overline{\vartheta}G_{T}
+ {\gamma}_{T}(\bold{U}\bold{T}\bold{D})|\vartheta|^2{F}_{T} L]+\varepsilon_{Z_2},
\end{aligned}
\end{equation}
where $G_{T}$ and $L$ are given in Table~\ref{apen_var}. Replacing $\bold{Q}_{i}$ with $\bold{Q}$ according to~(\ref{rank_one}), we obtain the first term of $\mathcal{Z}_{\vartheta}(\bold{U})$ as
\begin{equation}
\label{eq_z1}
\begin{aligned}
&Z_1=\E\sum_{i=1}^{M} \frac{|\vartheta|^2\omega^2\widetilde{d}^{2}_{i}\widetilde{b}_{i}^2 }{M^2}\Tr\bold{Q}_{i}\bold{U}\bold{C} \bold{D} \bold{Q}^{T}_{i}\bold{D}=
\\
&\frac{L |\vartheta|^2}{{\Delta}_{T}}  [{F}_{T}(\bold{U}\bold{T} \bold{D} )\theta{\gamma}_{T}\!+\!{\gamma}_{T}(\bold{U}\bold{T} \bold{D})(1\!-\!\vartheta {F}_{T})  ]\!+\!\varepsilon_{Z_1}.
\end{aligned}
\end{equation}

Combining the results in~(\ref{all_z1}),~(\ref{eq_z2}), and~(\ref{eq_z1}), the $\vartheta$ related terms in~(\ref{bias_init}) can be combined as
\begin{equation}
\begin{aligned}
&\mathcal{Z}_{\vartheta}(\bold{U})
=Z_{1}+Z_{2}+Z_{3,4,5,6}-\varepsilon_{u}=
\\
&\frac{1}{\Delta_{T}}[\overline{\vartheta}\underline{F}_{T}(\bold{D}\bold{T}\bold{U})(1\!-\!\vartheta F_{T})
\!+\!\vartheta{F}_{T}(\bold{U}\bold{T} \bold{D} )(1\!-\!\overline{\vartheta} \underline{F}_{T}) 
\\
&+\omega^2|\vartheta|^2 \gamma_{T}(\bold{U}\bold{T} \bold{D})\widetilde{\gamma}_{T}
- |\vartheta|^2\omega \gamma_{T}\widetilde{\mathcal{F}}_{T}(\bold{U})  ]+\varepsilon_{\theta}
\\
 &=\mathcal{Y}_{\vartheta}(\bold{U})+\varepsilon_{\theta}.
\end{aligned}
\end{equation}
Next, we will turn to the terms related to $\kappa$. By replacing $\widetilde{b}_{i}$, ${q}_{ii}$, $\widetilde{q}_{ii}$ with ${t}_{ii}$ and $\widetilde{t}_{ii}$ using~(\ref{q_t_bound}) and~(\ref{bound_e}), we can obtain
\begin{equation}
\begin{aligned}
&\mathcal{Z}_{\kappa}(\bold{U})
=
\E\sum_{i=1}^{M}\sum_{j=1}^{N}- \frac{\kappa\omega^{2}\widetilde{d}_{i}^{2}{d}_{j}^{2} \widetilde{b}^{2}_{i}\bold{a}_{i}^{H}\bold{Q}_{i}\bold{U}\bold{C}\bold{a}_{i}[\bold{Q}_{i}]^{2}_{jj}}{M^{2}(1+\alpha\widetilde{d}_{i})}
\\
&+  \frac{\kappa\omega^2 \widetilde{d}_{i}^2 d^{2}_{j} \widetilde{b}_{i}^2}{M^2}  \left[ \bold{Q}_{i}\right]_{jj} \left[ \bold{Q}_{i}\bold{U}\bold{C}  \right]_{jj}
\\
 &  =-\frac{\kappa\omega}{M}\Tr( \widetilde{\bold{D}}^2\mathrm{diag}(\widetilde{\bold{T}} )
\widetilde{\bold{R}}^{2} \bold{A}^{H}  \bold{T}\bold{U}\bold{T}\bold{A})\frac{1}{M}\Tr\bold{D}^2{\bold{S}}^2
\\
&+\frac{\kappa\omega^2}{M}\Tr( \mathrm{diag}({\bold{T}} )\bold{D}^2 \bold{T}\bold{U}\bold{T})\frac{1}{M}\Tr(\widetilde{\bold{D}}^2\widetilde{\bold{S}}^{2})
+\varepsilon_{\kappa}
\\
&=\mathcal{Y}_{\kappa}(\bold{U})+\varepsilon_{\kappa}.
\end{aligned}
\end{equation}
By combining the terms related to $\vartheta$ and $\kappa$, we have $\mathcal{Y}(\bold{U})=\mathcal{Y}_{\vartheta}(\bold{U})+\mathcal{Y}_{\kappa}(\bold{U})$ and 
recall $z=-\omega$. We complete the proof.
\end{IEEEproof}

\section{proof of Theorem~\ref{alt_exp}}
\label{alt_exp_proof}
\begin{IEEEproof}
First, we will compute the derivative of $(1-\vartheta F_{T})(1-\overline{\vartheta}\underline{F}_{T})- \omega^2|\vartheta|^2 {\gamma}_{T}{\widetilde{\gamma}}_{T}$ with respect to $\omega$. For that purpose, we obtain 
\begin{equation}
\label{ft_term}
\begin{aligned}
&\frac{\mathrm{d} {{F}}_{T}}{\mathrm{d} \omega}
=-\frac{1}{M}\Tr  \bold{D} \bold{T}^{T} (\bold{I}+\widetilde{\delta}\bold{D}+ \omega\widetilde{\delta}'_{\omega}\bold{D}
\\
&-{\delta}'_{\omega} \overline{\bold{A}}\widetilde{\bold{R}}^{2}\widetilde{\bold{D}} \bold{A}^{T} )   \bold{T}^{T} \overline{\bold{A}}\widetilde{\bold{D}}\widetilde{\bold{R}}^{2}\bold{A}^{H}\bold{T}
\\
&-\frac{1}{M}\Tr [ \bold{D} {\bold{T}^{T} \overline{\bold{A}}}\widetilde{\bold{D}}\widetilde{\bold{R}}^{2}\bold{A}^{H} \bold{T} 
(\bold{I}+\widetilde{\delta}\bold{D}+ \omega\widetilde{\delta}'_{\omega}\bold{D} 
\\
&
-\delta'_{\omega} {\bold{A}}\widetilde{\bold{R}}^{2}\widetilde{\bold{D}} \bold{A}^{H} )\bold{T}]
-\frac{2\delta' _{\omega}}{M} \Tr \bold{D}{\bold{T}}^{T}\overline{\bold{A}} \widetilde{\bold{R}}^{3} \widetilde{\bold{D}}^2\bold{A}^{H}\bold{T}
\\
&=-2(\widetilde{\delta}+\omega\widetilde{\delta}'_{\omega})F_{T}(\bold{D}\bold{T}\bold{D})
-2F_{T}(\bold{T}\bold{D})
\\
&-2\omega\delta'_{\omega} \underline{\widetilde{F}}_{T}(\widetilde{\bold{D}} \widetilde{\bold{T}} \widetilde{\bold{D}}). 
\end{aligned}
\end{equation}
Similarly, we have
\begin{equation}
\label{ft_term_1}
\begin{aligned}
\frac{\mathrm{d} \underline{{{F}}}_{T}}{\mathrm{d} \omega}
=&-2(\widetilde{\delta}+\omega\widetilde{\delta}'_{\omega})\underline{F}_{T}(\bold{D}\bold{T}\bold{D})
-2\underline{F}_{T}(\bold{D}\bold{T})
\\
&
-2\omega\delta'_{\omega} {{\widetilde{F}}}_{T}(\widetilde{\bold{D}} \widetilde{\bold{T}} \widetilde{\bold{D}}). 
\end{aligned}
\end{equation}

Given ${\bold{R}}=\widetilde{\delta}{\bold{D}} {\bold{R}}^{2}+{\bold{R}}^{2}$, $\widetilde{\gamma}_{T}$ can be decomposed as
\begin{equation}
\begin{aligned}
\label{gamma_de}
&\widetilde{\gamma}_{T}=\frac{1}{M}\Tr \widetilde{\bold{D}}\widetilde{\bold{T}}[\omega(\bold{I}+\delta\widetilde{\bold{D}})+\bold{A}^{H}\bold{R}\bold{A} ) ] \widetilde{\bold{T}}\widetilde{\bold{D}}{\widetilde{\bold{T}}}^{T}
\\
&=\omega\widetilde{\gamma}_{T}(\widetilde{\bold{D}}\widetilde{\bold{T}})
+  \omega\delta\widetilde{\gamma}_{T}(\widetilde{\bold{D}}\widetilde{\bold{T}}\widetilde{\bold{D}})
+ \widetilde{\delta} \widetilde{\mathcal{F}}_{T}({\bold{D}})+ \widetilde{\mathcal{F}}_{T}({\bold{I}}).
\end{aligned}
\end{equation}
Thus, we have
\begin{equation}
\label{G_term}
\begin{aligned}
&\frac{\mathrm{d}{\omega^{2}{\gamma}_{T}{\widetilde{\gamma}}_{T}}}{\mathrm{d} \omega}
=
-\frac{2\omega^{2}}{M}[\Tr \bold{D}\bold{T}^{2}\bold{D}{\bold{T}}^{T}
\\
&+(\widetilde{\delta}+\omega\widetilde{\delta}'_{\omega})  \Tr \bold{D}\bold{T} \bold{D}\bold{T} \bold{D}{\bold{T}}^{T}
\\
&-\delta'_\omega \Tr \bold{D}\bold{T} \bold{A}(\bold{I}+\delta \widetilde{\bold{D}})^{-2}\widetilde{\bold{D}}\bold{A}^{H} \bold{T} \bold{D}{\bold{T}^{T} }
  ]\widetilde{\gamma}_{T}
  \\
 & -\frac{2\omega^{2}}{M}[\Tr \widetilde{\bold{D}}\widetilde{\bold{T}}^{2}\widetilde{\bold{D}}{\widetilde{\bold{T}}}^{T}
+({\delta}+\omega {\delta}'_{\omega} )  \Tr \widetilde{\bold{D}}\widetilde{\bold{T}} \widetilde{\bold{D}}\widetilde{\bold{T}}\widetilde{ \bold{D}}{\widetilde{\bold{T}}}^{T}
\\
&-\widetilde{\delta}'_\omega \Tr \widetilde{\bold{D}}\widetilde{\bold{T}} \bold{A}(\bold{I}+\widetilde{\delta}{\bold{D}})^{-2}  {\bold{D}}\bold{A}^{H} \widetilde{\bold{T}} \widetilde{\bold{D}}{\widetilde{\bold{T}} }^{T}
  ]{\gamma}_{T}
 \\
 & +2\omega {\gamma}_{T}{\widetilde{\gamma}}_{T}
 \overset{\text{a}}{=}  - 2(\widetilde{\delta}+\omega\widetilde{\delta}'_{\omega})
 [\omega^2\gamma_{T}(\bold{D}\bold{T}\bold{D})\widetilde{\gamma}_{T}
\\
&
 -\omega \widetilde{\mathcal{F}}_{T}({\bold{D}}) \gamma_{T}]
 -2\omega \delta'_\omega[\omega^2\widetilde{\gamma}_{T}(\widetilde{\bold{D}}\widetilde{\bold{T}}\widetilde{\bold{D}} )\gamma_{T}
 \\
 &- \omega{\mathcal{F}}_{T}(\widetilde{\bold{D}}) \widetilde{\gamma}_{T}]
 -2 [\omega^2 \gamma_{T}(\bold{T}\bold{D})\widetilde{\gamma}_{T}- \omega\widetilde{\mathcal{F}}_{T}({\bold{I}})\gamma_{T}],
\end{aligned}
\end{equation}
where step $a$ follows from~(\ref{gamma_de}).

Notice that $F_{T}={\widetilde{F}_{T}}$. By~(\ref{ft_term}),~(\ref{ft_term_1}) and~(\ref{G_term}), we can conclude that 
\begin{equation}
\begin{aligned}
&\frac{\mathrm{d} (1-\vartheta F_{T})(1-\overline{\vartheta} \underline{F}_{T}) }{\mathrm{d}\omega}-\frac{|\vartheta|^2\mathrm{d}{\omega^{2}{\gamma}_{T}{\widetilde{\gamma}}_{T}}}{\mathrm{d} \omega}
\\
&=-(1-\vartheta F_{T})\frac{\overline{\vartheta}\mathrm{d} \underline{{{F}}}_{T}}{\mathrm{d} \omega} -(1-\overline{\vartheta} \underline{F}_{T}) \frac{\vartheta \mathrm{d} {F}_{T}}{\mathrm{d} \omega}
\\
&
-\frac{|\vartheta|^2\mathrm{d}{\omega^{2}{\gamma}_{T}{\widetilde{\gamma}}_{T}}}{\mathrm{d} \omega}
=2\Delta_{T}[ (\widetilde{\delta}+\omega\widetilde{\delta}'_{\omega})\mathcal{Y}_{\vartheta}(\bold{D})
\\
&
+\omega\delta'_{\omega} \widetilde{\mathcal{Y}}_{\vartheta}(\widetilde{\bold{D}}) 
+\mathcal{Y}_{\vartheta}(\bold{I}) ]
=2\Delta_{T} \mathcal{B}_{\vartheta}(z),
\end{aligned}
\end{equation}
where $\mathcal{Y}_{\vartheta}(\cdot)$, $\widetilde{\mathcal{Y}}_{\vartheta}(\cdot)$ are given in~(\ref{y_u_eq}). This implies that $\frac{-1}{2}\frac{\mathrm{d} \log (\Delta_{T})}{\mathrm{d} z}=\mathcal{B}_{\vartheta}(z)$.

Next, we will turn to the term $\mathcal{B}_{\kappa}(z)$ by considering the derivative of $\beta=\frac{\omega^2}{2}\Tr\bold{D}^2 \bold{S}^2\Tr\widetilde{\bold{D}}^2 \widetilde{\bold{S}}^2$ with respect to $\omega$.
By similar techniques in handling $\omega^2 \gamma_{T}\widetilde{\gamma}_{T}$, we have
\begin{equation}
\begin{aligned}
\frac{\mathrm{d} \beta}{\mathrm{d} \omega}
&=
\omega \Tr{\bold{D}}^2 {\bold{S}}^2 \Tr\widetilde{\bold{D}}^2 \widetilde{\bold{S}}^2 
-\omega^2\Tr\widetilde{\bold{D}}^2 \widetilde{\bold{S}}^2 
 \Tr[\bold{D}^2 \bold{S} 
  \times 
  \\
  & \mathrm{diag}(\bold{T} ( \bold{I}+\widetilde{\delta}\bold{D}+ \omega \widetilde{\delta}'_\omega\bold{D} 
 -\delta'_\omega \bold{A}\widetilde{\bold{R}}^{2}\widetilde{\bold{D}}  \bold{A}^{H} )\bold{T} )]
 \\
 & -\omega^2\Tr{\bold{D}}^2 {\bold{S}}^2 
 \Tr[\widetilde{\bold{D}}^2 \widetilde{\bold{S}}
  \times 
  \\
 & 
  \mathrm{diag}(\widetilde{\bold{T}} ( \bold{I}+{\delta}\widetilde{\bold{D}}+ \omega {\delta}'_{\omega}\widetilde{\bold{D}} 
 -\widetilde{\delta}'_{\omega} \bold{A}^{H}\bold{R}^{2}{\bold{D}}  \bold{A} )\widetilde{\bold{T}} )]
 \\
 & = -(\widetilde{\delta}+ \omega\widetilde{\delta}'_{\omega})[\omega^2\Tr\widetilde{\bold{D}}^2 \widetilde{\bold{S}}^2  \Tr\bold{D}^2 \bold{S} 
\bold{T}{\bold{D}}\bold{T} 
\\
&
 -\omega\Tr{\bold{D}}^2 {\bold{S}}^2 \Tr \widetilde{\bold{D}}^2 \widetilde{\bold{S}}
(\bold{I}+\delta\widetilde{\bold{D}})^{-2} \bold{A}^{H}  \bold{T}\bold{D}\bold{T}\bold{A}]  
\\
 & -  \omega {\delta}'_{\omega}[\omega^2\Tr{\bold{D}}^2 {\bold{S}}^2  \Tr\widetilde{\bold{D}}^2\widetilde{ \bold{S} }
\widetilde{\bold{T}}\widetilde{\bold{D}}\widetilde{\bold{T}} 
\\
&
-\omega\Tr\widetilde{\bold{D}}^2 \widetilde{\bold{S}}^2 \Tr {\bold{D}}^2  \bold{S}
(\bold{I}+\widetilde{\delta}{\bold{D}})^{-2} \bold{A} \widetilde{\bold{T}}\widetilde{\bold{D}}\widetilde{\bold{T}}\bold{A}^{H} ]
\\
&-[\omega^2\Tr\widetilde{\bold{D}}^2 \widetilde{\bold{S}}^2 
 \Tr\bold{D}^2 \bold{S}{\bold{T}}^2
 \\
 &
 -\omega \Tr{\bold{D}}^2 {\bold{S}}^2 \Tr \widetilde{\bold{D}}^2 \widetilde{\bold{S}}
(\bold{I}+\delta\widetilde{\bold{D}})^{-2} \bold{A}^{H}  \bold{T}^2\bold{A}].
\end{aligned}
\end{equation}
As a result, we have
\begin{equation}
\begin{aligned}
\mathcal{B}_{\kappa}(z)=\frac{\kappa \mathrm{d} \left(z^2  \Tr\bold{D}^2 \bold{S}^2\Tr\widetilde{\bold{D}}^2 \widetilde{\bold{S}}^2\right)}{2M^2 \mathrm{d}z}. 
\end{aligned}
\end{equation}
Therefore, the bias can be expressed as
\begin{equation}
\begin{aligned}
&\mathcal{M}(z)\xlongrightarrow[]{M\rightarrow \infty}\mathcal{B}_{\vartheta}(z)+\mathcal{B}_{\kappa}(z)
\\
&=
 \frac{1}{2} \frac{\mathrm{d} ( -\log\Delta_{T}
+\frac{ \kappa z^2}{M^2}\Tr\bold{D}^2 \bold{S}^2\Tr\widetilde{\bold{D}}^2 \widetilde{\bold{S}}^2  )
}{\mathrm{d}z}.
\end{aligned}
\end{equation}
\end{IEEEproof}

\section{proof of Proposition~\ref{linear_p}}
\label{lss_proof}
\begin{IEEEproof} First we show how to obtain $u_{+}$. $u_{+}$ comes from the fact 
\begin{equation}
\bold{H}\bold{H}^{H}\le 2(\bold{A}\bold{A}^{H}+\frac{1}{M}\bold{D}^{\frac{1}{2}}\bold{X}\widetilde{\bold{D}}\bold{X}^{H}\bold{D}^{\frac{1}{2}}),
\end{equation}
and 
\begin{equation}
\begin{aligned}
\|\frac{1}{M}\bold{D}^{\frac{1}{2}}\bold{X}\widetilde{\bold{D}}\bold{X}^{H}\bold{D}^{\frac{1}{2}})\| 
&
\le d_{max} \widetilde{d}_{max} \|\frac{1}{M}\bold{X}\bold{X}^{H}  \|  
\\
&\overset{a}{\le} d_{max} \widetilde{d}_{max} (1+\sqrt{c})^2,
\end{aligned}
\end{equation}
where the inequality $a$ holds true almost surely since the largest eigenvalue of $\frac{1}{M}\bold{X}\bold{X}^{H}$ is upper bounded by $(1+\sqrt{c})^2$ almost surely~\cite{bai1988note}. Therefore, the eigenvalues of $\bold{H}\bold{H}^{H}$ are contained in $[0, u_{+}]$ almost surely. Since $f$ is analytic in the region which contains $[0, u_{+}]$, we have 
\begin{equation}
\begin{aligned}
&\E \Tr f(\bold{H}\bold{H}^{H})=\E \sum_{i=1}^{N}f(\lambda_{i})
\\
&\overset{b}{=}\frac{1}{2\pi \jmath}\E  \sum_{i=1}^{N} \int_{\mathcal{C}_{i}} \frac{f(z)}{z-\lambda_{i}} \mathrm{d}z
\\
&=\frac{-1}{2\pi \jmath}\int_{\mathcal{C}} f(z) \E \Tr(\bold{H}\bold{H}^{H}-z\bold{I})^{-1}\mathrm{d}z
\\
&=\frac{-1}{2\pi \jmath}\int_{\mathcal{C}} f(z) \E \Tr\bold{Q}(z)\mathrm{d}z
\\
&=\frac{-1}{2\pi \jmath}\int_{\mathcal{C}} f(z) [\Tr\bold{T}(z)+\mathcal{M}(z) ] \mathrm{d}z
\\
&\xlongrightarrow[]{M \rightarrow \infty}  \mathcal{V}_{f}+\mathcal{B}_{f},
\end{aligned}
\end{equation}
where step $b$ follows from the Cauchy's integral formula and $\mathcal{C}_{i}$ is the contour containing $\lambda_{i}$ in the positive direction.
\end{IEEEproof}
\end{appendices}

%



%
%
\section*{Acknowledgment}
The authors would like to thank all reviewers and the editor for their time and efforts in reviewing our manuscript and their constructive comments.

\ifCLASSOPTIONcaptionsoff
  \newpage
\fi



\bibliographystyle{IEEEtran}
\bibliography{IEEEabrv,ref}

\begin{thebibliography}{10}
\providecommand{\url}[1]{#1}
\csname url@samestyle\endcsname
\providecommand{\newblock}{\relax}
\providecommand{\bibinfo}[2]{#2}
\providecommand{\BIBentrySTDinterwordspacing}{\spaceskip=0pt\relax}
\providecommand{\BIBentryALTinterwordstretchfactor}{4}
\providecommand{\BIBentryALTinterwordspacing}{\spaceskip=\fontdimen2\font plus
\BIBentryALTinterwordstretchfactor\fontdimen3\font minus
  \fontdimen4\font\relax}
\providecommand{\BIBforeignlanguage}[2]{{%
\expandafter\ifx\csname l@#1\endcsname\relax
\typeout{** WARNING: IEEEtran.bst: No hyphenation pattern has been}%
\typeout{** loaded for the language `#1'. Using the pattern for}%
\typeout{** the default language instead.}%
\else
\language=\csname l@#1\endcsname
\fi
#2}}
\providecommand{\BIBdecl}{\relax}
\BIBdecl

\bibitem{kamath2005asymptotic}
M.~A. Kamath and B.~L. Hughes, ``The asymptotic capacity of multiple-antenna
  {R}ayleigh-fading channels,'' \emph{{IEEE} Trans. Inf. Theory}, vol.~51,
  no.~12, pp. 4325--4333, Dec. 2005.

\bibitem{hachem2007deterministic}
W.~Hachem, P.~Loubaton, J.~Najim \emph{et~al.}, ``Deterministic equivalents for
  certain functionals of large random matrices,'' \emph{Ann. App. Probab.},
  vol.~17, no.~3, pp. 875--930, Jun. 2007.

\bibitem{wen2012deterministic}
C.-K. Wen, G.~Pan, K.-K. Wong, M.~Guo, and J.-C. Chen, ``A deterministic
  equivalent for the analysis of non-{G}aussian correlated {MIMO} multiple
  access channels,'' \emph{{IEEE} Trans. Inf. Theory}, vol.~59, no.~1, pp.
  329--352, Jan. 2012.

\bibitem{zhang2013capacity}
J.~Zhang, C.-K. Wen, S.~Jin, X.~Gao, and K.-K. Wong, ``On capacity of
  large-scale {MIMO} multiple access channels with distributed sets of
  correlated antennas,'' \emph{{IEEE} J. Sel. Areas Commun.}, vol.~31, no.~2,
  pp. 133--148, Feb. 2013.

\bibitem{zhang2021large}
X.~Zhang, X.~Yu, S.~Song, and K.~B. Letaief, ``{IRS}-aided {MIMO} systems over
  double-scattering channels: {Impact} of channel rank deficiency,'' accepted
  to \emph{{IEEE} Wireless Commun. Netw. Conf. (WCNC)}, Austin, TX, USA, April.
  2022.

\bibitem{hachem2008new}
W.~Hachem, O.~Khorunzhiy, P.~Loubaton, J.~Najim, and L.~Pastur, ``A new
  approach for mutual information analysis of large dimensional multi-antenna
  channels,'' \emph{{IEEE} Trans. Inf. Theory}, vol.~54, no.~9, pp. 3987--4004,
  Sep. 2008.

\bibitem{pastur2005simple}
L.~A. Pastur, ``A simple approach to the global regime of {Gaussian} ensembles
  of random matrices,'' \emph{Ukrainian Math. J.}, vol.~57, no.~6, pp.
  936--966, Jun. 2005.

\bibitem{dumont2010capacity}
J.~Dumont, W.~Hachem, S.~Lasaulce, P.~Loubaton, and J.~Najim, ``On the capacity
  achieving covariance matrix for {R}ician {MIMO} channels: an asymptotic
  approach,'' \emph{{IEEE} Trans. Inf. Theory}, vol.~56, no.~3, pp. 1048--1069,
  Mar. 2010.

\bibitem{hachem2012clt}
W.~Hachem, M.~Kharouf, J.~Najim, and J.~W. Silverstein, ``A {CLT} for
  information-theoretic statistics of non-centered {G}ram random matrices,''
  \emph{Random Matrices: Theory. Appl.}, vol.~1, no.~2, p. 1150010, Dec. 2012.

\bibitem{bai2008clt}
Z.~Bai and J.~W. Silverstein, ``{CLT} for linear spectral statistics of
  large-dimensional sample covariance matrices,'' \emph{Ann. Probab.}, vol.~32,
  no.~1, pp. 553--605, Jan. 2004.

\bibitem{fraidenraich2007mimo}
G.~Fraidenraich, O.~L{\'e}v{\^e}que, and J.~M. Cioffi, ``On the {MIMO} channel
  capacity for the dual and asymptotic cases over {H}oyt channels,''
  \emph{{IEEE} Commun. Lett.}, vol.~11, no.~1, pp. 31--33, Jan. 2007.

\bibitem{kammoun2010fluctuations}
A.~Kammoun, M.~Kharouf, W.~Hachem, J.~Najim, and A.~El~Kharroubi, ``On the
  fluctuations of the mutual information for non centered {MIMO} channels: The
  non {Gaussian} case,'' in \emph{Proc. {IEEE} Signal Process. Adv. Wireless
  Commun. Wkshps. (SPAWC Wkshps)}, Marrakech, Morocco, Jun. 2010, pp. 1--5.

\bibitem{bao2015asymptotic}
Z.~Bao, G.~Pan, and W.~Zhou, ``Asymptotic mutual information statistics of
  {MIMO} channels and {CLT} of sample covariance matrices,'' \emph{{IEEE}
  Trans. Inf. Theory}, vol.~61, no.~6, pp. 3413--3426, Jun. 2015.

\bibitem{hu2019central}
J.~Hu, W.~Li, and W.~Zhou, ``Central limit theorem for mutual information of
  large {MIMO} systems with elliptically correlated channels,'' \emph{{IEEE}
  Trans. Inf. Theory}, vol.~65, no.~11, pp. 7168--7180, Nov. 2019.

\bibitem{kammoun2012fluctuations}
A.~Kammoun, M.~Kharouf, R.~Couillet, J.~Najim, and M.~Debbah, ``On the
  fluctuations of the {SINR} at the output of the {W}iener filter for non
  centered channels: The non {Gaussian} case,'' in \emph{Proc. {IEEE} Int.
  Conf. Acoust., Speech and Signal Process. (ICASSP)}, Kyoto, Japan, Mar. 2012,
  pp. 3173--3176.

\bibitem{najim2016gaussian}
J.~Najim and J.~Yao, ``{Gaussian} fluctuations for linear spectral statistics
  of large random covariance matrices,'' \emph{Ann. App. Probab.}, vol.~26,
  no.~3, pp. 1837--1887, Jun. 2016.

\bibitem{banna2020clt}
M.~Banna, J.~Najim, and J.~Yao, ``A {CLT} for linear spectral statistics of
  large random information-plus-noise matrices,'' \emph{Stoch Process Their
  Appl.}, vol. 130, no.~4, pp. 2250--2281, Apr. 2020.

\bibitem{levin2011multi}
G.~Levin and S.~Loyka, ``From multi-keyholes to measure of correlation and
  power imbalance in {MIMO} channels: {Outage} capacity analysis,''
  \emph{{IEEE} Trans. Inf. Theory}, vol.~57, no.~6, pp. 3515--3529, May. 2011.

\bibitem{hoydis2010fluctuations}
J.~Hoydis, J.~Najim, R.~Couillet, and M.~Debbah, ``Fluctuations of the mutual
  information in large distributed antenna systems with colored noise,'' in
  \emph{Proc. 48th Annu. Allerton Conf. Communication, Control Computing
  (Allerton’10)}, Urbana-Champaign, IL, USA, Sep. 2010, pp. 240--245.

\bibitem{kammoun2009central}
A.~Kammoun, M.~Kharouf, W.~Hachem, and J.~Najim, ``A central limit theorem for
  the {SINR} at the {LMMSE} estimator output for large-dimensional signals,''
  \emph{{IEEE} Trans. Inf. Theory}, vol.~55, no.~11, pp. 5048--5063, Nov. 2009.

\bibitem{hachem2013bilinear}
W.~Hachem, P.~Loubaton, J.~Najim, and P.~Vallet, ``On bilinear forms based on
  the resolvent of large random matrices,'' in \emph{Annales de l'{IHP}
  Probabilit{\'e}s et statistiques}, vol.~49, no.~1, Feb. 2013, pp. 36--63.

\bibitem{kammoun2019asymptotic}
A.~Kammoun, L.~Sanguinetti, M.~Debbah, and M.-S. Alouini, ``Asymptotic analysis
  of {RZF} in large-scale {MU-MIMO} systems over {Rician} channels,''
  \emph{{IEEE} Trans. Inf. Theory}, vol.~65, no.~11, pp. 7268--7286, Nov. 2019.

\bibitem{sanguinetti2018theoretical}
L.~Sanguinetti, A.~Kammoun, and M.~Debbah, ``Theoretical performance limits of
  massive {MIMO} with uncorrelated {R}ician fading channels,'' \emph{{IEEE}
  Trans. Commun.}, vol.~67, no.~3, pp. 1939--1955, Mar. 2018.

\bibitem{jin2007ergodic}
S.~Jin, X.~Gao, and X.~You, ``On the ergodic capacity of rank-$1$
  {R}icean-fading {MIMO} channels,'' \emph{{IEEE} Trans. Inf. Theory}, vol.~53,
  no.~2, pp. 502--517, Feb. 2007.

\bibitem{bolcskei2003impact}
H.~Bolcskei, M.~Borgmann, and A.~J. Paulraj, ``Impact of the propagation
  environment on the performance of space-frequency coded {MIMO-OFDM},''
  \emph{{IEEE} J. Sel. Areas Commun.}, vol.~21, no.~3, pp. 427--439, Apr. 2003.

\bibitem{bohagen2009spherical}
F.~Bohagen, P.~Orten, and G.~E. Oien, ``On spherical vs. plane wave modeling of
  line-of-sight {MIMO} channels,'' \emph{{IEEE} Trans. Commun.}, vol.~57,
  no.~3, pp. 841--849, Mar. 2009.

\bibitem{dumont2006cth09}
J.~Dumont, P.~Loubaton, and S.~Lasaulce, ``On the capacity achieving transmit
  covariance matrices of {MIMO} correlated {Rician} channels: A large system
  approach,'' in \emph{Proc. {IEEE} Global Commun. Conf. (GLOBECOM)}, San
  Francisco, CA, USA, Apr. 2006, pp. 1--6.

\bibitem{girko2001theory}
V.~L. Girko, \emph{Theory of stochastic canonical equations}.\hskip 1em plus
  0.5em minus 0.4em\relax Dordrecht, The Netherland: Kluwer, 2001, vol. 535.

\bibitem{horn2012matrix}
R.~A. Horn and C.~R. Johnson, \emph{Matrix analysis}.\hskip 1em plus 0.5em
  minus 0.4em\relax Cambridge, U.K: Cambridge University Press, 2012.

\bibitem{dozier2007analysis}
R.~B. Dozier and J.~W. Silverstein, ``Analysis of the limiting spectral
  distribution of large dimensional information-plus-noise type matrices,''
  \emph{J. Multivariate Anal.}, vol.~98, no.~6, pp. 1099--1122, Jan. 2007.

\bibitem{kumar2010random}
S.~Kumar and A.~Pandey, ``Random matrix model for {N}akagami--hoyt fading,''
  \emph{{IEEE} Trans. Inf. Theory}, vol.~56, no.~5, pp. 2360--2372, May. 2010.

\bibitem{adali2011complex}
T.~Adali, P.~J. Schreier, and L.~L. Scharf, ``Complex-valued signal processing:
  {T}he proper way to deal with impropriety,'' \emph{{IEEE} Trans. Signal
  Process.}, vol.~59, no.~11, pp. 5101--5125, Nov. 2011.

\bibitem{silverstein1995empirical}
J.~W. Silverstein and Z.~Bai, ``On the empirical distribution of eigenvalues of
  a class of large dimensional random matrices,'' \emph{J. Multivariate Anal.},
  vol.~54, no.~2, pp. 175--192, Aug. 1995.

\bibitem{bai1988note}
Z.~D. Bai, J.~W. Silverstein, and Y.~Q. Yin, ``A note on the largest eigenvalue
  of a large dimensional sample covariance matrix,'' \emph{J. Multivariate
  Anal.}, vol.~26, no.~2, pp. 166--168, Aug. 1988.

\end{thebibliography}
%

%
%

%

%
\begin{IEEEbiographynophoto}{Xin Zhang} (Graduate Student Member, IEEE) received the B.Eng. degree in information engineering from Beijing University of Posts and Telecommunications (BUPT) in 2015, and the master's degree in electronic engineering from Tsinghua University in 2018. He is currently pursuing the Ph.D. degree in the Department of Electronic and Computer Engineering (ECE) at the Hong Kong University of Science and Technology (HKUST). His research interests include random matrix theory, information theory, and their applications in signal processing, communications, and learning.
\end{IEEEbiographynophoto}

\begin{IEEEbiographynophoto}{S.H. Song} (Senior Member, IEEE) is now an Assistant Professor jointly appointed by the Division of Integrative Systems and Design (ISD) and the Department of Electronic and Computer Engineering (ECE) at the Hong Kong University of Science and Technology (HKUST). His research is primarily in the areas of Wireless Communications and Machine Learning with current focus on Distributed Intelligence (Federated Learning), Machine Learning for Communications (Model and Data-driven Approaches), and Integrated Sensing and Communication.  He was named the Exemplary Reviewer for IEEE Communications Letter.
 
He is also interested in the research on Engineering Education and is now serving as an Associate Editor for the IEEE Transactions on Education. He has won several teaching awards at HKUST, including the Michael G. Gale Medal for Distinguished Teaching in 2018, the Best Ten Lecturers in 2013, 2015, and 2017, the School of Engineering Distinguished Teaching Award in 2012, the Teachers I Like Award in 2013, 2015, 2016, and 2017, and the MSc (Telecom) Teaching Excellent Appreciation Award for 2020-21. Dr. Song was one of the honorees of the Third Faculty Recognition at HKUST in 2021.
\end{IEEEbiographynophoto}






\end{document}